\documentclass[twocolumn,10pt,a4paper,romanappendices]{IEEEtran-v17} 
\let\labelindent\relax 
\usepackage{hhline}

\usepackage{amssymb}
\usepackage{amsfonts}
\usepackage{mathrsfs}
\usepackage{xspace}
\usepackage{bm}
\usepackage{upgreek}

\newcommand{\safemath}[2]{\newcommand{#1}{\ensuremath{#2}\xspace}}
\def\amsbb{\use@mathgroup \M@U \symAMSb}

\safemath{\bma}{\mathbf{a}}
\safemath{\bmb}{\mathbf{b}}
\safemath{\bmc}{\mathbf{c}}
\safemath{\bmd}{\mathbf{d}}
\safemath{\bme}{\mathbf{e}}
\safemath{\bmf}{\mathbf{f}}
\safemath{\bmg}{\mathbf{g}}
\safemath{\bmh}{\mathbf{h}}
\safemath{\bmi}{\mathbf{i}}
\safemath{\bmj}{\mathbf{j}}
\safemath{\bmk}{\mathbf{k}}
\safemath{\bml}{\mathbf{l}}
\safemath{\bmm}{\mathbf{m}}
\safemath{\bmn}{\mathbf{n}}
\safemath{\bmo}{\mathbf{o}}
\safemath{\bmp}{\mathbf{p}}
\safemath{\bmq}{\mathbf{q}}
\safemath{\bmr}{\mathbf{r}}
\safemath{\bms}{\mathbf{s}}
\safemath{\bmt}{\mathbf{t}}
\safemath{\bmu}{\mathbf{u}}
\safemath{\bmv}{\mathbf{v}}
\safemath{\bmw}{\mathbf{w}}
\safemath{\bmx}{\mathbf{x}}
\safemath{\bmy}{\mathbf{y}}
\safemath{\bmz}{\mathbf{z}}
\safemath{\bmzero}{\mathbf{0}}
\safemath{\bmone}{\mathbf{1}}

\bmdefine{\biad}{a}
\bmdefine{\bibd}{b}
\bmdefine{\bicd}{c}
\bmdefine{\bidd}{d}
\bmdefine{\bied}{e}
\bmdefine{\bifd}{f}
\bmdefine{\bigd}{g}
\bmdefine{\bihd}{h}
\bmdefine{\biid}{i}
\bmdefine{\bijd}{j}
\bmdefine{\bikd}{k}
\bmdefine{\bild}{l}
\bmdefine{\bimd}{m}
\bmdefine{\bind}{n}
\bmdefine{\biod}{o}
\bmdefine{\bipd}{p}
\bmdefine{\biqd}{q}
\bmdefine{\bird}{r}
\bmdefine{\bisd}{s}
\bmdefine{\bitd}{t}
\bmdefine{\biud}{u}
\bmdefine{\bivd}{v}
\bmdefine{\biwd}{w}
\bmdefine{\bixd}{x}
\bmdefine{\biyd}{y}
\bmdefine{\bizd}{z}

\bmdefine{\bixid}{\xi}
\bmdefine{\bilambdad}{\lambda}
\bmdefine{\bimud}{\mu}
\bmdefine{\bithetad}{\theta}
\bmdefine{\biphid}{\phi}
\bmdefine{\vecalpha}{\alpha}                 

\safemath{\bmia}{\biad}
\safemath{\bmib}{\bibd}
\safemath{\bmic}{\bicd}
\safemath{\bmid}{\bidd}
\safemath{\bmie}{\bied}
\safemath{\bmif}{\bifd}
\safemath{\bmig}{\bigd}
\safemath{\bmih}{\bihd}
\safemath{\bmii}{\biid}
\safemath{\bmij}{\bijd}
\safemath{\bmik}{\bikd}
\safemath{\bmil}{\bild}
\safemath{\bmim}{\bimd}
\safemath{\bmin}{\bind}
\safemath{\bmio}{\biod}
\safemath{\bmip}{\bipd}
\safemath{\bmiq}{\biqd}
\safemath{\bmir}{\bird}
\safemath{\bmis}{\bisd}
\safemath{\bmit}{\bitd}
\safemath{\bmiu}{\biud}
\safemath{\bmiv}{\bivd}
\safemath{\bmiw}{\biwd}
\safemath{\bmix}{\bixd}
\safemath{\bmiy}{\biyd}
\safemath{\bmiz}{\bizd}

\safemath{\bmxi}{\bixid}
\safemath{\bmlambda}{\bilambdad}
\safemath{\bmmu}{\bimud}
\safemath{\bmtheta}{\bithetad}
\safemath{\bmphi}{\biphid}

\safemath{\bA}{\mathbf{A}}
\safemath{\bB}{\mathbf{B}}
\safemath{\bC}{\mathbf{C}}
\safemath{\bD}{\mathbf{D}}
\safemath{\bE}{\mathbf{E}}
\safemath{\bF}{\mathbf{F}}
\safemath{\bG}{\mathbf{G}}
\safemath{\bH}{\mathbf{H}}
\safemath{\bI}{\mathbf{I}}
\safemath{\bJ}{\mathbf{J}}
\safemath{\bK}{\mathbf{K}}
\safemath{\bL}{\mathbf{L}}
\safemath{\bM}{\mathbf{M}}
\safemath{\bN}{\mathbf{N}}
\safemath{\bO}{\mathbf{O}}
\safemath{\bP}{\mathbf{P}}
\safemath{\bQ}{\mathbf{Q}}
\safemath{\bR}{\mathbf{R}}
\safemath{\bS}{\mathbf{S}}
\safemath{\bT}{\mathbf{T}}
\safemath{\bU}{\mathbf{U}}
\safemath{\bV}{\mathbf{V}}
\safemath{\bW}{\mathbf{W}}
\safemath{\bX}{\mathbf{X}}
\safemath{\bY}{\mathbf{Y}}
\safemath{\bZ}{\mathbf{Z}}

\safemath{\bZero}{\mathbf{0}}
\safemath{\bOne}{\mathbf{1}}
\safemath{\bDelta}{\mathbf{\Delta}}
\safemath{\bLambda}{\mathbf{\UpLambda}}
\safemath{\bPhi}{\mathbf{\Upphi}}
\safemath{\bSigma}{\mathbf{\Upsigma}}
\safemath{\bOmega}{\mathbf{\Upomega}}
\safemath{\bTheta}{\mathbf{\Uptheta}}

\bmdefine{\biAd}{A}
\bmdefine{\biBd}{B}
\bmdefine{\biCd}{C}
\bmdefine{\biDd}{D}
\bmdefine{\biEd}{E}
\bmdefine{\biFd}{F}
\bmdefine{\biGd}{G}
\bmdefine{\biHd}{H}
\bmdefine{\biId}{I}
\bmdefine{\biJd}{J}
\bmdefine{\biKd}{K}
\bmdefine{\biLd}{L}
\bmdefine{\biMd}{M}
\bmdefine{\biOd}{N}
\bmdefine{\biPd}{O}
\bmdefine{\biQd}{P}
\bmdefine{\biRd}{R}
\bmdefine{\biSd}{S}
\bmdefine{\biTd}{T}
\bmdefine{\biUd}{U}
\bmdefine{\biVd}{V}
\bmdefine{\biWd}{W}
\bmdefine{\biXd}{X}
\bmdefine{\biYd}{Y}
\bmdefine{\biZd}{Z}

\bmdefine{\biDelta}{\Delta}
\bmdefine{\biLambda}{\Lambda}
\bmdefine{\biPhi}{\Phi}
\bmdefine{\biSigma}{\Sigma}
\bmdefine{\biOmega}{\Omega}
\bmdefine{\biTheta}{\Theta}

\safemath{\bimA}{\biAd}
\safemath{\bimB}{\biBd}
\safemath{\bimC}{\biCd}
\safemath{\bimD}{\biDd}
\safemath{\bimE}{\biEd}
\safemath{\bimF}{\biFd}
\safemath{\bimG}{\biGd}
\safemath{\bimH}{\biHd}
\safemath{\bimI}{\biId}
\safemath{\bimJ}{\biJd}
\safemath{\bimK}{\biKd}
\safemath{\bimL}{\biLd}
\safemath{\bimM}{\biMd}
\safemath{\bimN}{\biNd}
\safemath{\bimO}{\biOd}
\safemath{\bimP}{\biPd}
\safemath{\bimQ}{\biQd}
\safemath{\bimR}{\biRd}
\safemath{\bimS}{\biSd}
\safemath{\bimT}{\biTd}
\safemath{\bimU}{\biUd}
\safemath{\bimV}{\biVd}
\safemath{\bimW}{\biWd}
\safemath{\bimX}{\biXd}
\safemath{\bimY}{\biYd}
\safemath{\bimZ}{\biZd}

\safemath{\bimDelta}{\biDelta}
\safemath{\bimLambda}{\biLambda}
\safemath{\bimPhi}{\biPhi}
\safemath{\bimSigma}{\biSigma}
\safemath{\bimOmega}{\biOmega}
\safemath{\bimTheta}{\biTheta}

\safemath{\setA}{\mathcal{A}}
\safemath{\setB}{\mathcal{B}}
\safemath{\setC}{\mathcal{C}}
\safemath{\setD}{\mathcal{D}}
\safemath{\setE}{\mathcal{E}}
\safemath{\setF}{\mathcal{F}}
\safemath{\setG}{\mathcal{G}}
\safemath{\setH}{\mathcal{H}}
\safemath{\setI}{\mathcal{I}}
\safemath{\setJ}{\mathcal{J}}
\safemath{\setK}{\mathcal{K}}
\safemath{\setL}{\mathcal{L}}
\safemath{\setM}{\mathcal{M}}
\safemath{\setN}{\mathcal{N}}
\safemath{\setO}{\mathcal{O}}
\safemath{\setP}{\mathcal{P}}
\safemath{\setQ}{\mathcal{Q}}
\safemath{\setR}{\mathcal{R}}
\safemath{\setS}{\mathcal{S}}
\safemath{\setT}{\mathcal{T}}
\safemath{\setU}{\mathcal{U}}
\safemath{\setV}{\mathcal{V}}
\safemath{\setW}{\mathcal{W}}
\safemath{\setX}{\mathcal{X}}
\safemath{\setY}{\mathcal{Y}}
\safemath{\setZ}{\mathcal{Z}}
\safemath{\emptySet}{\varnothing}

\safemath{\colA}{\mathscr{A}}
\safemath{\colB}{\mathscr{B}}
\safemath{\colC}{\mathscr{C}}
\safemath{\colD}{\mathscr{D}}
\safemath{\colE}{\mathscr{E}}
\safemath{\colF}{\mathscr{F}}
\safemath{\colG}{\mathscr{G}}
\safemath{\colH}{\mathscr{H}}
\safemath{\colI}{\mathscr{I}}
\safemath{\colJ}{\mathscr{J}}
\safemath{\colK}{\mathscr{K}}
\safemath{\colL}{\mathscr{L}}
\safemath{\colM}{\mathscr{M}}
\safemath{\colN}{\mathscr{N}}
\safemath{\colO}{\mathscr{O}}
\safemath{\colP}{\mathscr{P}}
\safemath{\colQ}{\mathscr{Q}}
\safemath{\colR}{\mathscr{R}}
\safemath{\colS}{\mathscr{S}}
\safemath{\colT}{\mathscr{T}}
\safemath{\colU}{\mathscr{U}}
\safemath{\colV}{\mathscr{V}}
\safemath{\colW}{\mathscr{W}}
\safemath{\colX}{\mathscr{X}}
\safemath{\colY}{\mathscr{Y}}
\safemath{\colZ}{\mathscr{Z}}

\safemath{\opA}{\mathbb{A}}
\safemath{\opB}{\mathbb{B}}
\safemath{\opC}{\mathbb{C}}
\safemath{\opD}{\mathbb{D}}
\safemath{\opE}{\mathbb{E}}
\safemath{\opF}{\mathbb{F}}
\safemath{\opG}{\mathbb{G}}
\safemath{\opH}{\mathbb{H}}
\safemath{\opI}{\mathbb{I}}
\safemath{\opJ}{\mathbb{J}}
\safemath{\opK}{\mathbb{K}}
\safemath{\opL}{\mathbb{L}}
\safemath{\opM}{\mathbb{M}}
\safemath{\opN}{\mathbb{N}}
\safemath{\opO}{\mathbb{O}}
\safemath{\opP}{\mathbb{P}}
\safemath{\opQ}{\mathbb{Q}}
\safemath{\opR}{\mathbb{R}}
\safemath{\opS}{\mathbb{S}}
\safemath{\opT}{\mathbb{T}}
\safemath{\opU}{\mathbb{U}}
\safemath{\opV}{\mathbb{V}}
\safemath{\opW}{\mathbb{W}}
\safemath{\opX}{\mathbb{X}}
\safemath{\opY}{\mathbb{Y}}
\safemath{\opZ}{\mathbb{Z}}
\safemath{\opZero}{\mathbb{O}}
\safemath{\identityop}{\opI}

\safemath{\veca}{\bma}
\safemath{\vecb}{\bmb}
\safemath{\vecc}{\bmc}
\safemath{\vecd}{\bmd}
\safemath{\vece}{\bme}
\safemath{\vecf}{\bmf}
\safemath{\vecg}{\bmg}
\safemath{\vech}{\bmh}
\safemath{\veci}{\bmi}
\safemath{\vecj}{\bmj}
\safemath{\veck}{\bmk}
\safemath{\vecl}{\bml}
\safemath{\vecm}{\bmm}
\safemath{\vecn}{\bmn}
\safemath{\veco}{\bmo}
\safemath{\vecp}{\bmp}
\safemath{\vecq}{\bmq}
\safemath{\vecr}{\bmr}
\safemath{\vecs}{\bms}
\safemath{\vecu}{\bmu}
\safemath{\vecv}{\bmv}
\safemath{\vecw}{\bmw}
\safemath{\vecx}{\bmx}
\safemath{\vecy}{\bmy}
\safemath{\vecz}{\bmz}

\safemath{\veczero}{\bmzero}
\safemath{\vecone}{\bmone}
\safemath{\vecxi}{\bmxi}
\safemath{\veclambda}{\bmlambda}
\safemath{\vecmu}{\bmmu}
\safemath{\vectheta}{\bmtheta}
\safemath{\vecphi}{\bmphi} 

\safemath{\matA}{\bA}
\safemath{\matB}{\bB}
\safemath{\matC}{\bC}
\safemath{\matD}{\bD}
\safemath{\matE}{\bE}
\safemath{\matF}{\bF}
\safemath{\matG}{\bG}
\safemath{\matH}{\bH}
\safemath{\matI}{\bI}
\safemath{\matJ}{\bJ}
\safemath{\matK}{\bK}
\safemath{\matL}{\bL}
\safemath{\matM}{\bM}
\safemath{\matN}{\bN}
\safemath{\matO}{\bO}
\safemath{\matP}{\bP}
\safemath{\matQ}{\bQ}
\safemath{\matR}{\bR}
\safemath{\matS}{\bS}
\safemath{\matT}{\bT}
\safemath{\matU}{\bU}
\safemath{\matV}{\bV}
\safemath{\matW}{\bW}
\safemath{\matX}{\bX}
\safemath{\matY}{\bY}
\safemath{\matZ}{\bZ}
\safemath{\matzero}{\bmzero}

\safemath{\matDelta}{\bDelta}
\safemath{\matLambda}{\bLambda}
\safemath{\matPhi}{\bPhi}
\safemath{\matSigma}{\bSigma}
\safemath{\matOmega}{\bOmega}
\safemath{\matTheta}{\bTheta}

\safemath{\matidentity}{\matI}
\safemath{\matone}{\matO}

\safemath{\rnda}{A}
\safemath{\rndb}{B}
\safemath{\rndc}{C}
\safemath{\rndd}{D}
\safemath{\rnde}{E}
\safemath{\rndf}{F}
\safemath{\rndg}{G}
\safemath{\rndh}{H}
\safemath{\rndi}{I}
\safemath{\rndj}{J}
\safemath{\rndk}{K}
\safemath{\rndl}{L}
\safemath{\rndm}{M}
\safemath{\rndn}{N}
\safemath{\rndo}{O}
\safemath{\rndp}{P}
\safemath{\rndq}{Q}
\safemath{\rndr}{R}
\safemath{\rnds}{S}
\safemath{\rndt}{T}
\safemath{\rndu}{U}
\safemath{\rndv}{V}
\safemath{\rndw}{W}
\safemath{\rndx}{X}
\safemath{\rndy}{Y}
\safemath{\rndz}{Z}

\safemath{\rveca}{\bimA}
\safemath{\rvecb}{\bimB}
\safemath{\rvecc}{\bimC}
\safemath{\rvecd}{\bimD}
\safemath{\rvece}{\bimE}
\safemath{\rvecf}{\bimF}
\safemath{\rvecg}{\bimG}
\safemath{\rvech}{\bimH}
\safemath{\rveci}{\bimI}
\safemath{\rvecj}{\bimJ}
\safemath{\rveck}{\bimK}
\safemath{\rvecl}{\bimL}
\safemath{\rvecm}{\bimM}
\safemath{\rvecn}{\bimN}
\safemath{\rveco}{\bomO}
\safemath{\rvecp}{\bimP}
\safemath{\rvecq}{\bimQ}
\safemath{\rvecr}{\bimR}
\safemath{\rvecs}{\bimS}
\safemath{\rvect}{\bimT}
\safemath{\rvecu}{\bimU}
\safemath{\rvecv}{\bimV}
\safemath{\rvecw}{\bimW}
\safemath{\rvecx}{\bimX}
\safemath{\rvecy}{\bimY}
\safemath{\rvecz}{\bimZ}

\safemath{\rvecxi}{\bmxi}
\safemath{\rveclambda}{\bmlambda}
\safemath{\rvecmu}{\bmmu}
\safemath{\rvectheta}{\bmtheta}
\safemath{\rvecphi}{\bmphi}

\safemath{\rmatA}{\bimA}
\safemath{\rmatB}{\bimB}
\safemath{\rmatC}{\bimC}
\safemath{\rmatD}{\bimD}
\safemath{\rmatE}{\bimE}
\safemath{\rmatF}{\bimF}
\safemath{\rmatG}{\bimG}
\safemath{\rmatH}{\bimH}
\safemath{\rmatI}{\bimI}
\safemath{\rmatJ}{\bimJ}
\safemath{\rmatK}{\bimK}
\safemath{\rmatL}{\bimL}
\safemath{\rmatM}{\bimM}
\safemath{\rmatN}{\bimN}
\safemath{\rmatO}{\bimO}
\safemath{\rmatP}{\bimP}
\safemath{\rmatQ}{\bimQ}
\safemath{\rmatR}{\bimR}
\safemath{\rmatS}{\bimS}
\safemath{\rmatT}{\bimT}
\safemath{\rmatU}{\bimU}
\safemath{\rmatV}{\bimV}
\safemath{\rmatW}{\bimW}
\safemath{\rmatX}{\bimX}
\safemath{\rmatY}{\bimY}
\safemath{\rmatZ}{\bimZ}

\safemath{\rmatDelta}{\bimDelta}
\safemath{\rmatLambda}{\bimLambda}
\safemath{\rmatPhi}{\bimPhi}
\safemath{\rmatSigma}{\bimSigma}
\safemath{\rmatOmega}{\bimOmega}
\safemath{\rmatTheta}{\bimTheta}

\pdfoutput=1

\newlength{\figwidth}
\usepackage{xcolor}
\definecolor{links}{rgb}{0.7,0,0}   
\definecolor{urls}{rgb}{0,0,0.8}    
\definecolor{cites}{rgb}{0,0,0.8}   

\usepackage[colorlinks,hyperindex,linkcolor=links,citecolor=cites,urlcolor=urls]{hyperref} 

\usepackage{tikz}
\usetikzlibrary{external}

\usepackage{pgfplots}
\usepackage[utf8]{inputenc}
\usepackage[nosort]{cite}        
\usepackage{url} 
\usepackage[intlimits]{amsmath}
\usepackage{bbm}
\usepackage{graphicx}
\usepackage{paralist}
\usepackage{fancyref}
\usepackage[stretch=16,shrink=16,step=4]{microtype}
\usepackage{amssymb}
\usepackage{amsfonts}
\usepackage{mathrsfs}
\usepackage{xspace}
\usepackage{bm}
\usepackage{fancyref}
\usepackage{textcomp} 
\usepackage{dsfont}
\usepackage{bbold}
\newcounter{MYtempeqncnt}

\makeatletter
\def\@IEEEinterspaceratioM{0.265}
\def\@IEEEinterspaceMINratioM{0.1651}
\def\@IEEEinterspaceMAXratioM{0.38}

\def\@IEEEinterspaceratioB{0.31}
\def\@IEEEinterspaceMINratioB{0.19}
\def\@IEEEinterspaceMAXratioB{0.38}
\@IEEEtunefonts
\makeatother
\def\cdf(#1)(#2)(#3){0.5*(1+(erf((#1-#2)/(#3*sqrt(2)))))}%

\newtheorem{definition}{Definition}
\newtheorem{remark}{Remark}
\newtheorem{theorem}{Theorem}
\newtheorem{lemma}[theorem]{Lemma}
\newtheorem{corollary}[theorem]{Corollary}

\providecommand{\ee}[1]{\exp\mathopen{}\left(#1\right)}

\providecommand{\eeBig}[1]{\exp\mathopen{}\Big(#1\Big)}

\providecommand{\e}[1]{e^{#1}}

\providecommand{\vectornorm}[1]{\left\lVert#1\right\rVert}

\providecommand{\vectornormbig}[1]{\big\lVert#1\big\rVert}

\providecommand{\indi}[1]{\mathds{1}\mathopen{}\left\{#1\right\}}

\providecommand{\indiBig}[1]{\mathds{1}\mathopen{}\Big\{#1\Big\}}

\providecommand{\vlflen}{\mathbb{n}}
\providecommand{\ellitoind}{\nu}

\newcommand{\const}{\mathbb{c}}

\newcommand{\E}[1]{\mathbb{E}\mathopen{}\left[#1\right]}
\newcommand{\Ebig}[1]{\mathbb{E}\mathopen{}\big[#1\big]}
\newcommand{\EBig}[1]{\mathbb{E}\mathopen{}\Big[#1\Big]}
\newcommand{\Ebigg}[1]{\mathbb{E}\mathopen{}\bigg[#1\bigg]}
\newcommand{\EBigg}[1]{\mathbb{E}\mathopen{}\Bigg[#1\Bigg]}
\newcommand{\EE}[2]{\mathbb{E}_{#1}\mathopen{}\left[#2\right]}

\newcommand{\EEbigg}[2]{\mathbb{E}_{#1}\mathopen{}\bigg[#2\bigg]}

\newcommand{\inff}[1]{\inf\mathopen{}\left\{#1\right\}}

\newcommand{\minn}[1]{\min\mathopen{}\left\{#1\right\}}
\newcommand{\minnBig}[1]{\min\mathopen{}\Big\{#1\Big\}}

\newcommand{\maxx}[1]{\max\mathopen{}\left\{#1\right\}}

\newcommand{\Va}[1]{\opV\!\mathrm{ar}\mathopen{}\left[#1\right]}
\newcommand{\Vaa}[2]{\opV\!\mathrm{ar}_{#1}\mathopen{}\left[#2\right]}

\newcommand{\Vaabigg}[2]{\opV\!\mathrm{ar}_{#1}\mathopen{}\bigg[#2\bigg]}

\newcommand{\farg}[1]{\mathopen{}\left( #1 \right)}
\newcommand{\fargBig}[1]{\mathopen{}\Big( #1 \Big)}

\newcommand{\fargBigg}[1]{\mathopen{}\Bigg( #1 \Bigg)}

\newcommand{\pr}[1]{\mathbb{P}\mathopen{}\left[#1\right]}
\newcommand{\prbig}[1]{\mathbb{P}\mathopen{}\big[#1\big]}
\newcommand{\prBig}[1]{\mathbb{P}\mathopen{}\Big[#1\Big]}
\newcommand{\prbigg}[1]{\mathbb{P}\mathopen{}\bigg[#1\bigg]}
\newcommand{\prBigg}[1]{\mathbb{P}\mathopen{}\Bigg[#1\Bigg]}

\newcommand{\prx}[2]{\mathbb{P}_{#1}\mathopen{}\left[#2\right]}
\newcommand{\prxbig}[2]{\mathbb{P}_{#1}\mathopen{}\big[#2\big]}

\newcommand{\prxbigg}[2]{\mathbb{P}_{#1}\mathopen{}\bigg[#2\bigg]}

\newcommand{\for}{\quad\text{for}\quad}

\newcommand{\Mf}{M^*_{\text{f}}}
\newcommand{\Mvf}{M^*_{\text{vf}}}

\newcommand{\vect}[1]{\boldsymbol{\mathbf{#1}}}

\newcommand{\diff}{\nabla}

\newcommand{\diffI}[2]{\left\langle \nabla I_{#1},#2\right\rangle}
\newcommand{\diffIbig}[2]{\big\langle \nabla I_{#1},#2\big\rangle}
\newcommand{\diffIBig}[2]{\Big\langle \nabla I_{#1},#2\Big\rangle}

\newcommand{\dd}{\mathop{}\!\mathrm{d}}

\newcommand{\Coopvar}{\Upsilon}
\newcommand{\sprob}{q}

\newcommand{\asym}{r}

\newlength\figureheight
\newlength\figurewidth
\newlength\smallfigureheight
\newlength\smallfigurewidth
\pgfplotsset{compat=newest}
\pgfplotsset{plot coordinates/math parser=false}
\usepgfplotslibrary{fillbetween}
\usetikzlibrary{patterns}
\usetikzlibrary{decorations.pathreplacing}
\usetikzlibrary{shapes,arrows, positioning, fit}
\tikzset{
    aligned pin/.style args={[#1]#2:#3}{
        pin={[%
            inner sep=0pt,%
            label={[%
                append after command={%
                    node[%
                        inner sep=0pt,%
                        at=(\tikzlastnode.#2),%
                        anchor=#1,%
                    ]{#3}%
                }%
            ]center:{}}%
        ]#2:{}}%
    }
}

\usepackage{ifthen}
\newboolean{finalmanuscript}
\setboolean{finalmanuscript}{false}

\ifthenelse{\boolean{finalmanuscript}}{}{\tikzexternalize[force remake,prefix=figs/]}

\begin{document}

\IEEEoverridecommandlockouts

\title{Common-Message Broadcast Channels with Feedback in the Nonasymptotic Regime: Full Feedback}
%
%
             %

%
\author{\thanks{The work of K. F. Trillingsgaard and P. Popovski  was supported by the European Research Council (ERC Consolidator Grant Nr. 648382 WILLOW) within the Horizon 2020 Program. The work of G. Durisi was supported by the Swedish Research Council under the grant 2016-03293. The material of this paper was presented in part at the 2017 IEEE International Symposium on Information Theory \cite{Trillingsgaard2017}.}Kasper Fløe Trillingsgaard\thanks{K. F. Trillingsgaard and P. Popovski are with the Department of Eletronic Systems, Aalborg University, 9220, Aalborg Øst, Denmark (e-mail: \{kft,petarp\}@es.aau.dk).}, \emph{Member, IEEE}, Wei Yang\thanks{W. Yang is with Qualcomm Technologies, Inc.,  San Diego, 92121, USA (e-mail: weiyang@qti.qualcomm.com).}, \emph{Member, IEEE},
Giuseppe Durisi\thanks{G. Durisi is with the Department of Electrical Engineering, Chalmers University of Technology, 41296, Gothenburg, Sweden (e-mail: durisi@chalmers.se).}, \emph{Senior Member, IEEE}, and
Petar Popovski, \emph{Fellow, IEEE}\ifthenelse{\boolean{finalmanuscript}}{\thanks{Copyright (c) 2018 IEEE. Personal use of this material is permitted.  However, permission to use this material for any other purposes must be obtained from the IEEE by sending a request to pubs-permissions@ieee.org.}}{}}
%
\maketitle
\thispagestyle{empty}

\begin{abstract}
We investigate the maximum coding rate achievable on a two-user broadcast channel for the case where a common message is transmitted with feedback using either fixed-blocklength codes or variable-length codes. For the fixed-blocklength-code setup, we establish nonasymptotic converse and achievability bounds. An asymptotic analysis of these bounds reveals that feedback improves the second-order term compared to the no-feedback case. In particular, for a certain class of antisymmetric broadcast channels, we show that the dispersion is halved. For the variable-length-code setup, we demonstrate that the channel dispersion is zero.
\end{abstract}

\begin{IEEEkeywords}
Broadcast channel with common-message, finite blocklength regime, full feedback, channel dispersion, variable-length coding.
\end{IEEEkeywords}

\IEEEpeerreviewmaketitle

\section{Introduction} 
\label{sec:introduction}
\IEEEPARstart{W}{e} consider a two-user common-message discrete-time memoryless broadcast channel (CM-DMBC) with full feedback.
When fixed-length codes are used, it is well-known that feedback does not improve capacity (as in the point-to-point setup), which is given by \cite{Gamal2011}
\begin{IEEEeqnarray}{rCl}
  C = \sup_P \min_{k\in\{1,2\}}I(P,W_k).\label{eq:C}
\end{IEEEeqnarray}
Here, $W_1$ and $W_2$ denote the component channels from the encoder to the two decoders and the supremum is over all input distributions $P$. 
In this paper, we show that feedback can improve the speed at which the maximum coding rate approaches capacity as the blocklength increases, and that the improvement in the speed of convergence differs depending on whether one uses fixed-length or variable-length codes.

For the no-feedback case, a CM-DMBC is equivalent to a compound channel, whose capacity was characterized in \cite{Blackwell59}, \cite{Wolfowitz62}, and whose second-order coding rate (i.e., the second-order expansion of the maximum coding rate in the limit of large blocklength) was found in~\cite{Polyanskiy}.
Specifically, it was proven in \cite{Polyanskiy} that the second-order term in the asymptotic expansion of the logarithm of the maximum number of codewords $M_{\text{no-f}}^*(n,\epsilon)$ (here ``no-f'' stands for no feedback) in the limit of large blocklength $n$ and for a fixed average error probability $\epsilon \in (0,1)$ is in general a function of both the \emph{dispersion}~\cite[Eq.~(222)]{Polyanskiy2010b} of the individual component channels and of the directional derivatives of their mutual informations computed at the capacity-achieving input distribution (CAID) $P^*$.

In contrast, when feedback is present, the capacity of a CM-DMBC differs in general from that of the compound channel \cite{Shrader2009}. This is because, for the compound channel with feedback, the transmitter can send a training sequence and learn the state of the compound channel via the feedback link. It can then adapt the input distribution to maximize the mutual information of that specific component channel. This communication scheme cannot be used in the CM-DMBC because both decoders are required to decode the message. 

In the fixed-length-code setup, it is known that feedback does not improve the second-order term for point-to-point discrete memoryless channels (DMCs) under certain symmetry conditions. 
Specifically, it was shown in~\cite{Polyanskiy2011} that feedback does not improve the second-order term for weakly input-symmetric DMCs. This result was extended to a broader class of DMCs in~\cite{feedback_improve}, where it was shown that the same holds when the conditional information variance is constant for all input symbols. 

When \emph{variable-length codes} are used, feedback is known to improve the speed at which the maximum coding rate converges to capacity for the point-to-point setup. 
This was first demonstrated by Burnashev \cite{Burnashev1976}, who proved that the error exponent with variable-length codes and full feedback is given by
\begin{IEEEeqnarray}{rCl}
  E(R) = \frac{\widetilde C_1}{\widetilde C}(\widetilde C - R)\label{eq:optimal_error_exponent}
\end{IEEEeqnarray}
for all $0<R<\widetilde C$. Here, $\widetilde C$ denotes the point-to-point capacity and $\widetilde C_1$ denotes the maximum relative entropy between conditional output distributions. Yamamoto~and~Itoh~\cite{Yamamoto1979} proposed a two-phase scheme that achieves the error exponent in \eqref{eq:optimal_error_exponent} and Berlin~\emph{et al.} \cite{Berlin2009} provided an alternative and simpler proof of the converse. 
For the CM-DMBC, Truong~and~Tan \cite{Truong2017} established upper and lower bounds on the error exponent using arguments similar to those of Burnashev, Yamamoto, and Itoh. The bounds in \cite{Truong2017} are tight when the broadcast channel is stochastically degraded.

For the regime of large average blocklength and fixed error probability, it was shown in~\cite{Polyanskiy2011} that there is no square-root penalty in the asymptotic expansion of the maximum coding rate achievable with variable-length codes for point-to-point DMCs, a result known as \emph{zero-dispersion}. 
Moreover, this fast convergence to capacity can be achieved using only \emph{stop feedback} (also known as \emph{decision feedback}). Namely, the feedback link is used only to stop the transmitter.
For CM-DMBCs, however,  stop feedback is not sufficient to achieve zero dispersion. 
Specifically, we recently showed that the asymptotic expansion of the maximum coding rate with variable-length codes and stop feedback contains a square-root penalty term, provided that some mild technical conditions are satisfied \cite{Trillingsgaard2018_VLF}. 

\subsubsection*{Contributions} 
We show that the presence of full feedback improves the second-order term in the asymptotic expansions of the maximal coding rate for a general class of CM-DMBCs.  Specifically, we prove nonasymptotic achievability and converse bounds on the maximal number of codewords $\Mf(n,\epsilon)$ and $\Mvf(\vlflen,\epsilon)$ (here ``f'' and ``vf'' stand for \emph{feedback} and \emph{variable-length plus feedback}, respectively) that can be transmitted over a CM-DMBC with feedback and with reliability $1-\epsilon$ using fixed-length codes and variable-length codes, respectively. Here, $n$ denotes the blocklength in the fixed-length setup whereas $\vlflen$ stands for the average blocklength in the variable-length setup. Through an asymptotic analysis of our bounds, we obtain the following results. 
\begin{itemize}
\item For the fixed-length case, we show that the second-order term depends on the directional derivatives of the mutual informations of the two component channels evaluated at the CAID through a channel-dependent parameter that we shall denote by $\eta\in(0,1)$. Specifically, we show that the second-order term is larger or equal to $\sqrt{\eta^2 V_1 + (1-\eta)^2 V_2}Q^{-1}(2\epsilon)$ (here, $\epsilon<1/2$), where $V_1$ and $V_2$ denote the conditional information variances of the two component channels evaluated at $P^*$. Furthermore, if the weighted sum of the conditional information variances given a specific input symbol $x$ takes the same value for all  $x$, the lower bound $\sqrt{\eta^2 V_1 + (1-\eta)^2 V_2}Q^{-1}(2\epsilon)$ is tight. For CM-DMBCs satisfying $V_1 = V_2$ and a certain ``antisymmetric property'' (which will be made precise in Definition~\ref{def:antisym} on page~\pageref{def:antisym}), the parameter $\eta$ is equal to $1/2$ and the second-order term simplifies to $\sqrt{V_1/2}Q^{-1}(2\epsilon)$, thereby demonstrating that, in this case, full feedback halves the dispersion compared to the no-feedback case. 
\item For the variable-length case, we show that full feedback yields zero dispersion. In light of our previous result for stop-feedback codes in \cite{Trillingsgaard2018_VLF}, this novel result shows that variable-length codes combined with full feedback attain a second-order term that is strictly better than the one of variable-length codes with stop-feedback. 
\end{itemize}
\subsubsection*{Intuition}
Feedback allows the encoder to compute the accumulated information density at both decoders and to adapt the input distribution accordingly.
Specifically, the encoder makes small adjustments to the input distribution in order to favor the decoder with the smallest information density and to drive both information densities close to their arithmetic mean. This strategy is helpful in both the fixed-length coding and variable-length coding setup as explained next.

Using this strategy in the fixed-length setup, one transforms the problem of computing the maximum coding rate into that of computing the $\epsilon$-quantile of the arithmetic mean of the two information densities.
The desired result follows because the arithmetic mean of the two information densities has variance $V_1/2$ when $V_1=V_2$.
Hence, the dispersion is halved.
Furthermore, the improvement from $Q^{-1}(\epsilon)$ to $Q^{-1}(2\epsilon)$ is achieved as follows: 
If the arithmetic mean of the information densities is below a suitably chosen threshold shortly before the end of the transmission,
the encoder changes the input distribution to either the CAID of $W_1$ or to the one of $W_2$.
In this way, it can ensure that at least one of the two decoders is successful with high probability. 

When variable-length codes are used, each decoder attempts decoding as soon as the information density at that decoder exceeds a certain threshold. Since the information densities at both decoders are driven towards their arithmetic mean using the above mentioned feedback scheme, both decoders attempt decoding at approximately the same time. This means that the stochastic overshoot of the arithmetic mean of the two information densities that results in the square-root penalty when fixed-length codes are used can be virtually eliminated by using variable-length codes. Hence, the dispersion is zero as in the single-user variable-length setup \cite{Polyanskiy2011}. Note that the zero-dispersion result depends crucially on the availability of full feedback at the encoder, which allows the encoder to steer the information densities between the transmitter and the two users towards their arithmetic mean. This is not possible if the encoder is only provided with stop feedback.

%


\paragraph*{Organization} In Section~\ref{sec:system_model_feed}, we introduce the system model. Our nonasymptotic bounds and asymptotic expansions for fixed-length codes with feedback and variable-length codes with feedback are presented in Section~\ref{sec:achievability_and_converse_bounds} and Section~\ref{sec:vlf_codes}, respectively. Section~\ref{sec:conclusion} concludes the paper.

\paragraph*{Notation}
We denote vectors by boldface letters; their entries are denoted by roman letters. Upper case, lower case, and calligraphic letters denote random variables (RVs), deterministic quantities, and sets, respectively. The cardinality of a set is denoted by ${|\cdot|}$. 
The inner product between two vectors $\vect{x}$ and $\vect{y}$ is denoted by $\langle \vect{x},\vect{y}\rangle$, and the $l_1$-norm and the $l_2$-norm are denoted by $\vectornorm{\cdot}_1$ and $\vectornorm{\cdot}_2$, respectively. The vectors $\vect{0}_d$ and $\vect{1}_d$ denote a $d$-dimensional column vectors whose entries are all zero and all one, respectively.
 We let $\mathbb{N}$, $\mathbb{Z}$, $\mathbb{Z}_+$, $\mathbb{R}$, and $\mathbb{R}_+$ denote the set of natural numbers, 
 the set of integers, the set of nonnegative integers, the set of real numbers, and the set of nonnegative real numbers, respectively. We also denote by $\mathbb{R}_0^d$, the set of $d$-dimensional real vectors whose entries sum to zero. We let $\const$ denote an arbitrary positive constant whose value may change at each occurrence and ${|\cdot|}_+$ denotes $\max\{0,\cdot\}$. Furthermore, we denote the base $e$ logarithm by $\log(\cdot)$. Throughout the paper, the index $k$ is used to designate one  of the two decoders. Hence, it always belongs to the set $\{1,2\}$, although this is sometimes not explicitly mentioned. We also set $\bar k=3 - k$. For a probability distribution $P$, we let $P^n$ be the joint probability distribution of the vector $[X_1,\ldots,X_n]$, whose entries $\{X_i\}$ are independently and identically distributed (i.i.d.) according to $P$. We let $\E{\cdot}$ and $\Va{\cdot}$ denote the expectation and the variance, respectively. For a probability distribution $Q$, we denote the expectation with respect to $Q$ by $\EE{Q}{\cdot}$. For a RV $X$ with probability distribution $P$ on $\mathcal{X}$ and a RV $Y$ which is the output of a channel transition matrix $W:\mathcal{X}\mapsto \mathcal{Y}$, we let $PW(\cdot)$ denote the induced output distribution on $\mathcal{Y}$; furthermore, we  denote the joint probability distribution of $(X,Y)$ by $P\times W$.
Given two probability distributions $P$ and $Q$ on a common measurable space, we define, for every $\alpha\in(0,1)$, the Neyman-Pearson function  $\beta_{1-\alpha}(P,Q)$ as the minimum type-II error probability of a binary hypothesis test between $P$ and $Q$ subject to the constraint that the type-I error probability does not exceed $\alpha$. Finally, for two functions $f(\cdot)$ and $g(\cdot)$, the notation $f(x) = \mathcal{O}(g(x))$, as $x\rightarrow \infty$, means that  $\lim_{x\rightarrow \infty} |f(x)/g(x)| < \infty$ and  $f(x)=o(g(x))$, as $x\rightarrow \infty$, means that $\lim_{x\rightarrow \infty} |f(x)/g(x)| = 0$.
 

\section{System Model} 
\label{sec:system_model_feed}
We consider a CM-DMBC with input alphabet $\mathcal{X}$ and output alphabets $\mathcal{Y}_1 = \mathcal{Y}_2 = \mathcal{Y}$.\footnote{The assumption $\mathcal{Y}=\mathcal{Y}_1=\mathcal{Y}_2$ comes without loss of generality.}
We assume that the channel outputs at any given time $i$ are conditionally independent given the input, namely
\begin{equation}\label{eq:factorization_feed}
  W(y_{1,i},y_{2,i}|x_i) \triangleq    W_1(y_{1,i}|x_i)W_2(y_{2,i}|x_i).
\end{equation}
The capacity of the CM-DMBC is given in~\eqref{eq:C}. We shall assume that it is achieved by $P^*$ and that this CAID is unique.\footnote{The assumption that  $P^*$ is unique is used in the proof of our converse results. Our achievability results can  be readily extended to CM-DMBCs with nonunique CAIDs. }
We denote the capacities of the two component channels by $C_1$ and $C_2$, respectively.

For every input distribution $P$ and every $n\in \mathbb{N}$, we denote the information density between the vectors $x^n$ and $y^n_k$ as
\begin{equation}
  \imath_{P, W_k}(x^n; y_k^n) \triangleq\sum_{i=1}^n \log \frac{W_k(y_{k,i}|x_i)}{PW_k(y_{k,i})}.\label{eq:inf_dens_def_feed}
\end{equation}
Furthermore, we let
\begin{IEEEeqnarray}{rCl}
I_k(P)\triangleq \EE{P\times W_k}{\imath_{P, W_k}(X; Y_k)}
\end{IEEEeqnarray}
  be the mutual information and
\begin{IEEEeqnarray}{rCl}
V_k(P)\triangleq \EE{P}{\Vaa{W_k}{\imath_{P, W_k}(X; Y_k)|X}}\label{eq:cond_inf_var_feed}
\end{IEEEeqnarray}
be the conditional information variance. 
Finally, we let $V_{k} \triangleq V_k(P^*)$ and we let  $\diffI{k}{\vect{v}}$ denote the directional derivative of the mutual information $I_k(\cdot)$ along the direction $\vect{v}\in\mathbb{R}_0^{|\mathcal{X}|}$  at the point~$P^*$
\begin{IEEEeqnarray}{rCl} 
  \diffI{k}{\vect{v}} \triangleq \sum_{x\in\mathcal{X}} v_x D(W_k(\cdot | x) || P_{Y_k}^* )\label{eq:differential_feed}
\end{IEEEeqnarray}
(the entries of $\vect{v}$ should sum to zero because the mutual information is only defined on the $|\mathcal{X}|$-dimensional probability simplex).\footnote{It is sometimes required in literature that $\vect{v}$ is a unit vector, but we do not assume this here.}
Here, $D(\cdot ||\cdot)$ denotes the relative entropy, $P_{Y_k}^* = P^*W_k(\cdot)$ is the capacity-achieving output distribution (CAOD). Without loss of generality, we have also assumed in \eqref{eq:differential_feed} that $\mathcal{X}=\{1,\ldots,N\}$ where $N=|\mathcal{X}|$.

Besides \eqref{eq:factorization_feed} and the uniqueness of $P^*$, we also assume that the channel laws satisfy the following conditions
\begin{enumerate}
  \item $P^*(x) > 0 $ for all $x\in\mathcal{X}$.
  \item $V_k > 0$ for $k\in\{1,2\}$.
  \item $C_k> C$ for $k\in\{1,2\}$.
\end{enumerate}

The following lemma provides a relation between $\diffI{1}{\cdot}$ and $\diffI{2}{\cdot}$ that will  be important in the asymptotic analyses of our converse and achievability bounds.
\begin{lemma}\label{lem:diffIkzero}
  There exists a unique constant $\eta\in(0,1)$ such that, for all $\vect{v} \in\mathbb{R}_0^{|\mathcal{X}|}$,
  \begin{IEEEeqnarray}{rCl}
    \eta \diffI{1}{\vect{v}}+(1-\eta) \diffI{2}{\vect{v}} = 0.\label{eq:diffIkzero}
  \end{IEEEeqnarray}
\end{lemma}
\begin{IEEEproof}
  Let $\overline{\diff}_k\in\mathbb{R}^{|\mathcal{X}|-1}$ be the gradient of $\diffIbig{k}{[\vect{\bar v},-\langle \vect{1}_{|\mathcal{X}|-1}, \vect{\bar v}\rangle]}$ with respect to $\vect{\bar v} \in\mathbb{R}^{|\mathcal{X}|-1}$. Note that the concavity of the mutual information and the assumption $C_k > C$ imply that $\overline{\diff}_k$ cannot be the zero vector.  Since $P^*$ maximizes $P\mapsto \min_k I_k(P)$, it follows that $\langle \overline{\diff}_1, \vect{\bar v}\rangle\langle \overline{\diff}_2, \vect{\bar v}\rangle \leq 0$ for all $\vect{\bar v}$. Suppose on the contrary that there exists a $\vect{\tilde v}\in\mathbb{R}^{|\mathcal{X}|-1}$ such that $\langle \overline{\diff}_1, \vect{\tilde v}\rangle\langle \overline{\diff}_2, \vect{\tilde v}\rangle > 0$. Then, by differentiability of $P\mapsto I_k(P)$ at $P^*$, there exists a sufficiently small constant $\xi\neq 0$ satisfying 
  \begin{IEEEeqnarray}{rCl}
\IEEEeqnarraymulticol{3}{l}{\min_k I_k(P^* + \xi [\vect{\tilde v}, -\langle \vect{1}_{|\mathcal{X}|-1}, \vect{\tilde v} \rangle])}\nonumber\\
 \qquad &\geq& \min_k\Big\{ I(P^*,W_k) +\frac{\xi}{2}\langle \overline{\diff}_k, \vect{\tilde v}\rangle \Big\} \\
 &>& \min_k I(P^*,W_k).\label{eq:lem1_contradiction1}
  \end{IEEEeqnarray}
  Note that the constant $\xi$ is positive if $\langle \overline{\diff}_k, \vect{\tilde v}\rangle > 0$ for $k\in\{1,2\}$ and negative if  $\langle \overline{\diff}_k, \vect{\tilde v}\rangle < 0$ for $k\in\{1,2\}$.
  Now, \eqref{eq:lem1_contradiction1} contradicts that $P^*$ maximizes $P\mapsto \min_k I(P,W_k)$, implying that $\langle \overline{\diff}_1, \vect{\bar v}\rangle\langle \overline{\diff}_2, \vect{\bar v}\rangle \leq 0$ for all $\vect{\bar v}$ as desired. 
  As a result, we have that
  \begin{IEEEeqnarray}{rCl}
  \frac{\overline{\diff}_1}{\vectornorm{\overline{\diff}_1}_2} + \frac{\overline{\diff}_2}{\vectornorm{\overline{\diff}_2}_2} = \vect{0}_{|\mathcal{X}|-1}.\label{eq:diffIkzero2}
  \end{IEEEeqnarray}
  Indeed, assume that 
  \begin{IEEEeqnarray}{rCl}
    \frac{\overline{\diff}_1}{\vectornorm{\overline{\diff}_1}_2}+\frac{\overline{\diff}_2}{\vectornorm{\overline{\diff}_2}_2} = \vect{w} \neq \vect{0}_{|\mathcal{X}|-1}. \label{eq:lem_contradiction}
  \end{IEEEeqnarray}
  Then, it follows from \eqref{eq:lem_contradiction} and the Cauchy-Schwarz inequality that
  \begin{IEEEeqnarray}{rCl}
\langle \overline{\diff}_1, \vect{w}\rangle = \vectornorm{\overline{\diff}_1}_2+\frac{\langle \overline{\diff}_1, \overline{\diff}_2 \rangle}{\vectornorm{\overline{\diff}_2}_2} \geq 0
  \end{IEEEeqnarray}
  and similarly that $\langle \overline{\diff}_2, \vect{w}\rangle \geq 0$. Here, the inequalities hold with equality if and only if \eqref{eq:diffIkzero2} holds, i.e., when $\overline{\diff}_1$ and $\overline{\diff}_2$ are linearly dependent and point in opposite directions. Hence, since $\vect{w}\neq \vect{0}_{|\mathcal{X}|-1}$, we conclude that $\langle \overline{\diff}_1, \vect{w}\rangle > 0$ and $\langle \overline{\diff}_2, \vect{w}\rangle > 0$. But this contradicts that $\langle \overline{\diff}_1, \vect{w}\rangle\langle \overline{\diff}_2, \vect{w}\rangle \leq 0$.
 Now, \eqref{eq:diffIkzero} follows from \eqref{eq:diffIkzero2} with $\eta = (1/\vectornorm{\overline{\diff}_1}_2)/(1/\vectornorm{\overline{\diff}_1}_2+1/\vectornorm{\overline{\diff}_2}_2)$. Uniqueness of $\eta$ follows trivially.
\end{IEEEproof}
We use \eqref{eq:diffIkzero} to define the channel-dependent constant $\eta\in(0,1)$ as follows:
\begin{IEEEeqnarray}{rCl}
  \eta = \frac{\diffI{2}{\vect{v}}}{\diffI{2}{\vect{v}}-\diffI{1}{\vect{v}}}.\label{eq:eta}
\end{IEEEeqnarray}
Here, $\vect{v}$ is an arbitrary vector in  $\mathbb{R}_0^{|\mathcal{X}|}$ satisfying $\diffI{1}{\vect{v}} \neq 0$. Note that by Lemma~\ref{lem:diffIkzero}, the value of $\eta$ does not depend on $\vect{v}$. For notational convenience, we set $\eta_1 = \eta$ and $\eta_2 = 1-\eta$ in the remainder of the paper.


Next, we define the notions of fixed-length feedback (FLF) codes and variable-length feedback (VLF) codes.
\begin{definition}
An $(n,M,\epsilon)$-FLF code for the CM-DMBC consists of:
\begin{enumerate}
\item $n$ encoding functions $f_i: \mathcal{M} \times \mathcal{Y}^{i-1}\times \mathcal{Y}^{i-1} \mapsto \mathcal{X}$, $i=1,\dots, n$ (possibly randomized),  mapping the message $J$, drawn uniformly from $ \mathcal{M}\triangleq \{1,\ldots,M\}$, and the past channel outputs $Y_1^{i-1},Y_2^{i-1}$ to the channel input $X_i = f_i(J, Y^{i-1}_1,Y^{i-1}_2)$.
\item Two decoders $g_{k}: \mathcal{Y}^n \mapsto  \mathcal{M}$ satisfying
\begin{equation}
 \prbig{g_{k}(Y_k^n)\not= J } \leq \epsilon,  \ \ \ \ \ k\in\{1,2\}.\label{eq:def_prob_error_feed}
\end{equation}
\end{enumerate}
\end{definition}
The maximum code size achievable  with blocklength $n$ and average error probability not exceeding $\epsilon$ is denoted by
 \begin{IEEEeqnarray}{rCl}
   \Mf(n, \epsilon) &\triangleq \maxx{M : \exists (n,M,\epsilon)\text{-FLF code}}.
 \end{IEEEeqnarray}
\begin{definition}\label{def:vlf}
An $(\vlflen, M,\epsilon)$-VLF code for the CM-DMBC consists of:
\begin{enumerate}
\item A RV $U\in\mathcal{U}$ with $|\mathcal{U}|\leq 3$, which is known at both the encoder and the decoders. 
\item A sequence of encoding functions   $f_i: \mathcal{U}\times  \mathcal{M} \times \mathcal{Y}^{i-1}\times \mathcal{Y}^{i-1} \mapsto \mathcal{X}$, each one mapping the message $J$, drawn uniformly at random from the set $\mathcal{M}$, the past channel outputs $Y_1^{i-1},Y_2^{i-1}$, and the auxiliary RV $U$ to the channel input $X_i = f_i(U,J, Y_1^{i-1},Y_2^{i-1})$.
\item Two nonnegative integer-valued RVs $\tau_1$ and $\tau_2$ that are stopping times with respect to the filtrations $\mathcal{F}_{1,i} \triangleq \sigma(U, Y_{1}^i)$ and $\mathcal{F}_{2,i} \triangleq \sigma(U, Y_{2}^i)$, respectively, and satisfy
\begin{IEEEeqnarray}{rCl}
  \Ebig{\max_k \tau_k}\leq \vlflen.\label{eq:avg_blocklength_const_feed}
\end{IEEEeqnarray}
\item A sequence of decoders $g_{k,i}: \mathcal{U}\times \mathcal{Y}^i \mapsto  \mathcal{M}$ satisfying
\begin{align}
  \prbig{J \not= g_{k,\tau_k}(U, Y_k^{\tau_k}) } \leq \epsilon, \qquad k\in\{1,2\}.\label{eq:vlf_def_prob_error}
\end{align}
\end{enumerate}
\end{definition}
The maximum code size achievable with average blocklength $\vlflen$ and average error probability not exceeding $\epsilon$ is denoted by
 \begin{IEEEeqnarray}{rCl}
   \Mvf(\vlflen, \epsilon) &\triangleq \maxx{M : \exists (\vlflen,M,\epsilon)\text{-VLF code}}.
 \end{IEEEeqnarray}
\begin{figure}[!t]
\begin{center}
\includegraphics[width=0.35\textwidth]{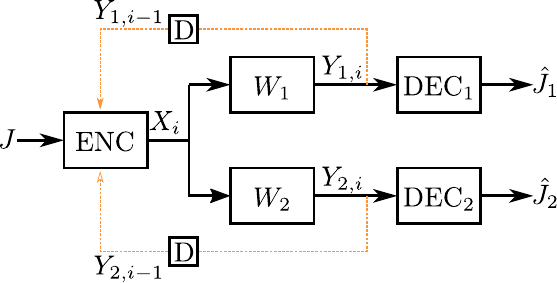}
\end{center}
\caption{The system model considered in this paper. Here, $\mathrm{D}$ stands for a delay of one unit of time.}
\label{fig:system_model2}
\end{figure}
 We depict our system model in Fig.~\ref{fig:system_model2}.

\section{Fixed-Length Feedback Codes} 
\label{sec:achievability_and_converse_bounds}
In this section, we present nonasymptotic achievability and converse bounds on $\Mf(n,\epsilon)$ for FLF codes. These bounds are then analyzed in the large-$n$ regime. Under some technical assumptions, the asymptotic expansions of our bounds are shown to match up to the second order. We then introduce a class of antisymmetric CM-DMBCs for which the second-order term in the asymptotic expansion of $\Mf(n,\epsilon)$ takes on a simple expression, which allows for an insightful comparison with the no-feedback case.
\subsection{Nonasymptotic bounds}
We first state a Verdú-Han-type converse bound for FLF codes. 
\begin{theorem}  \label{thm:nonasymptotic_converse}
Every $(n,M,\epsilon)$-FLF code  for the CM-DMBC satisfies\footnote{Recall that $\eta_1 = \eta$ and $\eta_2 = 1-\eta$, where $\eta$ was defined in \eqref{eq:eta}.}
 \begin{multline}
    \epsilon \geq \frac{1}{2}\prxbigg{J,Y_1^n,Y_2^n}{\sum_k \eta_k\sum_{i=1}^n \imath_{P^*,W_k}\Big({Y_{k,i};f_i(J,Y_1^{i-1},Y_2^{i-1})}\Big)
     \\ 
     \leq \log M - \lambda}-\e{-\lambda}\IEEEeqnarraynumspace\label{eq:converse}
  \end{multline}
  for every $\lambda>0$, where
   \begin{multline}
    \mathbb{P}_{J,Y_1^n,Y_2^n}(j, y_1^n,y_2^n)\\
     \triangleq\frac{1}{M} \prod_{i=1}^n \prod_k W_k(y_{k,i}|f_i(j,y_1^{i-1},y_2^{i-1})).\label{eq:Pmeas}
  \end{multline}
\end{theorem}
\begin{IEEEproof}
  Fix an $(n,M,\epsilon)$-FLF code with encoding functions $\{f_i\}$ and decoding functions $\{g_k\}$. Define the following auxiliary probability distributions on $\mathcal{M} \times \mathcal{Y}^{n}\times\mathcal{Y}^n$
  \begin{multline}
    \mathbb{Q}^{(k)}_{J,Y_1^n,Y_2^n}(j, y_1^n,y_2^n)\\ \triangleq \frac{1}{M} \prod_{i=1}^n W_{\bar k}(y_{\bar k,i}|f_i(j,y_1^{i-1},y_2^{i-1})) P^*W_k(y_{k,i}).\label{eq:Qmeas}
  \end{multline}
 Recall that $\bar k=1$ when $k=2$ and $\bar k =2$ when $k=1$. When $(J,Y_1^n,Y_2^n)\sim \mathbb{P}_{J,Y_1^n,Y_2^n}$, both decoders have an average error probability not exceeding $\epsilon$ because of \eqref{eq:def_prob_error_feed}. When $(J,Y_1^n,Y_2^n)\sim \mathbb{Q}_{J,Y_1^n,Y_2^n}^{(k)}$, the average error probability at decoder $k$ is $1-1/M$. The meta-converse theorem in \cite[Th.~27]{Polyanskiy2010b} and the inequality \cite[Eq.~(106)]{Polyanskiy2010b} then imply that 
  \begin{IEEEeqnarray}{rCl}
  \frac{1}{M} &\geq& \beta_{1-\epsilon}(\mathbb{P}_{J,Y_1^n,Y_2^n},\mathbb{Q}^{(k)}_{J,Y_1^n,Y_2^n} )\\
  &\geq& \frac{1}{\gamma}\left( \mathbb{P}_{J,Y_1^n,Y_2^n}\mathopen{}\left[ \log \frac{\dd \mathbb{P}_{J,Y_1^n,Y_2^n}}{\dd \mathbb{Q}^{(k)}_{J,Y_1^n,Y_2^n}} \leq \log \gamma \right] -\epsilon\right)\label{eq:metaconverse_res}\IEEEeqnarraynumspace
  \end{IEEEeqnarray}
  for every $\gamma>0$ and $k\in\{1,2\}$. Let 
\begin{equation}
   A_k(j,y_1^n,y_2^n) \triangleq \sum_{i=1}^n \imath_{P^*,W_k}(f_i(j,y^{i-1}_{1}, y_2^{i-1}); y_{k,i}).  
\end{equation} 
By setting  $\log \gamma \triangleq \log M -\lambda$ and by using \eqref{eq:Pmeas} and \eqref{eq:Qmeas} in \eqref{eq:metaconverse_res}, we obtain
  \begin{multline}
    \epsilon \geq  \max_k \mathbb{P}_{J,Y_1^n, Y_2^n}\mathopen{}\Bigl[A_k(J,Y_1^n,Y_2^n) \leq \log M - \lambda\Bigr] - \e{-\lambda}.\label{eq:converse2}
  \end{multline}  
The desired result \eqref{eq:converse} is obtained from \eqref{eq:converse2} through the following chain of inequalities:
\begin{IEEEeqnarray}{rCl}
 \IEEEeqnarraymulticol{3}{l}{\epsilon+\e{-\lambda}}\nonumber\\
 \quad &\geq& \frac{1}{2}\sum_{k}\mathbb{P}_{J,Y_1^n, Y_2^n}\mathopen{}\Bigl[A_k(J,Y_1^n,Y_2^n) \leq \log M - \lambda\Bigr]\label{eq:max_halfsum}\\
    &\geq&  \frac{1}{2}\mathbb{P}_{J,Y_1^n, Y_2^n}\mathopen{}\Biggl[\bigcup_{k}\Big\{A_k(J,Y_1^n,Y_2^n) \leq \log M - \lambda\Big\}\Biggr]\label{eq:union_bound_conv}\IEEEeqnarraynumspace\\
        &\geq&  \frac{1}{2}\mathbb{P}_{J,Y_1^n, Y_2^n}\mathopen{}\Biggl[\sum_k \eta_k A_k(J,Y_1^n,Y_2^n) \leq \log M - \lambda\Biggr].\label{eq:converse_asymp11} 
\end{IEEEeqnarray}
Here, \eqref{eq:max_halfsum} follows because $\max_k a_k \geq \frac{1}{2} \sum_k a_k$ for every pair $(a_1,a_2)\in\mathbb{R}_+^2$, \eqref{eq:union_bound_conv} follows from the union bound, and \eqref{eq:converse_asymp11} follows because $\sum_k \eta_k b_k \leq \zeta$ imply that either $b_1 \leq \zeta$ or $b_2\leq \zeta$ for every triple $(b_1,b_2,\zeta)\in\mathbb{R}^3$.
\end{IEEEproof}

Before stating our achievability bound, we need to introduce some notation. 
We let $S\geq 2$, $L$, and $m$ denote arbitrary integers. 
We also let $\mathcal{S}\triangleq \{1,\ldots,S\}$ and $\mathcal{L}\triangleq\{1,\ldots,L\}$. 
Furthermore, we index each element $\mathbb{x}$ of $\mathcal{X}^{S L m}$ as follows:
\begin{IEEEeqnarray}{rCl}
    \mathbb{x} &=& (\mathbb{x}_1(1),\ldots,\mathbb{x}_1(S),\ldots\ldots,\mathbb{x}_L(1),\ldots,\mathbb{x}_L(S)). \IEEEeqnarraynumspace
\end{IEEEeqnarray}
\begin{sloppypar}\noindent Here, $\mathbb{x}_{\ell}(s) = (\mathbb{x}_{\ell,1}(s),\ldots, \mathbb{x}_{\ell,m}(s))\in\mathcal{X}^m$.
For a given vector $b^{\ell} \in \mathcal{S}^\ell$, we let $\mathbb{x}^\ell(b^\ell) \triangleq (\mathbb{x}_1(b_1),\ldots,\mathbb{x}_\ell(b_\ell))$.  We use a similar notation for the elements in the set $\mathcal{Y}^{SLm}$. 
Finally, we define the minimum error probability achievable on an CM-DMBC using fixed-length codes with no feedback of blocklength $n_B$ and with $S^L$ codewords:
\begin{multline}
\epsilon^*(n_B,S^L)\\ \triangleq \min\{\epsilon: \exists (n_B,S^L,\epsilon)\text{ finite-length no-feedback code}\}.\label{eq:eps_nofeedback}
\end{multline}\end{sloppypar}

The intuition behind our achievability bound is as follows. The codebook consists of codewords of length $Lm$. Each codeword is divided into $L$ blocks of length $m$, which are of constant composition.
Within each block $\ell$,  feedback is used to compute the index $B_\ell$, which selects one out of the $S$ available subcodewords (the vectors $\mathbb{x}_{\ell}(B_{\ell})$, $B_{\ell}=1,\dots,S$). We then communicate the sequence $B^L$ to the decoders using a code with $S^L$ codewords of length $n_B$.
The following achievability bound for FLF codes is then a consequence of the achievability bound~\cite[Th.~3]{Polyanskiy} for compound DMCs without feedback. 

%
\begin{theorem}\label{thm:nonasympotic_achiev}
Let $P_1,\ldots,P_S$ be types of sequences in $\mathcal{X}^m$, let $\tau\in(0,\epsilon)$, let $\zeta >0$, and let $\{h_\ell\}_{\ell=1}^{L}$ be arbitrary mappings from $\mathcal{X}^{(\ell-1) m }\times\mathcal{Y}^{(\ell-1) m }\times \mathcal{Y}^{(\ell-1) m }$ to $\mathcal{S}$. 
Then
\begin{IEEEeqnarray}{rCl}
  \IEEEeqnarraymulticol{3}{l}{\log \Mf(L m + n_B,\epsilon+\epsilon^*(n_B,S^L))}\nonumber\\
   &\geq& \sup\mathopen{}\Bigl\{\gamma:  \max_k \prx{}{ \imath_k(\mathbb{x}^L(B^L); \mathbb{Y}_k^L(B^L)) \leq \gamma  } \nonumber\\
   &&\qquad\qquad\qquad\qquad\qquad\qquad\qquad{}<\epsilon-\tau- \e{-\zeta} \Bigr\}\nonumber\\
   &&{} +\log \frac{\tau}{2} - SL |\mathcal{X}|\log(1+m) -L\log S - \zeta.\label{eq:nonasymp_achievability}\IEEEeqnarraynumspace
\end{IEEEeqnarray}
Here, $\mathbb{x}\in\mathcal{X}^{S L m}$ is an arbitrary element in the set
\begin{equation}
  \mathbb{F} \triangleq \Big\{\mathbb{x}\in\mathcal{X}^{S L m}: \mathbb{x}_{\ell}(s)\text{ has type } P_s, \quad\forall s\in\mathcal{S},\, \forall\ell\in\mathcal{L}\}\label{eq:bbF}
\end{equation}
and $\mathbb{P}_{\mathbb{Y}_1, \mathbb{Y}_2,B^L}$ denotes the probability distribution of $(\mathbb{Y}_1,\mathbb{Y}_2,B^L)$ on $\mathbb{B} \times \mathbb{B}\times\mathcal{S}^L$ defined by
\begin{IEEEeqnarray}{rCl}
\IEEEeqnarraymulticol{3}{l}{\mathbb{P}_{\mathbb{Y}_1, \mathbb{Y}_2,B^L}(\mathbb{y}_1,\mathbb{y}_2,b^L)\triangleq}\nonumber\\
& & \prod_{\ell=1}^L\Bigg(\prod_{s=1}^S \prod_k W_k^m(\mathbb{y}_{{k,
  \ell}}(s)|\mathbb{x}_{\ell}(s)) \Bigg) \nonumber\\
  &&{}\quad\times \indi{h_{\ell}\Big(\mathbb{x}^{\ell-1}(b^{\ell-1}),\mathbb{y}_{1}^{\ell-1}(b^{\ell-1}),\mathbb{y}_{2}^{\ell-1}(b^{\ell-1})\Big) = b_{\ell}}\IEEEeqnarraynumspace
\end{IEEEeqnarray}
for $\mathbb{y}_{k,\ell}(b) \in\mathcal{Y}^{m}$.
 Finally, 
\begin{IEEEeqnarray}{rCl}
  \imath_k(\mathbb{x}_\ell(b_\ell);\mathbb{y}_{k,\ell}(b_\ell)) &\triangleq&   \sum_{i=1}^m \imath_{P_{b_\ell},W_k}(\mathbb{x}_{\ell,i}(b_\ell); \mathbb{y}_{k,\ell,i}(b_\ell)) \IEEEeqnarraynumspace
\end{IEEEeqnarray}
and $\imath_k(\cdot;\cdot)$ in \eqref{eq:nonasymp_achievability} is defined as follows:
\begin{IEEEeqnarray}{rCl}
    \imath_k(\mathbb{x}^L(b^L); \mathbb{y}_k^L(b^L)) \triangleq \sum_{\ell=1}^L \imath_k(\mathbb{x}_\ell(b_\ell);\mathbb{y}_{k,\ell}(b_\ell)).
\end{IEEEeqnarray}
\end{theorem}
\begin{IEEEproof}
See Appendix~\ref{app:achiev_proof_feed}.
\end{IEEEproof}
\begin{remark}
  The achievability bound in Theorem~\ref{thm:nonasympotic_achiev} holds also under the more stringent error probability constraint $\max_{j\in\mathcal{M}} \pr{g_k(Y_k^n)\neq j|J=j}\leq \epsilon$.
\end{remark}
\begin{remark}
The term ${-SL|\mathcal{X}|\log(1+m)}$ in \eqref{eq:nonasymp_achievability} represents the penalty due to the use of constant composition codes.
\end{remark}
\begin{remark}
We shall apply Theorem~\ref{thm:nonasympotic_achiev} in the following way: The types $P_1$ and $P_2$ are chosen close to $P^*$ but such that they slightly favor decoder $1$ and decoder $2$, respectively. The types $P_3, P_4$, and $P_5$ are set equal to $P^*$, $P_1^*$, and $P_2^*$, respectively. By the end of the $(\ell-1)$th block, for $\ell\in\{1,\ldots,L-1\}$, the encoder computes the information densities at the decoders and chooses $B_\ell = k$ if decoder $k$ has accumulated the smallest amount of information density. By the end of the $(L-1)$th block, the encoder chooses $B_L = 3$ if the arithmetic mean of the information densities at the decoders  exceeds a certain threshold and it sets  $B_L$ equal to $4$ or $5$, each with probability $1/2$, otherwise.
\end{remark}

\subsection{Asymptotic analysis}
By analyzing Theorem~\ref{thm:nonasymptotic_converse} and Theorem~\ref{thm:nonasympotic_achiev} in the large-blocklength limit, we will next establish asymptotic expansions for our converse and achievability bounds that match up to the second order provided that the assumption \eqref{eq:conv_cond} is satisfied. In particular, we show in the next theorem that in this case the channel dispersion \cite[Def.~1]{Polyanskiy2010b} is 
\begin{IEEEeqnarray}{rCl}
  V &\triangleq& \eta^2 V_1 +(1-\eta)^2 V_2.\label{eq:V_def}
\end{IEEEeqnarray}
\begin{theorem}\label{thm:asymp_variance_cond}
Suppose that 
\begin{multline}
\eta^2\Va{\imath_{P^*,W_1}(x; Y_1)|X=x}\\+(1-\eta)^2\Va{\imath_{P^*,W_2}(x; Y_2) | X= x}  = V\IEEEeqnarraynumspace\label{eq:conv_cond}
\end{multline}
for all $x\in\mathcal{X}$, where $V$ is defined in \eqref{eq:V_def}. Then, for every $\epsilon\in(0,1/2)$, we have
   \begin{IEEEeqnarray}{rCl}
  \log \Mf(n,\epsilon) = nC -\sqrt{n V}Q^{-1}(2\epsilon) + \mathcal{O}(n^{1/3}\log n).\IEEEeqnarraynumspace\label{eq:asymp_conv}
  \end{IEEEeqnarray}
\end{theorem}
\begin{IEEEproof}
  See Appendix~\ref{sec:asymp_converse} for the proof of the converse part and Appendix~\ref{sec:asymp_achievability} for the proof of the achievability part.
\end{IEEEproof}
\begin{remark}
The assumption $\epsilon\in(0,1/2)$ is crucial for~\eqref{eq:asymp_conv} to hold.
Indeed, when  $\epsilon>1/2$, one can achieve $\log \Mf(n,\epsilon) = n \min_k C_k + o(n)$  by using a standard point-to-point fixed-length code of rate $R<\min_k C_k$ and codewords generated i.i.d. according to $P^*_1$ (the CAID of $W_1$) with probability $1/2$ and another code of rate $R$ and codewords generated i.i.d. according to  $P^*_2$ (the CAID of $W_2$) with probability $1/2$.
This implies that the strong converse does not hold.
\end{remark}
\begin{remark}
The assumption \eqref{eq:conv_cond} is a multiuser analogue of the assumption in \cite[Th.~2]{feedback_improve} under which the authors of \cite{feedback_improve} show that feedback does not improve the second-order terms of point-to-point DMCs. The assumption \eqref{eq:conv_cond} is only needed in the proof of the converse part, and our achievability result continues to hold without it.
\end{remark}
 If \eqref{eq:conv_cond} does not hold, our bounds imply only the weaker asymptotic expansion given in the next theorem.
\begin{theorem}\label{thm:asymp_no_variance_cond}
  For every $\epsilon\in(0,1/2)$, there exists a positive constant $c$ such that
  \begin{multline}
 nC +\mathcal{O}(\sqrt{n}) = \log\Mf(n,\epsilon) \\\geq n C - \sqrt{n V}Q^{-1}(2\epsilon) + \mathcal{O}(n^{1/3}\log n).\IEEEeqnarraynumspace\label{eq:Mf_conclusion2} 
  \end{multline}
\end{theorem}
\begin{IEEEproof}
 See Appendix~\ref{sec:asymp_converse} for the proof of the converse part and Appendix~\ref{sec:asymp_achievability} for the proof of the achievability part.
\end{IEEEproof}
In the next section, we introduce a general class of CM-DMBCs for which the assumption \eqref{eq:conv_cond} holds, and the second-order term can be characterized using \eqref{eq:asymp_conv}.

If the CAID is nonunique and \eqref{eq:conv_cond} does not hold, then the second-order term can be further improved compared to what is reported in \eqref{eq:Mf_conclusion2} by using the feedback scheme in \cite{feedback_improve}, provided that the conditional information variance takes different values when evaluated for the different CAIDs. The intuition behind the feedback scheme proposed in \cite{feedback_improve} is as follows: The encoder starts by using the CAID that minimizes the conditional information variance. During the transmission, the encoder can use the feedback to track the information density accumulated at the decoder and thereby compute the conditional error probability given the realized channel uses. If the conditional error probability becomes sufficiently large, it eventually becomes favorable to change the input distribution to the CAID maximizing the conditional information variance.

\subsection{Antisymmetric CM-DMBCs}
We shall apply our results to the class of antisymmetric CM-DMBCs defined as follows.
\begin{definition}
\label{def:antisym}
A CM-DMBC is antisymmetric if $|\mathcal{X}|$ is even, $\mathcal{Y}$ can be decomposed in disjoint sets $\mathcal{\overline Y}_1,\ldots,\mathcal{\overline Y}_r$, and the channel transition matrices can be decomposed as follows:
\begin{IEEEeqnarray}{rCl}
  W_1 = \bigg[ \begin{array}{ccc} 
   p_{11}\overline W_{11}& \cdots & p_{1\asym}\overline W_{1\asym}  \\
   p_{21}\overline W_{21} &\cdots & p_{2\asym}\overline W_{2\asym}
  \end{array}
    \bigg] \label{eq:antisym_def1}
    \end{IEEEeqnarray}
     and
\begin{IEEEeqnarray}{rCl}
      W_2 = \bigg[ \begin{array}{ccc} 
   p_{21}\overline W_{21}& \cdots & p_{2\asym}\overline W_{2\asym}  \\
   p_{11}\overline W_{11} &\cdots & p_{1\asym}\overline W_{1\asym}
  \end{array}
    \bigg].\label{eq:antisym_def2}
\end{IEEEeqnarray}
Here, $\sum_{i=1}^\asym p_{ki}=1$ for $k\in\{1,2\}$ and $\{\overline W_{1i}\}$ and $\{\overline W_{2i}\}$ are weakly symmetric channel transition matrices\footnote{A channel transition matrix is weakly symmetric if the rows are permutations of each other and all column sums are equal. The CAID and CAOD of a weakly symmetric channel are uniform \cite[pp.~189--190]{Cover2012}.} with dimensions $\frac{|\mathcal{X}|}{2}\times |\mathcal{\overline Y}_i|$. For notational convenience, we let $W^{(1)}\triangleq [p_{11}\overline W_{11} ,\cdots,  p_{1\asym}\overline W_{1\asym}]$ and $W^{(2)}\triangleq [p_{21}\overline W_{21}, \cdots ,p_{2\asym}\overline W_{2\asym}]$.
\end{definition}
\begin{remark}\label{rem:antiZ}
  The CM-DMBC composed of antisymmetric Z-channels is an example of an antisymmetric CM-DMBC, where $r=2$, $\overline W_{11}=\overline W_{12}=\overline W_{21}=\overline W_{22} = 1$, $1-p_{11} =p_{12} = q$, and $p_{21} = 1-p_{22}=0$.
\end{remark}

We showed in  \cite{Trillingsgaard2017} that feedback halves the dispersion for the CM-DMBC  composed of antisymmetric Z-channels described in Remark~\ref{rem:antiZ}. The following corollary of Theorem~\ref{thm:asymp_variance_cond} and Theorem~\ref{thm:asymp_no_variance_cond} extends this result to the more general class of antisymmetric CM-DMBCs. 

Before stating the corollary, we shall briefly recall the intuition behind the half-dispersion result. The key point is that full feedback allows the encoder to compute the accumulated information densities at both decoders and make small adjustments in the input distribution in order to favor the decoder with the smallest accumulated information density. By doing so, both information densities are driven towards their arithmetic mean and the computation of the maximum coding rate roughly becomes equivalent to the computation of the $\epsilon$-quantile of the arithmetic mean of the two information densities. Now, because the encoder only makes small adjustments in the input distribution, the arithmetic mean of the information densities is well-approximated by a Gaussian distribution with mean $nC$ and variance $nV_1/2$. As a result, the dispersion is halved.
\begin{corollary}\label{cor:antisym}
For every antisymmetric CM-DMBC satisfying $\min_k C_k > C$ and for every $\epsilon\in(0,1/2)$, we have that
  \begin{IEEEeqnarray}{rCl}
    \log \Mf(n,\epsilon) = n C - \sqrt{\frac{n V_1}{2}} Q^{-1}(2\epsilon) + \mathcal{O}(n^{1/3}\log n).\IEEEeqnarraynumspace\label{eq:corollary}
  \end{IEEEeqnarray}
  \end{corollary}
\begin{IEEEproof}
  We need to show that, for every antisymmetric CM-DMBC,  we have that $\eta_1=\eta_2 = 1/2$, that $V_1 = V_2$, and that
  \begin{IEEEeqnarray}{rCl}
  \Va{\imath_{P^*,W_1}(x; Y_1)+ \imath_{P^*,W_2}(x; Y_2) | X= x}  = 2V_1\label{eq:anti_sym_var_cond}
  \end{IEEEeqnarray}
  for all $x\in\mathcal{X}$.
 We first show that $P^*$ is the uniform distribution on $\mathcal{X}$. In the following, we let $\alpha_P = \sum_{x=1}^{|\mathcal{X}|/2} P(x)$ and $\overline \alpha_P = 1-\alpha_P$ for every input distribution $P$ on $\mathcal{X}$. Note that
  \begin{IEEEeqnarray}{rCl}
    H(Y_1|X) &=& \alpha_P H(\mathbf{r}_1) + \overline\alpha_P H(\mathbf{r}_2)\label{eq:entr_Y1_X}\\
    H(Y_2|X) &=& \overline \alpha_P H(\mathbf{r}_1) + \alpha_P H(\mathbf{r}_2)\label{eq:entr_Y2_X}
  \end{IEEEeqnarray}
  where $\mathbf{r}_k$ denotes the first row of $W_k$.
  Define
  \begin{multline}
    \overline P_{Y_1}(y) \\= \left[\frac{\alpha_P p_{11}+\overline \alpha_P p_{21} }{|\mathcal{\overline Y}_1|} \vect{1}_{ |\mathcal{\overline Y}_1|},\ldots,\frac{\alpha_P p_{1\asym}+\overline \alpha_P p_{2\asym} }{|\mathcal{\overline Y}_\asym|} \vect{1}_{ |\mathcal{\overline Y}_\asym|}\right].
\end{multline}
It follows that \cite[Th.~2.6.4]{Cover2012}
\begin{IEEEeqnarray}{rCl}
0&\leq& D( P_{Y_1} || \overline P_{Y_1})\nonumber\\ &=&  \sum_{j=1}^\asym (\alpha_P p_{1j}+\overline \alpha_P p_{2j})\log \frac{|\mathcal{\overline Y}_j|}{ \alpha_P p_{1j}+\overline \alpha_P p_{2j} }-H(Y_1)\IEEEeqnarraynumspace 
\end{IEEEeqnarray}
which implies 
\begin{IEEEeqnarray}{rCl}
H(Y_1 ) \leq \sum_{j=1}^\asym (\overline\alpha_P p_{1j}+\alpha_P p_{2j})\log \frac{|\mathcal{\overline Y}_j|}{ \overline\alpha_P p_{1j}+ \alpha_P p_{2j} }.\label{eq:relentr_Y1_X}
\end{IEEEeqnarray}
Similarly,
\begin{IEEEeqnarray}{rCl}
H(Y_2 ) \leq \sum_{j=1}^\asym (\overline\alpha_P p_{1j}+\alpha_P p_{2j})\log \frac{|\mathcal{\overline Y}_j|}{ \overline\alpha_P p_{1j}+ \alpha_P p_{2j} }.\label{eq:relentr_Y2_X}
\end{IEEEeqnarray}
As a result of \eqref{eq:entr_Y1_X}--\eqref{eq:entr_Y2_X}, \eqref{eq:relentr_Y1_X}--\eqref{eq:relentr_Y2_X}, and because of symmetry, we have
\begin{multline}
  \min_k I_k(P)\\\leq \frac{1}{2} \left(\sum_{j=1}^\asym (p_{1j}+ p_{2j})\log \frac{2|\mathcal{\overline Y}_j|}{  p_{1j}+ p_{2j}} -H(\mathbf{r}_1) - H(\mathbf{r}_2)\right).
\end{multline}
This upper bound holds with equality when $P$ is uniform. Hence, we have shown that $P^*$ is the uniform distribution on $\mathcal{X}$, and consequently, we also have that $V_1=V_2$, that $P_{Y_1}^* = P_{Y_2}^*$ and that \eqref{eq:anti_sym_var_cond} holds. 

We next argue that $\eta_1 = \eta_2=1/2$.   Define the permutation $\pi$ of $\{1,\ldots,|\mathcal{X}|\}$ as follows: $\pi\triangleq[|\mathcal{X}/2|+1,\ldots, |\mathcal{X}|, 1,\ldots,|\mathcal{X}/2|]$.  We first show that for every $\vect{v}\in\mathbb{R}_0^{|\mathcal{X}|}$
\begin{IEEEeqnarray}{rCl}
\diffI{1}{\vect{v}_{\pi}} &=& \sum_{x\in\mathcal{X}} v_{\pi}(x) D(W_1(\cdot | x) || P_{Y_1}^* )\\
&=& \sum_{x\in\mathcal{X}} v_{\pi}(x) D(W_2(\cdot|\pi(x)) || P_{Y_2}^* )\\
&=& \sum_{x\in\mathcal{X}}  v(x) D(W_2(\cdot | x) || P_{Y_2}^* )\\
&=& \diffI{2}{\vect{v}}.\label{eq:perm_diff}
\end{IEEEeqnarray}
Here, $\vect{v}_{\pi}$ denotes the vector $\vect{v}$ whose entries are permuted according to $\pi$.
In particular,  since $\pi(\pi(x)) = x$ for all $x\in\mathcal{X}$, \eqref{eq:perm_diff} implies that 
\begin{IEEEeqnarray}{rCl}
\diffI{1}{\vect{v}+\vect{v}_{\pi}} = \diffI{2}{\vect{v}+\vect{v}_{\pi}}.
\end{IEEEeqnarray}
By the uniqueness of $P^*$ and by \eqref{eq:antisym_def1}--\eqref{eq:antisym_def2}, there cannot exist a vector $\vect{v}\in\mathbb{R}_0^{|\mathcal{X}|}$ such that $\diffI{1}{\vect{v}+\vect{v}_{\pi}}$ and $\diffI{2}{\vect{v}+\vect{v}_{\pi}}$ are simultaneously positive or simultaneously negative, and hence $\diffI{1}{\vect{v}+\vect{v}_{\pi}}$ must be equal to zero for all $\vect{v}\in\mathbb{R}_0^{|\mathcal{X}|}$. Therefore, by using \eqref{eq:perm_diff}, we conclude that
\begin{multline}
  0=\diffI{1}{\vect{v}+\vect{v}_{\pi}} = \diffI{1}{\vect{v}} + \diffI{1}{\vect{v}_{\pi}} \\=  \diffI{1}{\vect{v}} + \diffI{2}{\vect{v}}.\label{eq:zero_sum_innerprod}\IEEEeqnarraynumspace
\end{multline}
Since $\eta$ in \eqref{eq:diffIkzero} is unique, \eqref{eq:zero_sum_innerprod} implies that $\eta_1 = \eta_2 = 1/2$. The desired result is then established by invoking Theorem~\ref{thm:asymp_variance_cond} and Theorem~\ref{thm:asymp_no_variance_cond}.
\end{IEEEproof}

\begin{figure}[!t]
\begin{center}
    \includegraphics{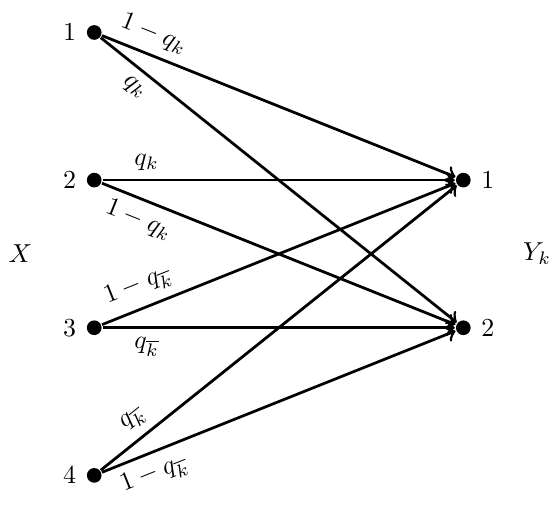}

\end{center}
\caption{An antisymmetric CM-DMBC composed of two binary symmetric channels. This figure illustrates the component channel $W_k$.}
\label{fig:parallel_bsc}
\end{figure}

\subsection{A specific CM-DMBC}
We shall next consider the antisymmetric CM-DMBC depicted in Fig.~\ref{fig:parallel_bsc}, which we shall use for evaluating numerically the accuracy of the asymptotic expansion provided in Corollary~\ref{cor:antisym}. For this specific class of antisymmetric CM-DMBC, we shall derive next an alternative achievability bound that yields tighter numerical results for small values of $n$ than the generic achievability bound in Theorem~\ref{thm:nonasympotic_achiev}. This bound, which is a random-coding union (RCU)-type of achievability bound, is based on a minimum distance mismatched decoder.
\begin{theorem}
\label{thm:rcu}
For the CM-DMBC depicted in Fig.~\ref{fig:parallel_bsc}, there exists an $(n,M,\epsilon)$-FLF code satisfying
\begin{IEEEeqnarray}{rCl}
  \epsilon \leq \sum_{t=0}^n P_{Z_1^{(n)}}(t) \minn{ 1, (M-1) \sum_{k=0}^t {{n}\choose{k}}2^{-n} }.\label{eq:rcu}
\end{IEEEeqnarray}
Here, $P_{Z_1^{(n)}}$ denotes the probability mass function of the RV $Z_1^{(n)}$ that is defined through the following recursion
\begin{IEEEeqnarray}{rCl}
Z_k^{(0)} &\triangleq& 0\\
  Z_{k}^{(i)} &\triangleq& Z_k^{(i-1)}\nonumber\\
  &&{} + \left\{ \begin{array}{ll} 
   E_{1}^{(i)} & \text{if } Z_1^{(i-1)} > Z_2^{(i-1)} \\
   & \text{or } \left(Z_1^{(i-1)} = Z_2^{(i-1)} \text{ and } \overline Z^{(i)} = 0\right) \\
   E_{2}^{(i)} & \text{if }Z_1^{(i-1)} < Z_2^{(i-1)} \\
   & \text{or } \left(Z_1^{(i-1)} = Z_2^{(i-1)} \text{ and } \overline Z^{(i)} = 1\right)
  \end{array}
  \right. \IEEEeqnarraynumspace
\end{IEEEeqnarray}
for $i\in\{1,\ldots,n\}$, where $E_k^{(i)} \sim \text{Bern}(q_k)$ and $\overline Z^{(i)}\sim\text{Bern}(1/2)$ are independent RVs. 
\end{theorem}
\begin{IEEEproof}
  See Appendix~\ref{proof:rcu}.
\end{IEEEproof}

\begin{figure}[!t]
  \begin{center}
    \setlength\smallfigureheight{7cm}
    \setlength\smallfigurewidth{8.65cm}
        \includegraphics{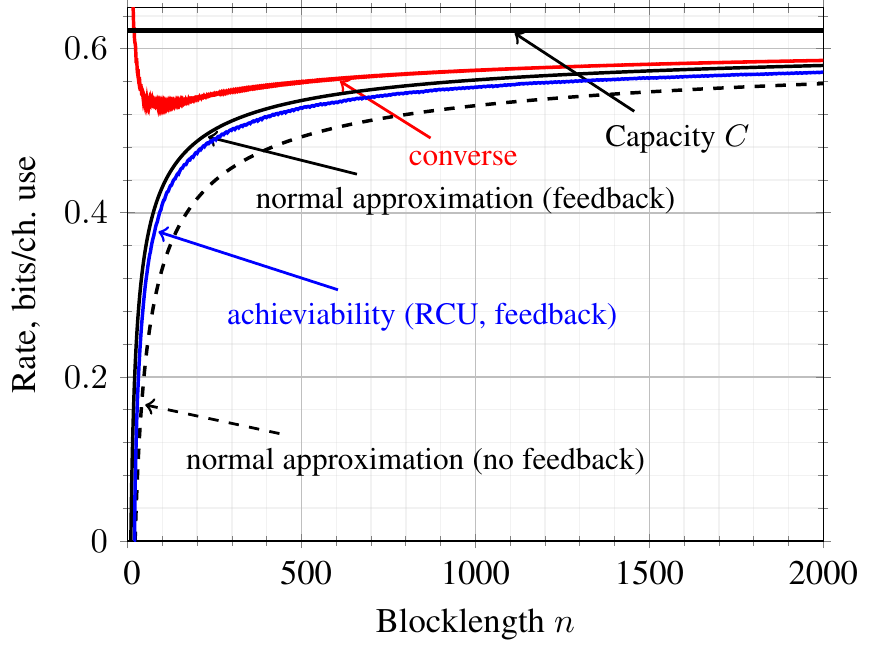}   
  \end{center}
  \caption{Comparison between the RCU-type achievability bound in \eqref{eq:rcu}, the converse bound in \eqref{eq:converse}, and the normal approximation in \eqref{eq:corollary} for the CM-DMBC depicted in Fig.~\ref{fig:parallel_bsc} with $q_1=0.05$, $q_2=0.10$, and $\epsilon=10^{-3}$. The normal approximation (feedback) is the expression in \eqref{eq:corollary} with the $\mathcal{O}(\cdot)$ term omitted. The normal approximation (no feedback) is the expression in \cite[Eq.~(12)]{Polyanskiy} with the $o(\sqrt{n})$ term omitted.}
  \label{fig:paral_bsc1}
\end{figure}
Let $W_1$ and $W_2$ be as depicted in Fig.~\ref{fig:parallel_bsc} with $q_1 = 0.05$ and $q_2=0.10$. This channel is an antisymmetric CM-DMBC and, hence, it satisfies the conditions of Corollary~\ref{cor:antisym}.  We plot the achievability bound given in Theorem~\ref{thm:rcu}, the converse bound given in Theorem~\ref{thm:nonasymptotic_converse}, and the first two terms of the asymptotic expansion (normal approximation) in Corollary~\ref{cor:antisym}. We observe that the normal approximation is an accurate proxy for the maximum coding rate for the considered range of $n$.
We would like to point out that the achievability scheme in Theorem~\ref{thm:rcu} is not capacity-achieving since it is based on minimum-distance decoding. Specifically, the capacity of the channel depicted in Fig.~\ref{fig:parallel_bsc} is given by $1-\frac{1}{2}\big(h(q_1)+h(q_2)\big) \approx 0.622$, where $h(x) \triangleq -x \log x - (1-x)\log(1-x)$ is the binary entropy function. In contrast, the rate achievable by minimum-distance decoding is only $1-h(\frac{1}{2}(q_1+q_2)) \approx 0.616 $. The plotted achievability bound is thus expected to be accurate only for small to moderate values of $n$. The generic achievability bound in Theorem~\ref{thm:nonasympotic_achiev} is not plotted since it is not accurate numerically for moderate values of $n$. This is because of the use of constant composition codes and because the sequence $B^L$ needs to be communicated explicitly.

\section{Variable-Length Feedback Codes}\label{sec:vlf_codes}
We first present a Fano-type converse bound  and a nonasymptotic achievability bound for VLF codes. We then show that these bounds imply that the second-order term in the asymptotic expansion of $\log \Mvf(\vlflen, \epsilon)$ is zero.
\subsection{Nonasymptotic bounds}
Our converse result for VLF codes is based on Lemma~\ref{lem:diffIkzero} and on \cite[Lem~1~and~Lem~2]{Burnashev1976}.
\begin{theorem} 
  Every $(\vlflen,M,\epsilon)$-VLF code with $0 < \epsilon \leq 1-1/M$ satisfies
\begin{IEEEeqnarray}{rCl}
  \log M \leq \frac{ \vlflen C+h(\epsilon)}{1-\epsilon}\label{eq:vlf_conv_statement}
\end{IEEEeqnarray}
where $h(x) \triangleq -x \log x - (1-x)\log(1-x)$ is the binary entropy function.
\label{thm:vlf_converse}
\end{theorem}  
\begin{IEEEproof}
See Appendix~\ref{proof:vlf_converse}.
\end{IEEEproof} 

Our nonasymptotic achievability bound for VLF codes leverages ideas similar to in Theorem~\ref{thm:nonasympotic_achiev} and in \cite[Th.~3]{Polyanskiy2011}.  The achievability scheme consists of two phases: a fixed-length transmission phase via an FLF code and a variable-length transmission phase via a variable-length stop-feedback code. 
\begin{theorem}
\label{thm:simple_achiev_vlf}
  Let $P_1,\ldots,P_S$ be probability distributions on $\mathcal{X}$, let $\gamma\geq 0$ and $0 \leq \sprob \leq 1$ be arbitrary scalars, and $\tau_{\text{max}}$ be an arbitrary integer larger than $Lm$. The stopping times $\tau_k$ and $\bar \tau_k$ for $k\in\{1,2\}$ are defined as follows:
  \begin{align}
  \tau_k &\triangleq \minn{ \inff{n \geq L m: S_{k,n}(X^n, Y^n_k;B^L)\geq \gamma },\tau_{\text{max}}}\label{eq:tau_k_def_feed}\\
  \bar \tau_k &\triangleq \minn{ \inff{n \geq L m: S_{k,n}(\bar X^n, Y^n_k;B^L)\geq \gamma },\tau_{\text{max}}}.
  \end{align}
  Here, we have defined  for $n\geq Lm$
  \begin{IEEEeqnarray}{rCl}
  S_{k,n}(x^n, y^n_k;b^L) &\triangleq& \sum_{\ell=1}^L \sum_{i=1}^m \imath_{P_{b_\ell},W_k}(x_{\ellitoind(\ell,i)}; y_{k,\ellitoind(\ell,i)})\nonumber\\
  &&{} + \sum_{i=L m +1}^{n}\imath_{P^*,W_k}(x_i; y_{k,i})\label{eq:vlf_Skn}
  \end{IEEEeqnarray}
  where $\ellitoind(\ell,i) \triangleq (\ell-1)m + i$, 
    and the joint probability distribution of $(X^n, \bar X^n, Y_1^n, Y_2^n,B^L)$ is
  \begin{IEEEeqnarray}{rCl}
    \IEEEeqnarraymulticol{3}{l}{\mathbb{P}_{X^n,\bar X^n, Y_{1}^n, Y_{2}^n,B^L}(x^n,\bar x^n,y_1^n,y_2^n,b^L)}\nonumber\\
    \quad&=& \left(\prod_{k}W_k^n(y_{k}^n|x^n)\right)\nonumber\\
    &&{}\times  \Bigg(\prod_{\ell=1}^L \left(\prod_{i=1}^m P_{b_\ell}(x_{\ellitoind(\ell,i)})P_{b_\ell}(\bar x_{\ellitoind(\ell,i)})\right)\nonumber\\
    &&\qquad\qquad\times \indiBig{b_\ell = h_{\ell}(x^{\ell-1},y_1^{\ell-1},y_2^{\ell-1})}\Bigg)\nonumber\\
    &&{}\times (P^*)^{n - Lm}(x_{Lm+1}^n)(P^*)^{n - Lm}(\bar x_{Lm+1}^n).\label{eq:joint_probability_dist_feed}
  \end{IEEEeqnarray}
Then, for every positive integer $M$ and $n_B$, there exists an $(\vlflen, M, \epsilon)$-VLF code with
\begin{IEEEeqnarray}{rCl}
    \vlflen &\leq& (1-\sprob) \Big(\EBig{\max_k \tau_k} + n_B\Big)\label{eq:achiev_Emax_feed}\\
    \epsilon &\leq&  \sprob + (1-\sprob)\epsilon^*(n_B, S^L)\nonumber\\
    &&{} +\max_k\mathopen\Big\{ (1-\sprob)(M-1)\pr{\tau_k\geq \bar\tau_k}\Big\}.\label{eq:achiev_Proberror1_feed}
\end{IEEEeqnarray}
\end{theorem}
\begin{remark}
  Following steps similar to those in \cite[Eqs.~(111)-(118)]{Polyanskiy2011}, one can further upper-bound $\epsilon$ in \eqref{eq:achiev_Proberror1_feed} as follows:
  \begin{IEEEeqnarray}{rCl}
    \epsilon &\leq& q + (1-q)\big(\epsilon^*(n_B,S^L) + \ee{-\gamma}\nonumber\\
    && \qquad\qquad\qquad\qquad{}  + \max_k\pr{\tau_k = \tau_{\text{max}}}\big).\IEEEeqnarraynumspace
  \end{IEEEeqnarray}
  Here, we have treated the event $\{\tau_k = \tau_{\text{max}}\}$ as an error.
\end{remark}
\begin{remark}
  Compared to the nonasymptotic achievability bound for VLF codes for DMCs reported in \cite{Polyanskiy2011}, in our bounds, the stopping times $\tau_k$ and $\bar \tau_k$ are lower- and upper-bounded by $Lm$ and $\tau_{\text{max}}$, respectively. These bounds simplify the asymptotic analysis of the achievability bound.
\end{remark}
\begin{IEEEproof}
  See Appendix~\ref{app:proof_vlf_achiev}.
\end{IEEEproof}

\subsection{Asymptotic analysis} 
\label{sec:vlf_asymptotic_analysis}
The following result reveals that one can achieve zero dispersion by using VLF codes.
\begin{theorem}\label{thm:asymp_vlf_achiev}
  For every $\epsilon\in(0,1)$, we have
  \begin{IEEEeqnarray}{rCl}
    \log \Mvf(\vlflen, \epsilon) \geq \frac{\vlflen C}{1-\epsilon} + \mathcal{O}(\vlflen^{1/3}\log \vlflen).\label{eq:vlf_asymp}
  \end{IEEEeqnarray}
\end{theorem}
\begin{IEEEproof}
  See Appendix~\ref{proof:asymp_vlf_achiev}.
\end{IEEEproof}

The intuition behind this result is as follows: The encoder uses a feedback scheme similar  to the one used for our fixed-blocklength result. This implies that the information densities at both decoders are driven towards their arithmetic mean. Each decoder sends a stop signal as soon as the information density at that decoder exceeds a threshold. Since the information densities at both decoders are well-approximated by their arithmetic mean, both decoders send a stop signal at approximately the same time, which yields zero dispersion

We would like to emphasize that we have previously shown that the dispersion is positive when only stop feedback is available \cite{Trillingsgaard2018_VLF}. Our result in Theorem~\ref{thm:asymp_vlf_achiev} implies that there is a fundamental difference between full feedback and stop feedback in terms of speed of convergence to capacity. To our knowledge, this is the first result in the literature on fixed-error asymptotics that explicitly shows that the speed of convergence to capacity is different with stop feedback and full feedback. For comparison, for the single-decoder case, both stop feedback and full feedback yield the zero dispersion.

\section{Conclusion}\label{sec:conclusion}
In this paper, we investigated the maximal coding rate for the CM-DMBC with feedback. When fixed-length codes are employed, we demonstrated that the second-order term is reduced by a factor of more than $1/\sqrt{2}$ under mild technical conditions. Under an additional symmetry condition, we also proved that the asymptotic expansions of the our converse and achievability bounds match up to the second order. This improved second-order term is in contrast to the point-to-point setup, where it is shown that the second-order term is unaltered under similar symmetry conditions \cite{feedback_improve}.
We also found that the second-order term vanishes under certain mild technical conditions if we allow the use of variable-length codes. This result together with the one previously reported in \cite{Trillingsgaard2018_VLF} implies that zero-dispersion is achievable when feedback is available and variable-length codes are used but that stop-feedback is not sufficient to attain it.

\appendices


\section{Proof of Theorem~\ref{thm:nonasympotic_achiev}}\label{app:achiev_proof_feed}
 
We consider feedback schemes in which the encoders can be represented as follows: There exist $M$ vectors in $\mathcal{X}^{S L m}$, denoted by $\{\mathbb{x}^{(1)},\ldots,\mathbb{x}^{(M)}\}$, such that
\begin{IEEEeqnarray}{rCl}
    f_i(j, Y^{i-1}_1, Y^{i-1}_2) &=& \mathbb{x}^{(j)}_{\ell(i), i - m(\ell(i)-1)}(B_{\ell(i)})\label{eq:simplfied_encoder}
\end{IEEEeqnarray}
for $i\in\{1,\ldots,L m\}$. Here, $\ell(i) \triangleq \lceil i/m \rceil$ and
\begin{IEEEeqnarray}{rCl}
B_\ell\triangleq h_{\ell}(X^{(\ell-1)m},Y_1^{(\ell-1)m},Y_2^{(\ell-1)m})
\end{IEEEeqnarray}
where $X^{(\ell-1)m}$ denotes the sequence of channel inputs transmitted up to time $(\ell-1)m$. 

In the remainder of the proof, we shall use $\mathbf{Y}_{k,\ell}$ to denote $(Y_{k,(\ell-1)m + 1},\ldots,Y_{k,\ell m})$ and we let $\mathbf{Y}_k$ denote $(\mathbf{Y}_{k,1},\ldots,\mathbf{Y}_{k,L})$.
The conditional probability distribution of $(\mathbf{Y}_1,\mathbf{Y}_2,B^L)$ given that $\mathbb{X}=\mathbb{x}\in\mathcal{X}^{S L m}$ is then given by
\begin{IEEEeqnarray}{rCl}
  \IEEEeqnarraymulticol{3}{l}{\mathbb{P}_{\mathbf{Y}_1,\mathbf{Y}_2,B^L | \mathbb{X}}(\mathbf{y}_1,\mathbf{y}_2,b^L|\mathbb{x})}\nonumber\\
  &=& \prod_{\ell=1}^L\bigg(\prod_k W_k^m(\mathbf{y}_{{k,
  \ell}}|\mathbb{x}_{\ell}(b_\ell)) \bigg)\nonumber\\
  &&\qquad{} \times  \indi{h_{\ell}\Big(\mathbb{x}^{\ell-1}(b^{\ell-1}),\mathbf{y}_{1}^{\ell-1},\mathbf{y}_{2}^{\ell-1}\Big) = b_{\ell}}.\IEEEeqnarraynumspace
\end{IEEEeqnarray}
Next, we shall apply the achievability bound for the compound channel with channel state information at the receiver reported in \cite[Th.~3]{Polyanskiy}. In particular, we let $\mathcal{X}^{S L m}$ be the input alphabet, $\mathbb{B}\triangleq\mathcal{Y}^{L m}\times \mathcal{S}^L$ be the output alphabets, and consider the compound channel with transition probability
\begin{IEEEeqnarray}{rCl}
  \IEEEeqnarraymulticol{3}{l}{\mathbb{P}_{\mathbf{Y}_k,B^L|\mathbb{X}}(\mathbf{y}_k,b^L|\mathbb{x})}\nonumber\\ &\triangleq& \sum_{\mathbf{y}_{\bar k}\in\mathcal{Y}^{L m}}\prod_{\ell=1}^L\bigg(\prod_k W_k^m(\mathbf{y}_{{k,
  \ell}}|\mathbb{x}_{\ell}(b_\ell)) \bigg) \nonumber\\
  &&\qquad\qquad\quad{} \times\indi{h_{\ell}\Big(\mathbb{x}^{\ell-1}(b^{\ell-1}),\mathbf{y}_{1}^{\ell-1},\mathbf{y}_{2}^{\ell-1}\Big) = b_{\ell}}\label{eq:abstract_channel}\IEEEeqnarraynumspace\\
  &=&  \left(\prod_{\ell=1}^L W_k^m(\mathbf{y}_{{k,
  \ell}}|\mathbb{x}_{\ell}(b_\ell)) \right) P_{B^L|\mathbb{X}}(b^L|\mathbb{x}).\label{eq:abstract_channel2}\IEEEeqnarraynumspace
\end{IEEEeqnarray}
Moreover, we set $\mathbb{F}$ as in \eqref{eq:bbF} and take as auxiliary channel
\begin{IEEEeqnarray}{rCl}
\mathbb{Q}_{\mathbf{Y}_k,B^L}(\mathbf{y}_k,b^L)&\triangleq& \frac{1}{S^L}\prod_{\ell=1}^L (P_{b_\ell}W_k)^m(\mathbf{y}_{k,\ell}).
\end{IEEEeqnarray}
It will turn out convenient to define also the following probability measure:
\begin{IEEEeqnarray}{rCl}
  \mathbb{Q}_{\mathbb{X}}(\mathbb{x}) &\triangleq& \prod_{\ell=1}^L \prod_{s=1}^S (P_{s})^m(\mathbb{x}_{\ell}(s)).\label{eq:PX}
\end{IEEEeqnarray}
Note that
\begin{IEEEeqnarray}{rCl}
  \IEEEeqnarraymulticol{3}{l}{\sum_{\mathbb{x}\in\mathcal{X}^{S L m}} \mathbb{Q}_{\mathbb{X}}(\mathbb{x}) \mathbb{P}_{\mathbf{Y}_k,B^L|\mathbb{X}}(\mathbf{y}_k,b^L|\mathbb{x})}\nonumber\\
  &\leq& \sum_{\mathbb{x}\in\mathcal{X}^{S L m}} \mathbb{Q}_{\mathbb{X}}(\mathbb{x}) \prod_{\ell=1}^L W_k^m(\mathbf{y}_{{k,
  \ell}}|\mathbb{x}_{\ell}(b_\ell)) \\
    &=& \prod_{\ell=1}^L (P_{b_\ell} W_k)^m(\mathbf{y}_{{k,\ell}})\\
    &=&     S^L \mathbb{Q}_{\mathbf{Y}_k,B^L}(\mathbf{y}_k,b^L).\label{eq:PYBx_QYB_rel}
\end{IEEEeqnarray}

We observe that this compound channnel is equivalent to a CM-DMBC with encoding functions satisfying \eqref{eq:simplfied_encoder}, provided that the decoders have knowledge about $B^L$. We can readily provide both decoders with knowledge of $B^L$ using a finite blocklength code with $n_B$ channel uses and an error probability $\epsilon^*(n_B,S^L)$ defined in \eqref{eq:eps_nofeedback} at the end of the transmission. Hence, we apply \cite[Th.~3]{Polyanskiy} and conclude that
\begin{IEEEeqnarray}{rCl}
  \Mf(Lm+n_B,\epsilon + \epsilon^*(n_B, S^L))\geq \frac{\kappa_\tau}{\beta_{1-\epsilon+\tau}} \label{eq:logMf_bound}
\end{IEEEeqnarray}
  where 
  \begin{IEEEeqnarray}{rCl}
    \beta_{1-\epsilon+\tau} &=& \sup_{\mathbb{x}\in\mathbb{F},k\in \{1,2\}} \beta_{1-\epsilon+\tau}(\mathbb{P}_{\mathbf{Y}_k,B^L|\mathbb{X}=\mathbb{x}},\mathbb{Q}_{\mathbf{Y}_k,B^L})\\
    \kappa_\tau &=& \inf_E \max_{k\in\{1,2\}} \mathbb{Q}_{\mathbf{Y}_k,B^L}[E]
  \end{IEEEeqnarray}
  and the infimum in the definition of $\kappa_\tau$ is with respect to all sets $E\subseteq \mathbb{B}$ satisfying 
  \begin{IEEEeqnarray}{rCl}
    \forall \mathbb{x}\in\mathbb{F},\exists k: \mathbb{P}_{\mathbf{Y}_k,B^L|\mathbb{X}=\mathbb{x}}[E]&\geq& \tau.\label{eq:E_cond}
  \end{IEEEeqnarray}
  We conclude the proof by providing an upper bound on $\beta_{1-\epsilon+\tau}$ and a lower bound on $\kappa_\tau$.

\paragraph*{Upper bound on $\beta_{1-\epsilon+\tau}$}
By symmetry, $\beta_{1-\epsilon+\tau}(\mathbb{P}_{\mathbf{Y}_k,B^L|\mathbb{X}=\mathbb{x}}, \mathbb{Q}_{\mathbf{Y}_k,B^L})$ takes the same value  for all $\mathbb{x}\in\mathbb{F}$. To upper-bound $\beta_{1-\epsilon+\tau}$, we apply \cite[Eq.~(103)]{Polyanskiy2010b} to obtain
\begin{multline}
  \log \beta_{1-\epsilon+\tau}  \leq  -{\sup\mathopen{}\Bigg\{ \gamma_0: } \\ \qquad\max_k \prx{\mathbb{x}}{ \log \frac{\mathbb{P}_{\mathbf{Y}_k,B^L|\mathbb{X}}(\mathbf{Y}_k, B^L|\mathbb{x})}{\mathbb{Q}_{\mathbf{Y}_k,B^L}(\mathbf{Y}_k, B^L)}<  \gamma_0 } \leq \epsilon-\tau \Bigg\}.\label{eq:log_beta}
\end{multline}
\begin{sloppypar}\noindent Here, $\mathbb{x}$ is an arbitrary element in $\mathbb{F}$. We further upper-bound \eqref{eq:log_beta} by lower-bounding $\log \frac{\mathbb{P}_{\mathbf{Y}_k,B^L|\mathbb{X}}(\mathbf{Y}_k, B^L|\mathbb{x})}{\mathbb{Q}_{\mathbf{Y}_k,B^L}(\mathbf{Y}_k, B^L)}$ as follows:\end{sloppypar}
\begin{IEEEeqnarray}{rCl}
  \IEEEeqnarraymulticol{3}{l}{\prx{\mathbb{x}}{\log \frac{\mathbb{P}_{\mathbf{Y}_k,B^L|\mathbb{X}}(\mathbf{Y}_k, B^L|\mathbb{x})}{\mathbb{Q}_{\mathbf{Y}_k,B^L}(\mathbf{Y}_k, B^L)} \geq \gamma_0}}\nonumber\\
  \quad &=& \prx{\mathbb{x}}{\imath_k(\mathbb{x}(B^L);\mathbf{Y}_k) +\log\farg{S^L P_{B^L|\mathbb{X}}(B^L|\mathbb{x})} \geq \gamma_0}\IEEEeqnarraynumspace\\
&\geq& \prxbig{\mathbb{x}}{\imath_k(\mathbb{x}(B^L);\mathbf{Y}_k) +\log \fargBig{S^LP_{B^L|\mathbb{X}}(B^L|\mathbb{x})} \geq \gamma_0\nonumber\\
&&\qquad {} , S^LP_{B^L|\mathbb{X}}(B^L|\mathbb{x}) \geq \e{-\zeta}}\IEEEeqnarraynumspace\\
&\geq& \prxbig{\mathbb{x}}{\imath_k(\mathbb{x}(B^L);\mathbf{Y}_k) -\zeta \geq \gamma_0\nonumber\\
&&\qquad{} , S^L P_{B^L|\mathbb{X}}(B^L|\mathbb{x}) \geq \e{-\zeta}}\\
&\geq& \prx{\mathbb{x}}{\imath_k(\mathbb{x}(B^L);\mathbf{Y}_k) - \zeta \geq \gamma_0} \nonumber\\
&&{}- \prx{\mathbb{x}}{S^L P_{B^L|\mathbb{X}}(B^L|\mathbb{x}) < \e{-\zeta}}\label{eq:achievability_second_last}\\
&\geq& \prx{\mathbb{x}}{\imath_k(\mathbb{x}(B^L);\mathbf{Y}_k) - \zeta \geq \gamma_0} - \e{-\zeta}.\label{eq:achievability_last}
\end{IEEEeqnarray}
Here, in \eqref{eq:achievability_second_last}, we have used that $\pr{\mathcal{A},\mathcal{B}} = \pr{\mathcal{A}} - \pr{\mathcal{A}\setminus \mathcal{B}} \geq \pr{\mathcal{A}} - \prbig{\mathcal{B}^\complement}$ for any events $\mathcal{A}$ and $\mathcal{B}$ with $\mathcal{B}^\complement$ being the complement of the set $\mathcal{B}$, and \eqref{eq:achievability_last} follows because 
\begin{IEEEeqnarray}{rCl}
  \IEEEeqnarraymulticol{3}{l}{\prx{\mathbb{x}}{P_{B^L|\mathbb{X}}(B^L|\mathbb{x}) < \e{-\zeta}/S^L}}\nonumber\\
   &=& \sum_{b^L\in\mathcal{S}^L} P_{B^L|\mathbb{X}}(b^L|\mathbb{x})\indi{ P_{B^L|\mathbb{X}}(b^L|\mathbb{x}) < \frac{\e{-\zeta}}{S^L}}\IEEEeqnarraynumspace\\
  &\leq& \sum_{b^L\in\mathcal{S}^L} \frac{\e{-\zeta}}{S^L} \indi{ P_{B^L|\mathbb{X}}(b^L|\mathbb{x}) < \frac{\e{-\zeta}}{S^L}}\IEEEeqnarraynumspace\\
&\leq& \e{-\zeta}.
\end{IEEEeqnarray}
\paragraph*{Lower bound on $\kappa_\tau$}
We follow steps similar to \cite[Eq.~(28)--(29)]{Polyanskiy}. Specifically, for any set $E$ satisfying \eqref{eq:E_cond}, we have
\begin{IEEEeqnarray}{rCl}
\sum_{k} \mathbb{P}_{\mathbf{Y}_k,B^L|\mathbb{X}=\mathbb{x}}[E] \geq \tau \indi{\mathbb{x}\in\mathbb{F}}.\label{eq:kappa_bound1}
\end{IEEEeqnarray}
By averaging \eqref{eq:kappa_bound1} over $\mathbb{X}\sim\mathbb{Q}_{\mathbb{X}}$ in \eqref{eq:PX} and by using \eqref{eq:PYBx_QYB_rel}, we obtain
\begin{IEEEeqnarray}{rCl}
\tau \mathbb{Q}_{\mathbb{X}}\mathopen{}[\mathbb{F}] &\leq& \sum_{k}\sum_{\mathbb{x}\in\mathcal{X}^{S L m}} \mathbb{Q}_{\mathbb{X}}(\mathbb{x}) \mathbb{P}_{\mathbf{Y}_k,B^L|\mathbb{X}=\mathbb{x}}[E]\\
&\leq& S^L\sum_{k} \mathbb{Q}_{\mathbf{Y}_k,B^L}(E).
\end{IEEEeqnarray}
This implies that
\begin{IEEEeqnarray}{rCl}
  \max_k \mathbb{Q}_{\mathbf{Y}_k,B^L}(E) \geq\frac{\tau \mathbb{Q}_{\mathbb{X}}(\mathbb{F})}{2 S^L } .
\end{IEEEeqnarray}
Since $E$ is arbitrary, we have shown that
\begin{IEEEeqnarray}{rCl}
\kappa_\tau \geq \frac{\tau \mathbb{Q}_{\mathbb{X}}(\mathbb{F})}{2 S^{L}}.
\end{IEEEeqnarray}
Next, by using that \cite[Lem~2.6]{Csiszar}
\begin{IEEEeqnarray}{rCl}
  \mathbb{Q}_{\mathbb{X}}(\mathbb{F}) &\geq& (1+m)^{-SL|\mathcal{X}|}
\end{IEEEeqnarray}
we conclude that
\begin{IEEEeqnarray}{rCl}
  \log \kappa_\tau &\geq& \log \frac{\tau}{2} - SL|\mathcal{X}|\log(1+m) -L \log S.\label{eq:log_kappa}
\end{IEEEeqnarray}
We establish the desired result \eqref{eq:nonasymp_achievability} by substituting \eqref{eq:achievability_last} and \eqref{eq:log_kappa} in \eqref{eq:logMf_bound}.

\section{Proof of Theorem~\ref{thm:asymp_variance_cond} (converse) and of Theorem~\ref{thm:asymp_no_variance_cond} (converse)}\label{sec:asymp_converse}

By Theorem~\ref{thm:nonasymptotic_converse}, every $(n,M,\epsilon)$-FLF satisfies
\begin{IEEEeqnarray}{rCl}
  \epsilon
        &\geq&  \frac{1}{2}\prx{J,Y_1^n,Y_2^n}{A(J,Y_1^n,Y_2^n) \leq \log M - \lambda}-\e{-\lambda}\label{eq:converse_asymp1} \IEEEeqnarraynumspace
\end{IEEEeqnarray} 
where
\begin{multline}
A(j,y_1^n,y_2^n) \triangleq \sum_k \eta_k\sum_{i=1}^n \imath_{P^*,W_k}\Big({y_{k,i};f_i(j,y_1^{i-1},y_2^{i-1})}\Big).
\end{multline}
Observe also that 
\begin{IEEEeqnarray}{rCl}
\IEEEeqnarraymulticol{3}{l}{\mathbb{E}_{\mathbb{P}_{Y_{1,i}, Y_{2,i}|Y_1^{i-1},Y_2^{i-1},J=j}}\mathopen{}\Bigg[}\nonumber\\ 
&&\qquad\qquad\sum_k\eta_k \log \frac{W_k(Y_{k,i}|f_i(j,Y^{i-1}_{1}, Y_2^{i-1}))}{P_{Y_k}^*(Y_{k,i})} \Bigg]\nonumber\\
\quad&=& \sum_k\eta_k D(W_k(\cdot|f_i(j,Y_1^{i-1},Y_2^{i-1}))|| P_{Y_k}^*)\\
&\leq& \max_{x\in\mathcal{X}}\mathopen{}\Big\{ \sum_k\eta_k D(W_k(\cdot|x)|| P_{Y_k}^*)\Big\}\\
&=& C+ \max_{x\in\mathcal{X}}\mathopen{}\Big\{ \sum_k\eta_k \diffI{k}{\mathbb{1}_x - P^*}   \Big\}\label{eq:conve_exp_diffI}\\
&=& C.\label{eq:conv_mean_C}
\end{IEEEeqnarray}
Here, in \eqref{eq:conve_exp_diffI}, we let $\mathbb{1}_x$ be the $|\mathcal{X}|$-dimensional vector that has a one in the $x$th entry and zeroes in all other entries and we have used \eqref{eq:differential_feed}. Furthermore, \eqref{eq:conv_mean_C} follows from an application of Lemma~\ref{lem:diffIkzero}.
\paragraph*{Proof of Theorem~\ref{thm:asymp_variance_cond} (converse)}
Next, we apply a Berry-Esseen-type central-limit theorem for martingales due to Bolthausen to estimate the probability $\prx{Y_1^n,Y_2^n|J=j}{ A(j,Y_1^n,Y_2^n) \leq \log M - \eta}$ \cite[p.~2]{ExactConvergenceMartingale} (see also \cite{Bolthausen}) in a manner similar to \cite[Th.~2]{feedback_improve}.
 \begin{theorem}\label{thm:martingale_berry_esseen}
  Let $0< \gamma <\infty$ and let $\{X_i\}_{i=1}^{n}, n\geq 2$, be a martingale difference sequence\footnote{A sequence $\{X_i\}_{i=1}^\infty$ is a martingale difference sequence with respect to the filtration $\{\mathcal{F}_i\}_{i=0}^\infty$ if it satisfies the following two conditions for all $t\in\mathbb{N}$: $X_i$ is $\mathcal{F}_i$-measurable  and $\E{X_i|\mathcal{F}_{i-1}}=0$ almost surely \cite{ExactConvergenceMartingale}.} with respect to the filtration $\{\mathcal{F}_i\}_{i=0}^n$ where $|X_i| \leq \gamma$ almost surely $i=1,\ldots,n$. Let $\sigma^2_i \triangleq \E{X_i^2|\mathcal{F}_{i-1}}$ and $\bar \sigma^2 \triangleq\E{X_i^2}$. Furthermore, assume that
  \begin{IEEEeqnarray}{rCl}
    s^2_n \triangleq \sum_{i=1}^n \bar \sigma^2_i= \sum_{i=1}^n \sigma^2_i 
  \end{IEEEeqnarray}
  almost surely. Then, there exists a constant $0<L(\gamma)<\infty$ that depends only on $\gamma$, so that
  \begin{IEEEeqnarray}{rCl}
    \sup_t \Bigg| \pr{\frac{1}{s_n}\sum_{i=1}^n X_i \leq t} - Q(-t) \Bigg| \leq \frac{ n L(\gamma)  \log n}{s^3_n}.\IEEEeqnarraynumspace
  \end{IEEEeqnarray}
\end{theorem}
To apply this result, we note that, for every $j\in\mathcal{M}$,
  \begin{IEEEeqnarray}{rCl}
   && \Bigg\{ \frac{1}{\sqrt{V}} \Bigg( \sum_k\eta_k \log \frac{W_k(Y_{k,i}|f_i(j,Y^{i-1}_{1}, Y_2^{i-1}))}{P_{Y_k}^*(Y_{k,i})} \nonumber\\
    && \quad {} - \EEbigg{\mathbb{P}_{Y_{1,i},Y_{2,i}|Y_{1}^{i-1},Y_{2}^{i-1},J=j}}{\nonumber\\
    &&\qquad\qquad\sum_k \eta_k \log \frac{W_k(Y_{k,i}|f_i(j,Y^{i-1}_{1}, Y_2^{i-1}))}{P_{Y_k}^*(Y_{k,i})}}  \Bigg) \Bigg\}_{i=1}^n\label{eq:mds_conv}\IEEEeqnarraynumspace
  \end{IEEEeqnarray}
  is a martingale difference sequence with respect to the filtration $\{\sigma(Y_1^{i},Y_2^{i})\}_{i=1}^n$ which is uniformly bounded because $W_k(\cdot|\cdot)$ and $P_{Y^*_k}(\cdot)$ only take a finite number of different values. Furthermore, the conditional variance $\sigma_i^2$ (defined as in Theorem~\ref{thm:martingale_berry_esseen}) of the RVs in \eqref{eq:mds_conv} is equal to $V$ because of \eqref{eq:conv_cond}. It then follows from Theorem~\ref{thm:martingale_berry_esseen} and from \eqref{eq:conv_mean_C} that 
\begin{IEEEeqnarray}{rCl}
  \IEEEeqnarraymulticol{3}{l}{\prx{Y_1^n,Y_2^n|J=j}{A(j,Y_1^n,Y_2^n) \leq \log M - \lambda} }\nonumber\\
  \qquad&\geq& Q\farg{\frac{n C - \log M + \lambda}{\sqrt{n V}}}- \frac{\const \log n}{\sqrt{n}}.\label{eq:converse_berry}\IEEEeqnarraynumspace
\end{IEEEeqnarray}
Here, the constant in the last term depends only on the component channels $\{W_k\}$.
Substituting \eqref{eq:converse_berry} into \eqref{eq:converse_asymp1}, we conclude that
\begin{IEEEeqnarray}{rCl}
\epsilon \geq \frac{1}{2}Q\farg{\frac{n C - \log M + \lambda}{\sqrt{n V}}}- \frac{\const \log n}{\sqrt{n}} - \e{-\lambda}.
\end{IEEEeqnarray}
By solving for $\log M$, by setting $\lambda = \log n$, and by performing a Taylor expansion of $Q^{-1}(\cdot)$ around $2\epsilon$, we obtain
\begin{IEEEeqnarray}{rCl}
  \IEEEeqnarraymulticol{3}{l}{\log M}\nonumber\\
   \quad &\leq&   n C-\sqrt{n V}Q^{-1}\mathopen{}\left(2\left(\epsilon + \frac{\const \log n}{\sqrt{n}} + \e{-\lambda}\right)\right)+\lambda\IEEEeqnarraynumspace\\ 
  &=& n C-\sqrt{n V}Q^{-1}\mathopen{}\left(2\epsilon\right)+\mathcal{O}(\log n)
\end{IEEEeqnarray}
which is the desired result.

\paragraph*{Proof of Theorem~\ref{thm:asymp_no_variance_cond} (converse)}
When \eqref{eq:conv_cond} is not satisfied, we resort to Chebyshev inequality \cite[Eq.~(3.1.1)]{Raginsky2014} instead of the central limit theorem. Specifically, since $\mathcal{X}$ and $\mathcal{Y}$ are finite-cardinality alphabets, there exists a constant $K<\infty$ such that
\begin{multline}
  \Va{\sum_{k}\eta_k \log\frac{W_k(Y_{k,i}|f_i(J,Y^{i-1}_{1}, Y_2^{i-1}))}{P_{Y_k}^*(Y_{k,i})}\bigg|J, Y_1^{i-1}, Y_2^{i-1}} \\\leq K\label{eq:conv_variance}
\end{multline}
for $i\in\{1,\ldots,t\}$.
It follows from \eqref{eq:converse_asymp1}, \eqref{eq:conv_mean_C}, \eqref{eq:conv_variance}, and Chebyshev inequality that 
\begin{IEEEeqnarray}{rCl}
  2(\epsilon + \e{-\lambda})& \geq& \prx{J,Y_1^n,Y_2^n}{A(J,Y_1^n,Y_2^n) \leq \log M -\lambda} \IEEEeqnarraynumspace\\
  &\geq& 1-\frac{K n}{(\log M - n C - \lambda)^2}.\IEEEeqnarraynumspace
\end{IEEEeqnarray}
By solving for $\log M$ and by setting $\lambda = \log n$, we obtain that for all $\epsilon\in(0,1/2)$ and for all sufficiently large $n$
\begin{IEEEeqnarray}{rCl}
  \log M &\leq& n C +\sqrt{\frac{K n}{1-2(\epsilon+1/n)}}+\log n\\ &=& n C +\mathcal{O}(\sqrt{n}).
\end{IEEEeqnarray}

\section{Proof of Theorem~\ref{thm:asymp_variance_cond} (achievability) and of Theorem~\ref{thm:asymp_no_variance_cond} (achievability)}\label{sec:asymp_achievability}
To establish that the right-hand side of \eqref{eq:Mf_conclusion2} in Theorem~\ref{thm:asymp_no_variance_cond} is achievable (which implies achievability in Theorem~\ref{thm:asymp_variance_cond} as well), we start by setting in Theorem~\ref{thm:nonasympotic_achiev} the parameter $S$ to $5$ (recall that $S$ controls the number of subcodewords available in each of the $L$ blocks).
Let now $\kappa$ be the smallest integer larger than $C^{-1}\log 5$. 
Furthermore, set the number of blocks and the number of channel uses per block in Theorem~\ref{thm:nonasympotic_achiev} to
\begin{IEEEeqnarray}{rCl}
L=L_n&\triangleq& \lfloor  n^{1/3} \rfloor \quad\text{  and  }\quad
  m=m_n \triangleq \lfloor n/L_n \rfloor-\kappa\label{eq:mn}. \IEEEeqnarraynumspace
\end{IEEEeqnarray}
We  use the remaining $\kappa L_n$ channel uses to communicate the sequence $B^{L_n}\in \{1,\dots,5\}^{L_n}$ to the decoders. 
Since $\kappa>C^{-1}\log 5$, the rate $\log 5/\kappa$ of the code  is smaller than $C$.
Hence, its error probability can be made to decay exponentially in $L_n$
\begin{IEEEeqnarray}{rCL}
  \epsilon^*(\kappa L_n,5^{L_n }) \leq \ee{ - \omega L_n }
\end{IEEEeqnarray}
for some positive constant $\omega$ \cite[Th.~10.10]{Csiszar}. In Theorem~\ref{thm:nonasympotic_achiev}, we  also set   $\tau= 1/\sqrt{n}$, $\zeta = n^{1/3}$.

Next, we specify the types $\{P_b\}_{b=1}^5$ in Theorem~\ref{thm:nonasympotic_achiev}. 
Let $\rho>0$  (a constant that we shall specify later) and let $\vect{\overline v}_0$ be an arbitrary nonzero vector in $\mathbb{R}_0^{|\mathcal{X}|}$ satisfying
\begin{IEEEeqnarray}{rCl}
\diffI{1}{\vect{\overline v}_0} = \rho.\label{eq:vectbarv}
\end{IEEEeqnarray}
We next define the following probability distributions
\begin{IEEEeqnarray}{rCl}
  \overline P^{(n)}_b &\triangleq& P^* - (-1)^b  \vect{\overline v}_0 n^{-1/3}, \quad b\in\{1,2\}\label{eq:Pb12_def}\\
  \overline  P_3^{(n)} &\triangleq& P^*,\quad \overline  P_4^{(n)} \triangleq P^*_1,\quad \overline  P_5^{(n)} \triangleq P^*_2.\label{eq:Pb345_def}
\end{IEEEeqnarray}
Since $P^*(x)>0$ for all $x\in\mathcal{X}$, the entries of $ \overline  P^{(n)}_b$ are nonnegative and sum to one for all sufficiently large $n$. Therefore, $ \overline  P^{(n)}_b$, $b\in\{1,2\}$,  are legitimate probability distributions that lie in the neighborhood of $P^*$.

Now, let $ P^{(n)}_b$ be the type $Q^{(n)}\in\mathcal{P}_{m_n}(\mathcal{X})$ that minimizes $\vectornormbig{\overline  P^{(n)}_b - Q^{(n)}}$ (recall that $\mathcal{P}_{m_n}(\mathcal{X})$ denotes the set of types of $m_n$-dimensional sequences and $\vectornorm{\cdot}$ denotes the Euclidean distance).
We then choose the types in Theorem~\ref{thm:nonasympotic_achiev} as $P_b=P_{b}^{(n)}$.

Observe that $\vectornormbig{ P_{b}^{(n)} - \overline P_{b}^{(n)}}_1 = \mathcal{O}(1/n)$ and that $I_k(\cdot)$ is differentiable. Hence, we have
\begin{IEEEeqnarray}{rCl}
 I_k(P_{b}^{(n)}) = I_k(\overline P_{b}^{(n)}) + \mathcal{O}(1/n)\label{eq:type_pdf_approximation}
\end{IEEEeqnarray}
for every $b\in\{1,\ldots, 5\}$.

The mappings $\{h_{\ell}\}_{\ell=1}^{L_n}$, which use the feedback to determine which subcodeword to transmit in each of the $L_n$ blocks, are chosen as follows: In the first $L_n-1$ blocks, we let the transmitter choose between subcodewords of type $P_{1}^{(n)}$ and of type $P_{2}^{(n)}$ depending on which decoder has the largest accumulated information density. 
This balances the information densities at the two decoders so that the difference is tightly concentrated around zero (as we shall see later).
In the last block, the transmitter chooses a subcodeword of type $P_{3}^{(n)}$ (which approximates the CAID $P^*$) if the arithmetic average of the information densities is above a suitably chosen threshold $\gamma_1$ (specified in~\eqref{eq:gamma1_def}). 
Otherwise, it chooses uniformly at random between the subcodeword of type $P_{4}^{(n)}$ (which approximates the CAID of $W_1$) and the one of type $P_{5}^{(n)}$ (which approximates the CAID of $W_2$).

Mathematically, we can express the chosen mappings $\{h_{\ell}\}_{\ell=1}^{L_n}$ as follows: $h_1=2$ and for $\ell=2,\dots, L_n-1$,
\begin{IEEEeqnarray}{rCl}
\IEEEeqnarraymulticol{3}{l}{h_{\ell}(\mathbb{x}^{\ell-1}(B^{\ell-1}),\mathbb{Y}_1^{\ell-1}(B^{\ell-1}),\mathbb{Y}_2^{\ell-1}(B^{\ell-1}))}\nonumber\\
 &=& 1 + \indiBig{\imath_1(\mathbb{x}^{\ell-1}(B^{\ell-1}); \mathbb{Y}_{1}^{\ell-1}(B^{\ell-1}))\nonumber\\
 &&\qquad\qquad\qquad{}\geq \imath_{2}(\mathbb{x}^{\ell-1}(B^{\ell-1}); \mathbb{Y}_{2}^{\ell-1}(B^{\ell-1}))}.\IEEEeqnarraynumspace\label{eq:hl1}
\end{IEEEeqnarray}
For $\ell=L_n$, we let $T$ be a uniformly distributed RV on $\{1,2\}$ and set
\begin{IEEEeqnarray}{rCl}
\IEEEeqnarraymulticol{3}{l}{h_{\ell}(\mathbb{x}^{\ell-1}(B^{\ell-1}),\mathbb{Y}_1^{\ell-1}(B^{\ell-1}),\mathbb{Y}_2^{\ell-1}(B^{\ell-1}))}\nonumber\\
&\triangleq & \begin{cases}
3,&\!\!\!\!\!\text{if } \sum_{k}\eta_k\imath_k(\mathbb{x}^{\ell-1}(B^{\ell-1}); \mathbb{Y}_{k}^{\ell-1}(B^{\ell-1}))  \geq \gamma_1\\
4, &\!\!\!\!\!\text{if } \sum_k \eta_k\imath_k(\mathbb{x}^{\ell-1}(B^{\ell-1}); \mathbb{Y}_{k}^{\ell-1}(B^{\ell-1}))< \gamma_1\\
&  \text{and }T = 1\\
5, &\!\!\!\!\!\text{if }\sum_k  \eta_k\imath_k(\mathbb{x}^{\ell-1}(B^{\ell-1}); \mathbb{Y}_{k}^{\ell-1}(B^{\ell-1})) < \gamma_1\\
&\text{and } T = 2.
\end{cases}\IEEEeqnarraynumspace\label{eq:hl2}
\end{IEEEeqnarray}
Roughly speaking, the threshold $\gamma_1$ is chosen so that $P_{3}^{(n)}$, $P_{4}^{(n)}$, and $P_{5}^{(n)}$ are used with probability $(1-2\epsilon)$, $\epsilon$, and $\epsilon$, respectively.

Using Theorem~\ref{thm:nonasympotic_achiev} with the parameters listed above, we obtain
\begin{IEEEeqnarray}{rCl}
 \IEEEeqnarraymulticol{3}{l}{\log \Mf(n,\epsilon)}\nonumber\\
 &\geq&\log \Mf(L_n (m_n + \kappa),\epsilon)\\
 &\geq& \sup\mathopen{}\Bigl\{ \gamma: \max_k \pr{\imath_k(\mathbb{x}(B^{L_n});\mathbb{Y}_k(B^{L_n}))< \gamma}
   \nonumber\\
   &&{}\qquad\qquad\qquad\quad\leq \epsilon  - \epsilon^*(\kappa L_n,S^{L_n})-  1/\sqrt{n} - \e{-n^{1/3}} \Bigr\} \nonumber\\
   && {} +\log\frac{1}{2\sqrt{n}}- S L_n|\mathcal{X}| \log(1+m_n)\nonumber\\
   &&{} - L_n\log S - n^{1/3}\IEEEeqnarraynumspace \\
    &\geq&  \sup\mathopen{}\Bigl\{ \gamma: \max_k \pr{\imath_k(\mathbb{x}(B^{L_n});\mathbb{Y}_k(B^{L_n}))< \gamma} \nonumber\\
    &&\qquad\qquad\qquad\qquad\qquad\qquad\qquad{}\leq \epsilon - c_1 n^{-1/2}\log n \Bigr\} \nonumber\\
   &&{}+ \mathcal{O}(n^{1/3} \log n)\label{eq:Mf_lower_bound}
\end{IEEEeqnarray}
for some $c_1>0$. 
In \eqref{eq:Mf_lower_bound}, we used that
\begin{IEEEeqnarray}{rCl}
\IEEEeqnarraymulticol{3}{l}{\epsilon^*(\kappa L_n, S^{L_n})+1/\sqrt{n} + \e{-n^{1/3}}}\nonumber\\
\qquad &\leq& \e{-\omega L_n}+1/\sqrt{n} + \e{-n^{1/3}}\\
& \leq& c_1 n^{-1/2}\log n
\end{IEEEeqnarray}
for sufficiently large $n$.
To conclude the proof, we show that
\begin{IEEEeqnarray}{rCl}
    \gamma &\triangleq& L_n m_n C - \sqrt{(L_n-1)m_n  V}\nonumber\\
    &&\qquad\qquad{} \times Q^{-1}\fargBig{2\epsilon  - 2\sqrt{2}(c_1+c_2) n^{-1/2}\log n }\IEEEeqnarraynumspace\label{eq:gamma_def_feed}
\end{IEEEeqnarray}
(the constant $c_2>0$ will be defined shortly) satisfies
\begin{equation}
  \pr{\imath_k(\mathbb{x}(B^{L_n});\mathbb{Y}_k(B^{L_n})) < \gamma}
     \leq \epsilon- c_1 n^{-1/2}\log n\label{eq:achiev_lower_bound_conclusion}
\end{equation}
for all sufficiently large $n$. The desired result~\eqref{eq:Mf_conclusion2} then follows by using~\eqref{eq:achiev_lower_bound_conclusion} in~\eqref{eq:Mf_lower_bound} to further lower-bound $\log \Mf(n,\epsilon)$, by observing that $(L_n-1)m_n = n +\mathcal{O}(n^{2/3})$, and by performing a Taylor expansion of $Q^{-1}(\cdot)$ in~\eqref{eq:gamma_def_feed} around $2\epsilon$.
\paragraph*{Proof of \eqref{eq:achiev_lower_bound_conclusion}}
To simplify the notation, we set
\begin{IEEEeqnarray}{rCl}
Z_{k,\ell,i}^{(b)} &\triangleq& \imath_{P_{b}^{(n)},W_k}(\mathbb{x}_{\ell,i}(b); \mathbb{Y}_{k,\ell,i}(b)) \\
\overline Z_{k,\ell_1,\ell_2} &\triangleq& \sum_{\ell=\ell_1}^{\ell_2} \sum_{i=1}^{m_n} Z_{k,\ell,i}^{(B_{\ell})}.
\end{IEEEeqnarray}
The RVs $Z_{k,\ell,i}^{(b)}$ are independent for all $k$, $b$, $\ell$, and $i$.
Note also that $\overline Z_{k,1,L_n}=\imath_k(\mathbb{x}(B^{L_n});\mathbb{Y}_k(B^{L_n}))$. 
Hence,  the left-hand side of~\eqref{eq:achiev_lower_bound_conclusion} can be rewritten as $\pr{\overline Z_{k,1,L_n} < \gamma}$.

Let $\ell(i)\triangleq \lceil i/m_n\rceil$. It will turn out convenient to define the following random quantities as well:
\begin{IEEEeqnarray}{rCl}
Z^*_i &\triangleq&  \sum_{k\in\{1,2\}} \eta_k \imath_{P^*,W_k}\mathopen{}\Big(\mathbb{x}_{\ell(i),i - (\ell(i)-1)m_n}(B_\ell);\nonumber\\
&&\qquad\qquad\qquad\qquad{} \mathbb{Y}_{k,\ell(i),i - (\ell(i)-1)m_n}(B_\ell)\Big)\label{eq:Zstar_i} \IEEEeqnarraynumspace\\
    \overline Z^* &\triangleq& \sum_{i=1}^{(L_n-1)m_n}Z_i^* \IEEEeqnarraynumspace\\
    A &\triangleq&\sum_{\ell=1}^{L_n-1}\sum_{i=1}^{m_n} \sum_{k\in\{1,2\}}\eta_k\log \frac{ P_{B_\ell}^{(n)}W_k(\mathbb{Y}_{k,\ell,i}(B_\ell))}{P_{Y_k}^*(\mathbb{Y}_{k,\ell,i}(B_\ell))}\\
      E &\triangleq&  \overline Z_{1,1,L_n-1}-\overline Z_{2,1,L_n-1}.
\end{IEEEeqnarray}
In the remainder of the proof, we shall make use of the following decomposition:
\begin{equation}\label{eq:decomposition}
  \overline Z_{k,1,L_n-1}= \overline Z^* - A -\eta_{\bar k}(-1)^k  E.
\end{equation}
We next show that $\overline Z^*$ is accurately approximated by a normal distribution with mean $(L_n -1)m_n C$ and variance $(L_n-1) m_n  V$, whereas  both $A$ and $E$ are of order $n^{1/3}$.
To prove these claims, we rely on the Berry-Esseen-type central limit theorem for martingales given in Theorem~\ref{thm:martingale_berry_esseen}, and also on the Hoeffding inequality~\cite{Hoeffding1963}, on the Azuma-Hoeffding inequality \cite[Th.~2.2.1]{Raginsky2014}, and a stabilization lemma, which we state in Lemma~\ref{lem:concentration_overall}. 
\begin{lemma}[Hoeffding inequality \cite{Hoeffding1963}]
If $X_1,\ldots,X_n$ are independent and $a_i \leq X_i \leq b_i$, for $i=1,\ldots,n$, then for $t> 0$,
\begin{multline}
  \pr{\sum_{i=1}^n (X_i- \E{X_i}) \geq n t} \leq \ee{-\frac{2n^2 t^2}{\sum_{i=1}^n (b_i-a_i)^2}}.
\end{multline}
\end{lemma}
\begin{lemma}[Azuma-Hoeffding inequality \cite{Raginsky2014}]\label{lem:azuma}
  Let $\{X_i, \mathcal{F}_i\}_{i=0}^n$ be a real-valued martingale sequence. Suppose that there exist nonnegative real numbers $d_1,\ldots,d_n$, such that $|X_i - X_{i-1}|\leq d_k$ almost surely for all $k\in\{1,\ldots,n\}$. Then, for every $t>0$, 
  \begin{IEEEeqnarray}{rCl}
    \pr{|X_n - X_0|> t} \leq 2\ee{-\frac{t^2}{2\sum_{i=1}^n d_i^2}}.
  \end{IEEEeqnarray}
\end{lemma}
\begin{lemma}[Stabilization lemma]\label{lem:concentration_overall}
Let $\{X_{\ell}^{(1)}\}$ and $\{X_{\ell}^{(2)}\}$ be i.i.d. RVs with mean $\mu_1>0>\mu_2$ satisfying 
\begin{IEEEeqnarray}{rCl}
 \pr{X_{\ell}^{(b)} \geq v} &\leq& \e{- \beta |v-\mu_b|_+^2 }\label{eq:stab_lem_Xellb_ineq1}
 \end{IEEEeqnarray}
and
 \begin{IEEEeqnarray}{rCl}
 \pr{X_{\ell}^{(b)} \leq v} &\leq& \e{- \beta |\mu_b-v|_+^2 }\label{eq:stab_lem_Xellb_ineq2}
\end{IEEEeqnarray}
for all $b\in\{1,2\}$ and all $\ell\in\mathbb{N}$. 
Define the sequence $\{Y_\ell\}$ as follows:  $Y_1 = X_1^{(2)}$ and
\begin{IEEEeqnarray}{rCl}
Y_\ell = \left\{\begin{array}{ll} 
Y_{\ell-1} +  X_\ell^{(1)} & \for Y_{\ell-1} < 0\\
 Y_{\ell-1} +  X_\ell^{(2)} & \for Y_{\ell-1} \geq 0\label{eq:stab_lem_Yell}
 \end{array}\right.
\end{IEEEeqnarray}
for $\ell\geq 2$.
Let $c\geq 1$ satisfy 
\begin{IEEEeqnarray}{rCl}
\min\{\mu_1,-\mu_2\} \geq \sqrt{\frac{\pi}{\beta}}\ee{\frac{c^2}{4}}.\label{eq:concentration_cond}
\end{IEEEeqnarray}
Then
\begin{IEEEeqnarray}{rCl}
  \pr{|Y_\ell| \geq v} \leq 2\e{-c\sqrt{\beta} (v-\mu_1+\mu_2)}.\label{eq:concentration1}
\end{IEEEeqnarray}
\end{lemma}
The proof of this lemma is delegated to Appendix~\ref{app:concentration_proof}.

Let now $c_3>0$ be an arbitrary constant and define the thresholds
\begin{IEEEeqnarray}{rCl}
\gamma_1 &\triangleq& \gamma - m_n C + n^{1/3}\log n \label{eq:gamma1_def} \IEEEeqnarraynumspace\\
    \gamma_2 &\triangleq& (L_n-1)m_n C - c_3 \sqrt{n}\log n.\label{eq:gamma2_def}
\end{IEEEeqnarray} 
Here, $\gamma$ is given in~\eqref{eq:gamma_def_feed}.
Roughly speaking, $\gamma_1$ is chosen so as to be the $2\epsilon$-quantile of $\overline Z^*-A$ for large $n$, whereas  $\gamma_2$ is needed to upper-bound the probability of the (rare) event that $\overline Z^*-A$ is far smaller than its mean. 
Note also that $\gamma_1 > \gamma_2$ for sufficiently large $n$. 
Define the four disjoint events
\begin{IEEEeqnarray}{rCl}
        \mathcal{E}_{1,m} &\triangleq& \Big\{ \overline Z^*-A\in [\gamma_2,\gamma_1),T=m \Big\}, \quad m\in\{1,2\}\IEEEeqnarraynumspace\label{eq:defE1k}\\
    \mathcal{E}_2 &\triangleq& \Big\{ \overline Z^*  - A \geq \gamma_1 \Big\}\label{eq:E2_def}\\
     \mathcal{E}_3&\triangleq& \Big\{ \overline Z^* -A < \gamma_2 \Big\}.
\end{IEEEeqnarray}
Using the events defined above, we obtain the following upper bound by an application of the union bound and the inequality $\pr{A \cap B} \leq \pr{A}$, which holds for any two events $A$ and $B$,
\begin{IEEEeqnarray}{rCl}
\IEEEeqnarraymulticol{3}{l}{\pr{\overline Z_{k,1,L_n} < \gamma}}\nonumber\\
&=& \prBig{\Big(\mathcal{E}_{1,k} \cap \big\{\overline Z_{k,1,L_n} < \gamma\big\}\Big)\cup \Big(\mathcal{E}_{1,\bar k} \cap \big\{\overline Z_{k,1,L_n} < \gamma\big\}\Big)\nonumber\\
&&{} \cup \Big(\mathcal{E}_2 \cap \big\{\overline Z_{k,1,L_n} < \gamma\big\}\Big)\cup \Big(\mathcal{E}_3 \cap \big\{\overline Z_{k,1,L_n} < \gamma\big\}\Big) } \IEEEeqnarraynumspace \\
&\leq& \pr{\mathcal{E}_{1,\bar k} }\nonumber\\
&&{}+\prbig{\mathcal{E}_{1,k} \cap \big\{\overline Z_{k,1,L_n} < \gamma\big\}}+ \prbig{\mathcal{E}_2 \cap \big\{\overline Z_{k,1,L_n} < \gamma\big\}}\nonumber\\
&&{}+\pr{\mathcal{E}_3 }.\IEEEeqnarraynumspace\label{eq:i1_bound}
\end{IEEEeqnarray}
In the following, we shall upper-bound each of these four probabilities. 
It turns out that only $\prbig{\mathcal{E}_{1,\bar k}}$ yields a nonvanishing contribution, which converges to $\epsilon$ from below at a rate $\mathcal{O}(n^{-1/2}\log n)$ as $n\rightarrow \infty$. The remaining three probabilities all vanish at a rate of $\mathcal{O}(1/n)$. This implies that there exists a constant $c_1 > 0$ such that the inequality \eqref{eq:achiev_lower_bound_conclusion} holds for all sufficiently large $n$.

\paragraph*{Bound on $\prbig{\mathcal{E}_{1,\bar k}}$} 
\label{par:bounding_prbig_mathcal_e1_bar_k}
We first establish a bound on the upper-tail probability of $A$ in~\eqref{eq:defE1k}.
For $b\in\{1,2\}$, we have
\begin{IEEEeqnarray}{rCl}
\IEEEeqnarraymulticol{3}{l}{D(P_{b}^{(n)}W_k|| P_{Y_k}^*)}\nonumber\\
  &=& D(W_k ||P_{Y_k}^*| \overline P_{b}^{(n)})-I_k( \overline P_{b}^{(n)})+\mathcal{O}(1/n)\label{eq:sum_eta_Ik}\\ 
 &=& \Big(D(W_k ||P_{Y_k}^*| P^*) -(-1)^b \diffI{k}{\vect{\overline v}_0}n^{-1/3}\Big) \nonumber\\
 &&{}- \Big(I_k(P^*) -(-1)^b \diffI{k}{\vect{\overline v}_0}n^{-1/3}\Big)+\mathcal{O}(n^{-2/3})\label{eq:sum_eta_Ik2}\IEEEeqnarraynumspace\\
  &=& \mathcal{O}(n^{-2/3}).\label{eq:sum_eta_Ik3}
\end{IEEEeqnarray}
Here, \eqref{eq:sum_eta_Ik} follows from the Golden formula $I(P_X, P_{Y|X}) = D(P_{Y|X}||Q_Y|P_X) - D(P_Y||Q_Y)$ and from \eqref{eq:type_pdf_approximation}; \eqref{eq:sum_eta_Ik2} follows from the differentiability of $I_k(\cdot)$ and of $D(W_k||P_{Y_k}^* | \cdot)$ at $P^*$, by Taylor-expanding $I_k(\cdot)$ and $D(W_k||P_{Y_k}^*|\cdot)$ around $P^*$, and by using \eqref{eq:Pb12_def}; finally, \eqref{eq:sum_eta_Ik3} follows because $D(W_k||P_{Y_k}^*|P^*)=I_k(P^*)=C$ for $k\in\{1,2\}$.

Next, it follows from \eqref{eq:Pb12_def} and from the differentiability of $\log(1+x)$ around $x=0$  that there exists a constant $c_4>0$ such that
\begin{IEEEeqnarray}{rCl}
\IEEEeqnarraymulticol{3}{l}{\Bigg|\sum_k \eta_k\log \frac{P_{b}^{(n)}W_k(y)}{P_{Y_k}^*(y)}\Bigg|}\nonumber\\
&=& \Bigg|\sum_k \eta_k\log\fargBig{1- n^{-1/3}(-1)^b\frac{ \sum_{x\in\mathcal{X}} \overline v_{0,x} W_k(y|x)}{P_{Y_k}^*(y)}}\Bigg|\nonumber\\
& \leq& c_4 n^{-1/3}\label{eq:PbWky_div_Py}
\end{IEEEeqnarray}
for sufficiently large $n$. 
Hence, for sufficiently large $n$,
\begin{IEEEeqnarray}{rCl}
  \IEEEeqnarraymulticol{3}{l}{\prBig{ A\geq n^{1/3}\log n}}\nonumber\\
  &=& \prBigg{\sum_{\ell=1}^{L_n-1}\sum_{i=1}^{m_n}\sum_k \eta_k \log \frac{P_{B_\ell}^{(n)}W_k(\mathbb{Y}_{k,\ell,i}(B_\ell))}{P^*_{Y_k}(\mathbb{Y}_{k,\ell,i}(B_\ell))} \nonumber\\
  && \qquad\qquad\qquad\qquad\qquad\qquad\qquad \geq n^{1/3}\log n}\IEEEeqnarraynumspace\label{eq:symmetry_used}\\
  &\leq& \ee{-\frac{\big|n^{1/3}\log n - \const \  n^{-2/3} (L_n-1) m_n \big|_+^2}{2c_4^2 n^{-2/3} (L_n-1)m_n}}\label{eq:PYQY_hoeffding}\IEEEeqnarraynumspace\\
  &\leq& \ee{-\const n^{1/3}\log^2 n}\label{eq:PYQY_hoeffding2} \\
  &=&\mathcal{O}\farg{\frac{1}{n}}.\label{eq:Alnmn_bound}
\end{IEEEeqnarray}
Here, \eqref{eq:PYQY_hoeffding} follows from the Azuma-Hoeffding inequality (see Lemma~\ref{lem:azuma}) applied to the bounded martingale difference sequence \eqref{eq:mds_full}, shown in the top of the next page,\begin{figure*}[!t]
\normalsize
\setcounter{MYtempeqncnt}{\value{equation}}
\setcounter{equation}{161}
\begin{IEEEeqnarray}{rCl}
&&\Bigg\{\sum_k \eta_k \log \frac{ P_{B_{\ell(i)}}^{(n)}W_k(\mathbb{Y}_{k,\ell(i),i-(\ell(i)-1)m_n}(B_{\ell(i)}))}{P^*_{Y_k}(\mathbb{Y}_{k,\ell(i),i-(\ell(i)-1)m_n}(B_{\ell(i)}))}- \EBigg{\sum_k \eta_k \log \frac{ P_{B_{\ell(i)}}^{(n)}W_k(\mathbb{Y}_{k,\ell(i),i-(\ell(i)-1)m_n}(B_{\ell(i)}))}{P^*_{Y_k}(\mathbb{Y}_{k,\ell(i),i-(\ell(i)-1)m_n}(B_{\ell(i)}))} \Bigg| \mathcal{\overline F}_{i-1}}\Bigg\}_{i=1}^{(L_n-1)m_n}\label{eq:mds_full}
\end{IEEEeqnarray}

\setcounter{equation}{\value{MYtempeqncnt}}
\hrulefill
\vspace*{4pt}
\end{figure*}\addtocounter{equation}{1} 
with respect to the filtration 
\begin{IEEEeqnarray}{rCl}
&&\mathcal{\overline F}_i\triangleq \sigma\mathopen\Big\{B_1,\ldots,B_{\ell(i+1)},\nonumber\\
&&\quad{}\mathbb{Y}_{1,1},\ldots,\mathbb{Y}_{1,\ell(i)-1},\mathbb{Y}_{1,\ell(i),1},\ldots,\mathbb{Y}_{1,\ell(i),i-(\ell(i)-1)m_n},\nonumber\\
&&\quad{}\mathbb{Y}_{2,1},\ldots,\mathbb{Y}_{2,\ell(i)-1},\mathbb{Y}_{2,\ell(i),1},\ldots,\mathbb{Y}_{2,\ell(i),i-(\ell(i)-1)m_n}\Big\}.\IEEEeqnarraynumspace
\end{IEEEeqnarray}
In the application of the Azuma-Hoeffding inequality (see Lemma~\ref{lem:azuma}), we also need \eqref{eq:sum_eta_Ik3}, which implies \eqref{eq:mds_upper_bound}, shown in the top of the next page.\begin{figure*}[!t]
\normalsize
\setcounter{MYtempeqncnt}{\value{equation}}
\setcounter{equation}{163}
\begin{equation}
\sum_{i=1}^{(L_n-1)m_n} \EBigg{\sum_k \eta_k \log \frac{ P_{B_{\ell(i)}}^{(n)}W_k(\mathbb{Y}_{k,\ell(i),i-(\ell(i)-1)m_n}(B_{\ell(i)}))}{P^*_{Y_k}(\mathbb{Y}_{k,\ell(i),i-(\ell(i)-1)m_n}(B_{\ell(i)}))} \Bigg| \mathcal{\overline F}_{i-1}}
\leq \const (L_n -1)m_n  n^{-2/3}.\label{eq:mds_upper_bound}
\end{equation}

\setcounter{equation}{\value{MYtempeqncnt}}
\hrulefill
\vspace*{4pt}
\end{figure*}\addtocounter{equation}{1}

Next, we shall invoke the Berry-Esseen-type central limit theorem for martingales given in Theorem~\ref{thm:martingale_berry_esseen}. Specifically, we define the increasing filtration 
\begin{IEEEeqnarray}{rCl}
  \mathcal{F}_i \triangleq \sigma(Z_1^*, \ldots,Z_i^*,B_1,\ldots,B_{\ell(i+1)})
\end{IEEEeqnarray}
and the sequence
\begin{IEEEeqnarray}{rCl}
D_i &\triangleq& \sqrt{\frac{V}{V^{(n)}_{B_{\ell(i)}}}} \big(Z_i^* - \E{Z_i^*|\mathcal{F}_{i-1}}\big)
\end{IEEEeqnarray}
where
\begin{IEEEeqnarray}{rCl}
  V_{b}^{(n)} &\triangleq& \EEbigg{P_{b}^{(n)}\times W_k}{\Vaabigg{W_k}{\sum_k \eta_k\imath_{P^*, W_k}(X;Y_k) \bigg|X}}\IEEEeqnarraynumspace\\
  & =&  V + \mathcal{O}(n^{-1/3}).\label{eq:Vb_approx}\IEEEeqnarraynumspace
\end{IEEEeqnarray}
One can readily verify that $\{D_i\}$ is a martingale difference sequence since $D_i$ is $\mathcal{F}_i$-measurable ($B_{\ell(i)}$ is $\mathcal{F}_{i-1}$-measurable, which implies that $\E{D_i|\mathcal{F}_{i-1}} = \sqrt{V/V_{B_{\ell(i)}^{(n)}}} \E{Z_i^*- \E{Z_i^*|\mathcal{F}_{i-1}}|\mathcal{F}_{i-1}} = 0$). Furthermore, $|D_i|$ is bounded. Next, for all $\ell\in\{1,\ldots,L_n-1\}$, because $\mathbb{x}_\ell(b)$ is of constant composition for $b\in\{1,2\}$ and $\ell\in\{1,\ldots,L_n\}$, it follows that 
\begin{IEEEeqnarray}{rCl}
  \IEEEeqnarraymulticol{3}{l}{\sum_{i=(\ell-1)m_n+1}^{\ell m_n} \E{Z_i^*|F_{i-1}}}\nonumber\\
  \qquad &=& m_n \sum_k \eta_k D(W_k||P_{Y_k}^* |  P_b^{(n)})\\ 
  &=& m_n \sum_k \eta_k \big(C + \diffIbig{k}{ P_b^{(n)}-P^*}\big)\label{eq:EZistar1}\\ 
  &=& m_n C.\label{eq:ED_F_sum}
\end{IEEEeqnarray}
Here, \eqref{eq:EZistar1} follows because $D(W_k||P_{Y_k}^* | P)$ is linear in $P\in\mathcal{P}(\mathcal{X})$ and \eqref{eq:ED_F_sum} follows from Lemma~\ref{lem:diffIkzero}. It also follows that
\begin{IEEEeqnarray}{rCl}
  \IEEEeqnarraymulticol{3}{l}{\sum_{i=(\ell-1)m_n + 1}^{\ell m_n}\E{D_i^2|\mathcal{F}_{i-1}}}\nonumber\\
   &=& \frac{ V}{V_{B_{\ell}}^{(n)}}\sum_{i=(\ell-1)m_n + 1}^{\ell m_n} \E{\left(Z_i^* -\E{Z_i^*|\mathcal{F}_{i-1}}\right)^2\Big|\mathcal{F}_{i-1}}\IEEEeqnarraynumspace\\
  &=& m_n  V.
\end{IEEEeqnarray}
Recall here the definition of $ V$ in \eqref{eq:V_def} and of $Z_i^*$ in \eqref{eq:Zstar_i}.
Moreover, \eqref{eq:Vb_approx}  implies that there exists a constant $d_1>0$ such that
\begin{IEEEeqnarray}{rCl}
   \frac{ V}{V_{b}^{(n)}} \leq \frac{1}{1 - d_1 n^{-1/3}}\label{eq:Va_div_Vbn_approx}
\end{IEEEeqnarray}
for all sufficiently large $n$ and for $b\in\{1,2\}$.

We next upper-bound $\prbig{\mathcal{E}_{1,\bar k}}$ in \eqref{eq:defE1k_used}--\eqref{eq:PrE1kbark}, shown in the top of the next page. 
 Here, \eqref{eq:defE1k_used} follows from \eqref{eq:defE1k}, from \eqref{eq:Alnmn_bound}, and because $T$ is uniform on $\{1,2\}$ and is independent of $(\mathbb{Y}_1,\mathbb{Y}_2)$; \eqref{eq:defE1k_used2} follows from \eqref{eq:ED_F_sum}; and \eqref{eq:Va_div_Vbn_use} follows from \eqref{eq:Va_div_Vbn_approx}. By applying Theorem~\ref{thm:martingale_berry_esseen} to the martingale difference sequence $\{D_i\}$ in \eqref{eq:Va_div_Vbn_use}, we observe that, for sufficiently large $n$,  there exists a constant $c_2>0$ satisfying\addtocounter{equation}{3} 
 \begin{IEEEeqnarray}{rCl}
 \prbig{\mathcal{E}_{1,\bar k}} &\leq&   \frac{1}{2}Q\fargBigg{\frac{ (L_n-1) m_n C -\gamma_1 - n^{1/3}\log n}{\sqrt{ V (L_n-1) m_n (1-d_1 n^{-1/3})} }} \nonumber\\
 &&{}+ \frac{\sqrt{2}c_2 \log n}{\sqrt{n}}\label{eq:berry_used}\IEEEeqnarraynumspace \\
&\leq& \epsilon - \frac{\sqrt{2}c_1 \log n}{\sqrt{n}}\label{eq:PrE1kbark}\IEEEeqnarraynumspace
\end{IEEEeqnarray}
where \eqref{eq:PrE1kbark} follows from \eqref{eq:gamma_def_feed} and \eqref{eq:gamma1_def}.
\begin{figure*}[!t]
\normalsize
\setcounter{MYtempeqncnt}{\value{equation}}
\setcounter{equation}{174}
\begin{IEEEeqnarray}{rCl}
\prbig{\mathcal{E}_{1,\bar k}}&\leq&     \frac{1}{2} \pr{ \overline Z^*  < \gamma_1+n^{1/3}\log n }+ \mathcal{O}(1/n)\label{eq:defE1k_used}\\
 &\leq& \frac{1}{2}\pr{  \sum_{i=1}^{(L_n-1)m_n} \frac{Z_i^* - \E{Z_i^*|\mathcal{F}_{i-1}}}{\sqrt{1 - d_1 n^{-1/3}}} < \frac{\gamma_1 + n^{1/3}\log n - (L_n -1)m_n C}{\sqrt{1 - d_1 n^{-1/3}}} }+ \mathcal{O}(1/n)\label{eq:defE1k_used2}\\
 &\leq& \frac{1}{2}\prBigg{ \frac{1}{\sqrt{ V (L_n - 1)m_n}} \sum_{i=1}^{(L_n-1)m_n} D_i < \frac{\gamma_1 + n^{1/3}\log n - (L_n -1)m_n C}{\sqrt{ V (L_n -1)m_n (1 - d_1 n^{-1/3})}} }+ \mathcal{O}(1/n).\label{eq:Va_div_Vbn_use}\IEEEeqnarraynumspace
 \end{IEEEeqnarray}

\setcounter{equation}{\value{MYtempeqncnt}}
\hrulefill
\vspace*{4pt}
\end{figure*}

\paragraph*{Bound on $\prbig{\mathcal{E}_{1,k} \cap \big\{\overline Z_{k,1,L_n} < \gamma\big\}}$} 
\label{par:bound_on}
We shall first show that $E$ in~\eqref{eq:decomposition} is sufficiently concentrated around $0$. 
Specifically, let 
\begin{IEEEeqnarray}{rCl}
  F_\ell^{(b)} 
  \triangleq \frac{1}{2}\sum_{i=1}^{m_n} (Z_{1,\ell,i}^{(b)}-Z_{2,\ell,i}^{(b)})
\end{IEEEeqnarray}
for $b\in\{1,2\}$.
The  $\big\{F_\ell^{(b)}\big\}$ are independent RVs and $E = \sum_{\ell=1}^{L_n-1}F_\ell^{(B_{\ell})}$.
It follows from \eqref{eq:vectbarv}--\eqref{eq:Pb12_def}, \eqref{eq:type_pdf_approximation}, and from Taylor's theorem that there exist convergent sequences\footnote{We say that a sequence is convergent if its limit exists and is finite.} $\{\xi_{b}^{(n)}\}$ such that
\begin{IEEEeqnarray}{rCl}
 I_1( P_{b}^{(n)}) - I_2( P_{b}^{(n)})&=&  -2(-1)^{b}(\rho  n^{-1/3} +\xi_{b}^{(n)} n^{-2/3}).\IEEEeqnarraynumspace\label{eq:convergent_seq_used}
\end{IEEEeqnarray} 
Consequently, since the $\{\mathbb{x}_{\ell,i}(b)\}_{i=1}^{m_n}$ are of constant composition, it follows that  $\Ebig{F_\ell^{(b)}} = \mu_{b}^{(n)}$, 
where 
 \begin{multline}
 \mu_{b}^{(n)} \\ \triangleq -(-1)^{b}(\rho   m_n  n^{-1/3} +\xi_{b}^{(n)} m_n n^{-2/3}) = \mathcal{O}(n^{1/3}).
 \end{multline}
Since $F_\ell^{(b)}$ is a sum of independent RVs, it follows from Hoeffding inequality that
\begin{IEEEeqnarray}{rCl}
\prbig{F^{(b)}_\ell\geq v} \leq \e{ -\frac{1}{2m_n c_5^2}\left|v -  \mu_{b}^{(n)}\right|_+^2}
\end{IEEEeqnarray}
and 
\begin{IEEEeqnarray}{rCl}
\prbig{F^{(b)}_\ell \leq v} \leq \e{ -\frac{1}{2m_n c_5^2}\left|\mu_{b}^{(n)}-v\right|_+^2 }.
\end{IEEEeqnarray}
Here, $c_5\geq \max_{\ell,i}\maxx{ 1,|Z_{1,\ell,i}-Z_{2,\ell,i}|}$.
We now choose $\rho$ so that  $\rho > \sqrt{4\pi  c_5^2}\e{c_5^2}$.
Next, we apply Lemma~\ref{lem:concentration_overall} with $X_\ell^{(b)}=F^{(b)}_\ell$, $\beta = 1/(2m_n c_5^2)$, and $c=2c_5$ and conclude that, for sufficiently large $n$,
\begin{multline}
    \prbig{|E| \geq n^{1/3}\log n} 
    \\\leq 2\e{- \sqrt{\frac{2}{m_n}} ( n^{1/3}\log n - \mu_{1}^{(n)} + \mu_{2}^{(n)}) }\leq 3/n.\label{eq:ELnm1_bound}
\end{multline}
In the last step,  we used that $\xi_n m_n n^{-2/3} = \mathcal{O}(1)$.

Define 
\begin{IEEEeqnarray}{rCl}
\eta_{\text{max}} = \max_k \eta_k.\label{eq:eta_max}
\end{IEEEeqnarray}
It now follows that there exists a constant $c_6>0$ such that
\begin{IEEEeqnarray}{rCl}
    \IEEEeqnarraymulticol{3}{l}{\pr{\mathcal{E}_{1,k}\cap\left\{\overline Z_{k,1,L_n} < \gamma\right\}}}\nonumber\\
     &\leq& \prbig{\mathcal{E}_{1,k}\cap  \big\{\left|E\right| \leq n^{1/3}\log n,\,\overline Z_{k,1,L_n} < \gamma\big\}}+ 3/n\label{eq:E111}\\
     &\leq&  \pr{\overline Z_{k,L_n,L_n} < \gamma - \gamma_2 + \eta_{\text{max}} n^{1/3}\log n|B_{L_n} = 3+k}\nonumber\\
     &&{}+3/n
      \label{eq:E112}\\
      &\leq& \eeBig{-\frac{c_6}{m_n} \left( m_n (C_k-C) -2c_3 \sqrt{n}\log n  \right)^2 }+ 3/n\IEEEeqnarraynumspace\label{eq:E113}\\
    &=& \mathcal{O}\farg{{1}/{n}}. \label{eq:i1_bound3}
\end{IEEEeqnarray}
Here, \eqref{eq:E111} follows from of \eqref{eq:ELnm1_bound};  \eqref{eq:E112} follows because if $\mathcal{E}_{1,k}$ occurs then $B_{L_n}=3+k$ (see \eqref{eq:hl2}), because $\overline Z^*-A\geq \gamma_2$, and because $\overline Z_{k,1,L_n} = \overline Z^*-A -\eta_{\bar k}(-1)^{k}  E+\overline Z_{k,L_n,L_n}$. Finally, \eqref{eq:E113} follows from \eqref{eq:gamma_def_feed}, from \eqref{eq:gamma2_def}, and from Hoeffding inequality applied to $\overline Z_{k,L_n,L_n}$
\paragraph*{Bound on $\prbig{\mathcal{E}_2 \cap \big\{\overline Z_{k,1,L_n} < \gamma\big\}}$} 
The upper bound on $\prbig{\mathcal{E}_2 \cap \big\{\overline Z_{k,1,L_n} < \gamma\big\}}$ follows by an argument similar to the one we used to establish \eqref{eq:i1_bound3}.  Specifically,
\begin{IEEEeqnarray}{rCl}
  \IEEEeqnarraymulticol{3}{l}{\prbig{\mathcal{E}_2 \cap \big\{\overline Z_{k,1,L_n} < \gamma\big\}}}\nonumber\\
    &\leq& \prBig{\overline Z^*-A\geq \gamma_1,|E|\leq n^{1/3}\log n,\overline Z_{k,1,L_n} < \gamma}\nonumber\\
    &&{} +3/n\label{eq:PrE2union1}\\
&=& \prBig{\overline Z_{k,1,L_n-1}+ \eta_{\bar k}(-1)^k E\geq \gamma_1\nonumber\\
&&\quad ,|E|\leq n^{1/3}\log n,\overline Z_{k,1,L_n} < \gamma,B_{L_n}=3}+3/n\IEEEeqnarraynumspace\label{eq:PrE2_2}\\
&\leq& \prBig{\overline Z_{k,1,L_n-1}\geq \gamma_1 - \eta_{\text{max}} n^{1/3}\log n\nonumber\\
&&\qquad\qquad\qquad\qquad,\overline Z_{k,1,L_n} < \gamma,B_{L_n}=3} +3/n\IEEEeqnarraynumspace\label{eq:PrE2_3}\\
&\leq& \prBig{\overline Z_{k,L_n,L_n} < \gamma - \gamma_1 + \eta_{\text{max}} n^{1/3}\log n,B_{L_n}=3} \nonumber\\
&&{}+3/n\IEEEeqnarraynumspace\label{eq:PrE2_4}\\
&\leq& \prBig{\overline Z_{k,L_n,L_n} < m_n C - (1-\eta_{\text{max}})n^{1/3}\log n |B_{L_n}=3} \nonumber\\
&&{}+3/n.\IEEEeqnarraynumspace\label{eq:PrE2_5}
\end{IEEEeqnarray} 
Here, \eqref{eq:PrE2union1} follows from \eqref{eq:E2_def}, from \eqref{eq:ELnm1_bound}, and from the inequality $\pr{A} \leq \pr{A,B} + \pr{B^{\complement}}$ for events $A$ and $B$ with  $B^\complement$ being the complement of $B$; \eqref{eq:PrE2_2} follows from \eqref{eq:decomposition} and because $\overline Z^* - A\geq \gamma_1$ implies $B_{L_n}=3$ (see \eqref{eq:hl2} and \eqref{eq:E2_def}); \eqref{eq:PrE2_2} follows because $\max_k \eta_k \leq \eta_{\text{max}}$; \eqref{eq:PrE2_4} follows because $\overline Z_{k,1,L_n}-\overline Z_{k,1,L_n-1} = \overline Z_{k,L_n,L_n}$; and \eqref{eq:PrE2_5} follows from \eqref{eq:gamma_def_feed} and \eqref{eq:gamma1_def}.

Finally, we apply Hoeffding inequality to the right-hand side of \eqref{eq:PrE2_5} to obtain the desired result:
\begin{IEEEeqnarray}{rCl}
\prbig{\mathcal{E}_2 \cap \big\{\overline Z_{k,1,L_n} < \gamma\big\}} &\leq& \ee{-\frac{\const n^{2/3}\log^2 n }{m_n}} +3/n\label{eq:PrE2_6}\IEEEeqnarraynumspace\\
&=& \mathcal{O}(1/n).
\end{IEEEeqnarray}
In the application of Hoeffding inequality, we have used that $\eta_{\text{max}}<1$ and that the conditional expectation of $\overline Z_{k,L_n,L_n}$ given $B_{L_n}=3$ is $m_n C$.
\paragraph*{Bound on $\pr{\mathcal{E}_3 }$}
The upper bound on $\pr{\mathcal{E}_3 }$ follows by an application of Azuma-Hoeffding inequality (see Lemma~\ref{lem:azuma}) applied to the martingale difference sequence $\big\{Z_i^* - \E{Z_i^*|\mathcal{F}_{i-1}}\big\}_{i=1}^{(L_n-1)m_n}$ for the filtration 
\begin{IEEEeqnarray}{rCl}
\{\mathcal{F}_i = \sigma(Z_1^*,\ldots,Z_i^*, B_1,\ldots,B_{\ell(i+1)})\}_i. 
\end{IEEEeqnarray}
Specifically,
\begin{IEEEeqnarray}{rCl}
  \IEEEeqnarraymulticol{3}{l}{ \pr{\mathcal{E}_3 } }\nonumber\\
  &=& \prBigg{\overline Z^* -A < (L_n-1)m_n C -c_3\sqrt{n}\log n}\label{eq:PrE3_Abound0}\\
  &\leq&\prBigg{\sum_{i=1}^{(L_n-1)m_n}\Big(Z^*_i -\E{Z_i^*|\mathcal{F}_{i-1}}\Big) \nonumber\\
  &&\qquad\qquad{}< -c_3\sqrt{n}\log n+n^{1/3}\log n}+\mathcal{O}(1/n)\label{eq:PrE3_Abound}\IEEEeqnarraynumspace\\
    &\leq& \ee{-\frac{\const (c_3 \sqrt{n}\log n -n^{1/3}\log n)}{(L_n-1)m_n}}+\mathcal{O}(1/n) \label{eq:PrE3_Abound3}\IEEEeqnarraynumspace\\
    &=& \mathcal{O}(1/n).
\end{IEEEeqnarray}
In  \eqref{eq:PrE3_Abound0}, we have used \eqref{eq:gamma2_def}, \eqref{eq:PrE3_Abound} follows from \eqref{eq:Alnmn_bound}, and \eqref{eq:PrE3_Abound3} follows from Azuma-Hoeffding inequality.

\section{Proof of Theorem~\ref{thm:rcu}}\label{proof:rcu}
We first construct an auxiliary channel that embeds our feedback scheme. To this end, we write the input $X_i$ of the channel as
\begin{IEEEeqnarray}{rCl}
  X_i = \overline X_i + 2 B_i\label{eq:simple_cmdmbc_processor_relation}
\end{IEEEeqnarray}
where $i\in\{1,\ldots,n\}$, $\overline X_i \in\{1,2\}$, and 
\begin{IEEEeqnarray}{rCl}
  B_i &=& \left\{ \begin{array}{ll}
  0 & \text{if } d(\overline X^{i-1}, Y_1^{i-1}) > d(\overline X^{i-1}, Y_2^{i-1}) \\
  1 & \text{if } d(\overline X^{i-1}, Y_1^{i-1}) < d(\overline X^{i-1}, Y_2^{i-1})\\
\overline Z^{(i)} & \text{if } d(\overline X^{i-1}, Y_1^{i-1}) = d(\overline X^{i-1}, Y_2^{i-1}).
  \end{array}
    \right.
\end{IEEEeqnarray}
Here, $d(\cdot,\cdot)$ denotes the Hamming distance.
Through \eqref{eq:simple_cmdmbc_processor_relation}, we define the auxiliary component channels $P_{Y_k^n|\overline X^n} = P_{X^n|\overline X^n} \circ P_{Y^n_k |X^n}$. We note that, by symmetry, the component channels $P_{Y_1^n|\overline X^n}$ and $P_{Y_2^n|\overline X^n}$ are identical, i.e., $P_{Y_1^n|\overline X^n}(y^n|\overline x^n) = P_{Y_2^n|\overline X^n}(y^n|\overline x^n)$ for all $\overline x^n\in\{1,2\}^n$. Hence, the average error probabilities at both decoders are the same. We next apply the RCU bound under minimum distance decoding to the component channel $P_{Y_1^n|\overline X^n}$. 

The remaining steps follow closely the proof of the RCU bound in \cite[Th.~16]{Polyanskiy2010b}.
Consider first the average error probability of a binary codebook $\mathbf{\underline C} = (\vect{C}(1),\ldots,\vect{C}(M)) \in\{1,2\}^{M \times n}$ with a minimum Hamming distance decoder averaged over all randomly generated codebooks with i.i.d. entries uniformly distributed on $\{1,2\}$:
\begin{multline}
 \E{\epsilon_{\mathbf{\underline C}}} 
 \leq \frac{1}{M} \sum_{m=1}^M\EBigg{ \prBigg{\bigcup_{j=1; j\neq m}^M \Big\{d(\vect{C}(j), Y_1^n) \\\leq d(\vect{C}(m), Y_1^n)\Big\}\bigg| \overline X^n = \vect{C}(m)}}.\label{eq:rcu_bound1}
\end{multline}
The symmetry in the probability distribution of the codewords in $\mathbf{\underline C}$ implies that all terms in the summation on the right-hand side of \eqref{eq:rcu_bound1} are identical. Hence, we obtain the inequalities \eqref{eq:epsC1}--\eqref{eq:epsC3}, shown in the top of the next page.\begin{figure*}[!t]
\normalsize
\setcounter{MYtempeqncnt}{\value{equation}}
\setcounter{equation}{205}
\begin{IEEEeqnarray}{rCl}
 \E{\epsilon_{\mathbf{\underline C}}}
  &\leq& \prx{\overline X^n(1),\ldots,\overline X^n(M), Y_1^n}{\bigcup_{j=2}^M \left\{d(\overline X^n(j), Y_1^n) \leq d(\overline X^n(1), Y_1^n)\right\}}\label{eq:epsC1}\\
 &=&\EEbigg{\overline X^n(1), Y_1^n}{\prxbigg{\overline X^n(2),\ldots,\overline X^n(M)}{\bigcup_{j=2}^M \left\{d(\overline X^n(j), Y_1^n) \leq d(\overline X_1^n, Y_1^n)\right\}\Big| \overline X_1^n,Y_1^n}}\\
 &\leq& \EE{\overline X^n(1), Y_1^n}{\minn{1, (M-1)\prx{\overline X^n(2)}{d(\overline X^n(2), Y_1^n) \leq d(\overline X^n(1), Y_1^n)\Big| \overline X^n(1), Y_1^n}}}\label{eq:union_bound_and_one}\\
 &=& \sum_{t=0}^n P_{Z_1^{(n)}}(t) \minn{1, (M-1)\pr{d(\overline X^n(2), Y_1^n) \leq t}}.\label{eq:epsC3}
\end{IEEEeqnarray}

\setcounter{equation}{\value{MYtempeqncnt}}
\hrulefill
\vspace*{4pt}
\end{figure*}\addtocounter{equation}{4} 
Here, 
\begin{multline}
  \mathbb{P}_{\overline X^n(1),\ldots,\overline X^n(M), Y_1^n}(\overline x^n(1),\ldots,\overline x^n(M),y_1^n)\\= 2^{-n M} P_{Y^n_1|\overline X^n(1)}(y_1^n|\overline x^n(1))
\end{multline}
and, in \eqref{eq:union_bound_and_one}, we have bounded the probability by one and used the union bound.
The desired result follows by noting that $d(\overline X^n(2), Y_1^n)$ is a Binomial RV with parameters $n$ and $1/2$. Indeed, $\overline X^n(2)$ is uniformly distributed  on $\{1,2\}^n$, independent of $\overline X^n(1)$ and $Y_1^n$.

\section{Proof of Theorem~\ref{thm:vlf_converse}}\label{proof:vlf_converse}
We first establish the result for deterministic VLF codes with $|\mathcal{U}|$ = 1. At the end of the proof, we show that \eqref{eq:vlf_conv_statement} holds also when  $|\mathcal{U}|>1$ by using the same argument as in \cite[Th.~4]{Polyanskiy2011}.

It will turn out convenient in the remainder of the proof to use the notations of conditional entropy and conditional mutual information given a sigma-algebra $\mathcal{F}$. For two random variables $Z$ and $T$ belonging to discrete spaces $\mathcal{Z}$ and $\mathcal{T}$, these two quantities are defined as follows:
\begin{IEEEeqnarray}{rCl} 
  \mathcal{H}(Z | \mathcal{F}) &\triangleq& -\sum_{z\in\mathcal{Z}} \pr{Z=z|\mathcal{F}}\log \pr{Z=z|\mathcal{F}}\\
  \mathcal{I}(Z ;T |\mathcal{F}) &\triangleq& \nonumber\\
  \IEEEeqnarraymulticol{3}{l}{ \sum_{(z,t)\in\mathcal{Z}\times \mathcal{T}} \pr{Z=z,T=t|\mathcal{F}} \log \frac{\pr{Z=z,T=t|\mathcal{F}}}{\pr{Z=z|\mathcal{F}}\pr{T=t|\mathcal{F}}}.}\IEEEeqnarraynumspace
\end{IEEEeqnarray}
As in \cite{Truong2017}, we simply write $Y^n$ in place of $\sigma(Y^n)$ when $\mathcal{F}=\sigma(Y^n)$. 

Fix an arbitrary $(\vlflen,M,\epsilon)$-VLF code with $0\leq \epsilon \leq 1 -1/M$ and $|{\cal U}| = 1$. Now, by \cite[Lem.~1]{Burnashev1976} (see also \cite[Lem.~3]{Truong2017}), we have that
\begin{IEEEeqnarray}{rCl}
  \E{\mathcal{H}(J | Y_k^{\tau_k})} \leq h(\epsilon) + \epsilon \log M.
\end{IEEEeqnarray}
Here, the expectation is with respect to $Y_1^\infty, Y_2^\infty, \tau_1$ and $\tau_2$.
Using that $\log M = H(J) = \E{\mathcal{H}(J|Y_k^0)}$ and that $\min_{k\in\{1,2\}} a_k \leq \sum_{k}\eta_k a_k$ for all $(a_1,a_2)\in\mathbb{R}_+^2$, we obtain
\begin{IEEEeqnarray}{rCl} 
  \IEEEeqnarraymulticol{3}{l}{(1-\epsilon)\log M}\nonumber\\\quad &\leq& \min_k \Big\{ \E{\mathcal{H}(J|Y_k^0) - \mathcal{H}(J|Y_k^{\tau_k})} \Big\} + h(\epsilon)\\ 
  &\leq& \sum_k \eta_k \Big( \E{\mathcal{H}(J|Y_k^0) - \mathcal{H}(J|Y_k^{\tau_k})} \Big) + h(\epsilon).\label{eq:fano_result}
\end{IEEEeqnarray}
We continue from \eqref{eq:fano_result} by using the telescoping sum $\sum_{i=1}^{\tau_k} \Big(\mathcal{H}(J|Y_k^{i-1}) - \mathcal{H}(J|Y_k^{i})\Big) = \mathcal{H}(J|Y_k^0) - \mathcal{H}(J|Y_k^{\tau_k})$ to obtain the following chain of inequalities:
\begin{IEEEeqnarray}{rCl}
  \IEEEeqnarraymulticol{3}{l}{(1-\epsilon) \log M-h(\epsilon)}\nonumber\\
  &\leq& \sum_k \eta_k \Ebigg{\sum_{i=1}^{\tau_k} \big(\mathcal{H}(J|Y_k^{i-1}) - \mathcal{H}(J|Y_k^{i})\big)} \label{eq:expectations_are_with_respect_to_tau}\\
    &=& \sum_k \eta_k \Ebigg{\sum_{i=1}^{\infty}\indi{\tau_k \geq i} \left(\mathcal{H}(J|Y_k^{i-1}) - \mathcal{H}(J|Y_k^{i})\right)} \\
&=& \sum_{i=1}^{\infty} \Ebigg{\sum_k \eta_k\indi{\tau_k \geq i} \left(\mathcal{H}(J|Y_k^{i-1}) - \mathcal{H}(J|Y_k^{i})\right)}. \label{eq:tonelli_used2}\IEEEeqnarraynumspace
\end{IEEEeqnarray}
Here, \eqref{eq:tonelli_used2} follows from Fubini's theorem \cite[Th.~18.3]{billingsley} because $\sum_{i=1}^\infty |\indi{\tau_k \geq i}(\mathcal{H}(J|Y_k^{i-1})-\mathcal{H}(J|Y_k^{i}))| \leq \tau_k \log M$ implies that $\indi{\tau_k \geq i}(\mathcal{H}(J|Y_k^{i-1})-\mathcal{H}(J|Y_k^{i}))$ is integrable. Next, we apply to \eqref{eq:tonelli_used2} the law of total expectation, where the outer expectation is with respect to $(Y_{1}^{i-1},Y_{2}^{i-1})$ and the inner expectation is with respect to $(Y_{1,i},Y_{2,i})$, and use that the event $\{\tau_k\geq i\}$ is $\mathcal{F}_{k,i-1}$-measurable to obtain 
\begin{IEEEeqnarray}{rCl}
\IEEEeqnarraymulticol{3}{l}{(1-\epsilon) \log M-h(\epsilon)}\nonumber\\
&\leq& \sum_{i=1}^{\infty} \Ebigg{\Ebigg{\sum_k \eta_k\indi{\tau_k \geq i} \nonumber\\
&&\qquad\quad{}\times\left(\mathcal{H}(J|Y_k^{i-1}) - \mathcal{H}(J|Y_k^{i})\right)\Big| Y_{1}^{i-1},Y_{2}^{i-1}}} \label{eq:law_of_total_expectation}\IEEEeqnarraynumspace\\
&=& \sum_{i=1}^{\infty} \Ebigg{\sum_k \eta_k\indi{\tau_k \geq i} \nonumber\\
&&{}\qquad\quad{}\times \EBig{\mathcal{H}(J|Y_k^{i-1}) - \mathcal{H}(J|Y_k^{i})\Big| Y_{1}^{i-1},Y_{2}^{i-1}}} \label{eq:tauk_is_measurable}\\
&=& \sum_{i=1}^{\infty} \Ebigg{\sum_k \eta_k\indi{\tau_k \geq i}  \mathcal{I}(J; Y_{k,i}|Y_k^{i-1})} \\
&=&  \sum_{i=1}^{\infty}\bigg(\pr{\tau_1 \geq i,\tau_2\geq i}\nonumber\\
&&\qquad\quad{}\times\Ebigg{ \sum_k \eta_k \mathcal{I}(J;Y_{k,i}|Y_k^{i-1}) \bigg|\tau_1 \geq i,\tau_2\geq i} \nonumber\\
&&\qquad{}+ \sum_k  \eta_k\pr{\tau_k \geq i,\tau_{\bar k}< i}\nonumber\\
&&\qquad\qquad\quad{}\times \E{ \mathcal{I}(J;Y_{k,i}|Y_k^{i-1}) \Big|\tau_k \geq i,\tau_{\bar k}< i} \bigg).\label{eq:vlf_sum_bound}\IEEEeqnarraynumspace
\end{IEEEeqnarray}

Now, the Markov chain $(Y_1^{i-1},Y_2^{i-1}, J) - X_i - (Y_{1,i},Y_{2,i})$ and the data-processing inequality imply that
\begin{IEEEeqnarray}{rCl}
\mathcal{I}(J;Y_{k,i}|Y_k^{i-1}) &\leq& \mathcal{I}(X_i;Y_{k,i}|Y_k^{i-1}).\label{eq:dataprocessing}
\end{IEEEeqnarray}
We also note that the chain of inequalities  \eqref{eq:vlf_conv_lemma1_used_0}--\eqref{eq:vlf_conv_lemma1_used} in the top of the next page hold.\begin{figure*}[!t]
\normalsize
\setcounter{MYtempeqncnt}{\value{equation}}
\setcounter{equation}{223}
\begin{IEEEeqnarray}{rCl}
  \Ebigg{\sum_k \eta_k \mathcal{I}(X_i;Y_{k,i}|Y_k^{i-1})\bigg|\tau_1 \geq i ,\tau_2 \geq i}
   &=& \sum_k \eta_k \E{I_k\farg{P_{X_i|Y_{1}^{i-1},Y_2^{i-1}}\farg{\cdot|Y_1^{i-1},Y_2^{i-1}}} \Big| \tau_1 \geq i,\tau_2\geq i}\label{eq:vlf_conv_lemma1_used_0}\\
   &\leq& \sum_k \eta_k I_k\farg{\E{P_{X_i|Y_{1}^{i-1},Y_2^{i-1}}\farg{\cdot|Y_1^{i-1},Y_2^{i-1}}\Big| \tau_1 \geq i,\tau_2\geq i}}\label{eq:vlf_conv_jensen}\\
   &\leq& 
    C + \sum_k\eta_k\diffIBig{k}{\E{P_{X_i|Y_{1}^{i-1},Y_2^{i-1}}\farg{\cdot|Y_1^{i-1},Y_2^{i-1}}\Big| \tau_1 \geq i,\tau_2\geq i}-P^*}
    \label{eq:vlf_conv_concavity_of_mut_inf3}\IEEEeqnarraynumspace\\
   &=& C.\label{eq:vlf_conv_lemma1_used}
\end{IEEEeqnarray}

\setcounter{equation}{\value{MYtempeqncnt}}
\hrulefill
\vspace*{4pt}
\end{figure*}\addtocounter{equation}{4} 
Here, \eqref{eq:vlf_conv_jensen} follows from Jensen's inequality and from the concavity of mutual information, \eqref{eq:vlf_conv_concavity_of_mut_inf3} follows again from the concavity of mutual information, and \eqref{eq:vlf_conv_lemma1_used} follows from Lemma~\ref{lem:diffIkzero}. Similarly, we also have that $\mathcal{I}(X_i,Y_{k,i}|Y_k^{i-1}) \leq C_k$. Substituting \eqref{eq:dataprocessing} and \eqref{eq:vlf_conv_lemma1_used} in \eqref{eq:vlf_sum_bound}, we find that
\begin{IEEEeqnarray}{rCl}
\IEEEeqnarraymulticol{3}{l}{(1-\epsilon)\log M-h(\epsilon)}\nonumber\\ &\leq& \sum_{i=1}^{\infty}\pr{\tau_1 \geq i,\tau_2\geq i}C \nonumber \\
&&{}+ \sum_k \eta_k C_k \sum_{i=1}^\infty \pr{\tau_k \geq i,\tau_{\bar k}< i}\\
&=&  \EBig{\min_k \tau_k}C + \sum_k \Ebig{|\tau_k - \tau_{\bar k}|_+}  \eta_k C_k\\
&\leq&  \EBig{\min_k \tau_k}C +  \Ebig{|\tau_k - \tau_{\bar k}|}\max_k \eta_k C_k  \label{eq:vlf_conv2}\\
&=&  \EBig{\max_k \tau_k}C +  \Ebig{|\tau_k - \tau_{\bar k}|}\Big(\max_k \eta_k C_k-C\Big)  \IEEEeqnarraynumspace\\
&\leq& \vlflen C. \label{eq:vlf_conv}
\end{IEEEeqnarray}
In \eqref{eq:vlf_conv2}, we have used that $\sum_k \Ebig{|\tau_k - \tau_{\bar k}|_+} =\Ebig{|\tau_k - \tau_{\bar k}|} $ and in \eqref{eq:vlf_conv}, we have used \eqref{eq:avg_blocklength_const_feed} and that
\begin{IEEEeqnarray}{rCl}
  \eta_1 C_1 &\leq& \eta_1 C + \eta_1 \diffI{1}{P_1^* - P^*}\label{eq:concavity_of_mut_inf}\\
  &=& \eta_1 C - \eta_2 \diffI{2}{P_1^* - P^*}\label{eq:lem1_used}\\
  &\leq& \eta_1 C + \eta_2 C\label{eq:concavity_of_mut_inf2}\\
  &=& C{}
\end{IEEEeqnarray}
where \eqref{eq:concavity_of_mut_inf} follows from the concavity of mutual information, \eqref{eq:lem1_used} follows from Lemma~\ref{lem:diffIkzero}, and \eqref{eq:concavity_of_mut_inf2} follows because, by the concavity of mutual information, 
$C + \diffI{2}{P^*_1-P^*} \geq I_2(P_1^*) \geq 0$. The same argument shows that $\eta_2 C_2 \leq C$.

Now, consider an arbitrary $(\mathbb{n},M,\epsilon)$-VLF code with $0\leq \epsilon \leq 1 -1/M$ but without a constraint on $|\mathcal{U}|$. Then, the argument outlined above shows that
\begin{IEEEeqnarray}{rCl}
  \log M &\leq &C \Ebig{\max_k \tau_k\big| U=u} \nonumber\\
  &&{}+ \sum_k \eta_k \Big[h\mathopen{}\Big(\prbig{J \neq \hat J_k|U=u}\Big)\nonumber\\
  &&\qquad\qquad\quad{} + \prbig{J \neq \hat J_k|U=u} \log M \Big]\IEEEeqnarraynumspace
\end{IEEEeqnarray}
for every $u\in\mathcal{U}$.
By taking expectations with respect to $U$ on both sides and by applying Jensen's inequality to the binary entropy function, we obtain
\begin{IEEEeqnarray}{rCl}
  \IEEEeqnarraymulticol{3}{l}{\log M}\nonumber\\
   \quad&\leq &C \Ebig{\max_k \tau_k}\nonumber\\
  &&{} + \sum_k \eta_k \Big(h\mathopen{}\Big(\prbig{J \neq \hat J_k}\Big) + \prbig{J \neq \hat J_k} \log M \Big)\IEEEeqnarraynumspace\\
  &\leq &C \Ebig{\max_k \tau_k} + h(\epsilon) + \epsilon \log M. \label{eq:var_conv_final}
\end{IEEEeqnarray}
Here, we have also used that $\prbig{J\neq \hat J_k}\leq \epsilon  \leq 1-1/M$, that $h(x) + x \log M $ is an increasing function of $x\in [0,1-1/M]$,  and that $\eta_1 + \eta_2 = 1$.
Solving \eqref{eq:var_conv_final} for $\log M$ establishes the desired result in \eqref{eq:vlf_conv_statement}.

\section{Proof of Theorem~\ref{thm:simple_achiev_vlf}}\label{app:proof_vlf_achiev}
The proof follows closely the proofs of \cite[Th.~3]{Polyanskiy2011} and \cite[Th.~1]{Trillingsgaard2018_VLF}. We let $\Coopvar$ be a Bernoulli RV with $\pr{\Coopvar=1}=q$ and let its probability mass function be given by $P_\Upsilon$. The RV $U$ has the following domain $\mathcal{U}$ and probability mass function $P_U$
\begin{IEEEeqnarray}{rCl}
  \mathcal{U}&\triangleq& \{0,1\}\times \underbrace{\mathcal{X}^{m}\times\cdots\times \mathcal{X}^{m}}_{S Lm\text{ times}} \times \mathcal{X}^{\infty}\\
  P_U &\triangleq& P_\Upsilon\times \underbrace{P_1\times \cdots\times P_1}_{Lm\text{ times}}\times\cdots \times \underbrace{P_\Upsilon\times \cdots\times P_\Upsilon}_{Lm\text{ times}}\nonumber\\
  &&{} \times (P^*)^\infty.
\end{IEEEeqnarray}
The realization of $U=u$ implicitly defines codebooks $\{\mathbf{C}^{(u)}_{j,\ell}(b)\in\mathcal{X}^m\}$ and $\{\mathbf{\bar C}^{(u)}_j\in\mathcal{X}^\infty\}$ for $(j,b,\ell)\in\mathcal{M}\times\{1,2\}\times\{1,\ldots,L\}$. The encoding functions are defined as 
\begin{IEEEeqnarray}{rCl}
  f_{\ellitoind(\ell,i)}(u, j, y_1^{\ellitoind(\ell,i-1)},y_2^{\ellitoind(\ell,i-1)})&&\nonumber\\
  \IEEEeqnarraymulticol{3}{l}{\qquad\qquad\triangleq \Big(\mathbf{C}^{(u)}_{j,\ell}\Big(h_{\ell}\Big(x^{\ellitoind(\ell,0)},y_1^{\ellitoind(\ell,0)},y_2^{\ellitoind(\ell,0)}\Big)\Big)\Big)_{i}} \IEEEeqnarraynumspace\\
  f_{L m + n_B + i}(u, j,\cdot,\cdot) &\triangleq&(\mathbf{\bar C}^{(u)}_{j})_i
\end{IEEEeqnarray}
for $\ell\in\{1,\ldots,L\}$, $i\in\{1,\ldots,m\}$, and $i\in\{1,\ldots,\infty\}$
(recall that $\ellitoind(\ell,i) = (\ell-1)m + i$). At the time indices $L m + 1,\ldots,L m + n_B$, the transmitter sends $B^L$ using a fixed-blocklength code. 
To keep the notation compact, we omit the superscript $(u)$ in the remaining part of the proof. 

At time $n\geq L m+n_B$, decoder $k$ computes the information densities
\begin{IEEEeqnarray}{rCl}
  \IEEEeqnarraymulticol{3}{l}{A_{k,n}(j) }\nonumber\\
  &\triangleq& \sum_{\ell=1}^L \sum_{i=1}^m \imath_{P_{B_\ell},W_k}\left(f_{\ellitoind(\ell,i)}(U,Y_1^{\ellitoind(\ell,i)-1},Y_2^{\ellitoind(\ell,i)-1}); Y_{k,\ellitoind(\ell,i)}\right)\nonumber\\
  &&{} + \sum_{i=L m +n_B+1}^{n}\imath_{P^*,W_k}(f_{i}(U,j,\cdot,\cdot); Y_{k,i})\label{eq:Skn_def_feed}
\end{IEEEeqnarray}
for $j\in \mathcal{M}$. Define the stopping times
\begin{IEEEeqnarray}{rCl}
  \tau_k(j) &=& \indi{\Coopvar=0}\minnBig{\inff{n\geq Lm+n_B:A_{k,n}(j)\geq \gamma }\nonumber\\
  &&{}\qquad\qquad\qquad\qquad\qquad\qquad,\tau_{\text{max}} + n_B}\IEEEeqnarraynumspace 
\end{IEEEeqnarray}
and let $\tau^*_k$ be the time at which decoder $k$ makes the final decision:
\begin{IEEEeqnarray}{rCl}
  \tau^*_k \triangleq \min_{j\in\mathcal{M}}\tau_k(j).\label{eq:tau_min_feed}
\end{IEEEeqnarray}
The output of decoder $k$ at time $\tau^*_k$ is 
\begin{IEEEeqnarray}{rCl}
  g_{k,\tau^*_k}(U,Y^{\tau_k^*}_k,B^L) \triangleq \max\{j\in\mathcal{M}: \tau_k(j)=\tau^*_k\}.\label{eq:simple_achiev_decoder_feed}
\end{IEEEeqnarray}
When $\Coopvar=1$, we have $\tau_k^*=0$, and hence the decoder $g_{k,\tau_k^*}(U,Y^{\tau_k^*}_k)$ outputs $M$. The average blocklength is then given by
\begin{IEEEeqnarray}{rCl}
  \IEEEeqnarraymulticol{3}{l}{\EBig{\max_k \tau^*_k}}\nonumber\\
  \quad &=& (1-q)\EBig{\max_k \tau^*_k | \Coopvar= 0} \\
 &\leq& (1-q)\frac{1}{M}\sum_{j=1}^M \EBig{\max_k\tau_k(j)|J=j,\Coopvar=0} \label{eq:simple_achiev_Emax_1_feed}\\
&=& (1-q)\E{\max\tau_k(1)|J=1,\Coopvar=0}\label{eq:simple_achiev_Emax_sym_feed}\\
&=& (1-q)\Big(\EBig{\max_k\tau_k}+n_B\Big)\label{eq:simple_achiev_Emax_3_feed}
\end{IEEEeqnarray}
where the expectation is over $J$, $Y_1^{\infty}$, $Y_2^{\infty}$, and $U$ ($B^L$ is deterministic given these RVs). Here, \eqref{eq:simple_achiev_Emax_1_feed} follows from \eqref{eq:tau_min_feed}; \eqref{eq:simple_achiev_Emax_sym_feed} follows from symmetry; and \eqref{eq:simple_achiev_Emax_3_feed} follows from the definition of $\tau_k$ in \eqref{eq:tau_k_def_feed}.
We next bound the average error probability:
\begin{IEEEeqnarray}{rCl}
  \IEEEeqnarraymulticol{3}{l}{\pr{ g_{k,\tau^*_k}(U, Y^{\tau^*_k}_k,B^L)\not= J} }\nonumber\\
    &\leq& \sprob + (1-\sprob)\pr{ g_{k,\tau^*_k}(U, Y^{\tau^*_k}_k,B^L) \not=  J| \Upsilon=0} \nonumber\\
    &&{}+(1-q)\epsilon^*(n_B,S^L)\\
   &\leq& \sprob + (1-\sprob) \pr{ g_{k,\tau^*_k}(U, Y^{\tau^*_k}_k,B^L)\not= 1 | J=1,\Upsilon=0} \nonumber\\
   &&{}+(1-q)\epsilon^*(n_B,S^L).\label{eq:simple_achiev_error_1_feed}
   \end{IEEEeqnarray}
Here, \eqref{eq:simple_achiev_error_1_feed} follows from \eqref{eq:simple_achiev_decoder_feed}. We proceed with bounding probability term in \eqref{eq:simple_achiev_error_1_feed}:
\begin{IEEEeqnarray}{rCl}
\IEEEeqnarraymulticol{3}{l}{\pr{ g_{k,\tau^*_k}(U, Y^{\tau^*_k}_k,B^L)\not= 1 | J=1,\Upsilon=0} }\nonumber\\
\qquad&\leq& \pr{\tau_k(1) \geq \tau^*_k|\Upsilon=0}\\
&=&  \prBigg{ \bigcup_{j=2}^M \{\tau_k(1)\geq\tau_k(j)  \} \Bigg|\Upsilon=0}\\
&\leq& (M-1)\pr{\tau_k(1)\geq\tau_k(2)|\Upsilon=0 } \IEEEeqnarraynumspace\\
&=& (M-1)\pr{\tau_k\geq\bar \tau_k}.\label{eq:simple_achiev_error_final_feed}
\end{IEEEeqnarray}
Here, \eqref{eq:simple_achiev_error_final_feed} follows by noting that, given $J=1$, the RVs 
\begin{IEEEeqnarray}{rCl}
\left(A_{k,n}(1), A_{k,n}(2), \tau_k(1),\tau_k(2)\right)
\end{IEEEeqnarray}
where the RVs $\{A_{k,n}(j)\}$ are defined in \eqref{eq:Skn_def_feed}, have the same joint distribution as 
\begin{IEEEeqnarray}{rCl}
\left(A_{k,n}(X^n;Y_k^n), A_{k,n}(\bar X^n;Y_k^n), \tau_k+n_B,\bar \tau_k+n_B\right).\IEEEeqnarraynumspace
\end{IEEEeqnarray}
We conclude the proof by noting that \eqref{eq:simple_achiev_Emax_3_feed}, \eqref{eq:simple_achiev_error_1_feed}, and \eqref{eq:simple_achiev_error_final_feed} imply that the tuple $(U,h_\ell,f_n, \{g_{k,n}\}, \{\tau_k^*\})$ defines an $(\vlflen,M,\epsilon)$-VLSF code satisfying \eqref{eq:achiev_Emax_feed} and \eqref{eq:achiev_Proberror1_feed}.

\section{Proof of Theorem~\ref{thm:asymp_vlf_achiev}}\label{proof:asymp_vlf_achiev}
We let  $\bar\vlflen$ be a variable related to the average blocklength $\vlflen$ (it is roughly equal to $(1-\epsilon)\bar\vlflen$), and we analyze the achievability bound in Theorem~\ref{thm:simple_achiev_vlf} in the limit $\bar\vlflen\rightarrow \infty$. We set $S=2$ and let $\kappa$ be the smallest integer strictly larger than $C^{-1}\log 2$. We also set 
\begin{IEEEeqnarray}{rCl}
L=L_{\bar\vlflen}&\triangleq& \lfloor  (\bar\vlflen- \sqrt{\bar\vlflen}\log \bar\vlflen)^{1/3} \rfloor\\
m=m_{\bar\vlflen} &\triangleq& \Big\lfloor  \frac{(\bar\vlflen- \sqrt{\bar\vlflen}\log \bar\vlflen)^{1/3}}{L_{\bar\vlflen} }\Big\rfloor \IEEEeqnarraynumspace\\
\bar\vlflen_- &\triangleq& L_{\bar\vlflen}m_{\bar\vlflen}\\
\tau_{\text{max}}=\bar\vlflen_+ &\triangleq& \lfloor\bar\vlflen + \sqrt{\bar\vlflen}\log \bar\vlflen\rfloor
\end{IEEEeqnarray}
and $n_B \triangleq \kappa L_{\bar\vlflen}=\mathcal{O}(\bar\vlflen^{1/3})$. Note that $\bar\vlflen_+ - \bar\vlflen_- = \mathcal{O}(\sqrt{\bar\vlflen}\log \bar\vlflen)$. As in the proof of Theorem~\ref{thm:asymp_no_variance_cond}, we have that
\begin{IEEEeqnarray}{rCl}
  \epsilon^*(n_B,2^{L_{\bar\vlflen}}) \leq \ee{-\omega L_{\bar\vlflen}}\triangleq \epsilon_{\bar\vlflen}^* 
\end{IEEEeqnarray}
for some positive constant $\omega$. In Appendix~\ref{sec:proof_of_tau_k_nplus}, we prove that there exists a constant $c_1$ such that, for all sufficiently large $\bar\vlflen$,
\begin{IEEEeqnarray}{rCl}
  \max_k\pr{\tau_k = \bar\vlflen_+} \leq \frac{c_1}{\bar\vlflen}.\label{eq:asymp_vlf_tauk_vlflen}
\end{IEEEeqnarray}

Next, we let $\vect{\overline v}_0\in\mathbb{R}^{|\mathcal{X}|}_0$ be an arbitrary vector satisfying $\diffI{1}{\vect{\overline v}_0} > 0 > \diffI{2}{\vect{\overline v}_0}$. We define the probability distributions $P_1$ and $P_2$ as follows
\begin{IEEEeqnarray}{rCl}
  P_b^{(\vlflen)} \triangleq P^* - (-1)^b \vect{\overline v}_0 n^{-1/3}
\end{IEEEeqnarray}
for $b\in\{1,2\}$. For sufficiently large $\vlflen$, $P_b^{(\bar\vlflen)}$ are in the neighborhood of $P^*$ and are legitimate probability distributions (recall that $P^*(x)>0, \forall x\in\mathcal{X}$). 
Moreover, for $\ell\in\{1,\ldots,L_{\bar\vlflen}\}$, we recursively define $\{h_\ell\}_{\ell=1}^{L_{\bar\vlflen}}$ and $\{B_\ell\}_{\ell=1}^{L_{\bar\vlflen}}$ as follows
\begin{IEEEeqnarray}{rCl}
\IEEEeqnarraymulticol{3}{l}{h_{\ell}(X^{\ellitoind(\ell,0)},Y_1^{\ellitoind(\ell,0)},Y_2^{\ellitoind(\ell,0)})}\nonumber\\
\qquad\qquad\qquad&\triangleq& 1 + \indiBig{ \overline Z_{1,\ellitoind(\ell,0)} \geq  \overline Z_{2,\ellitoind(\ell,0)}}\IEEEeqnarraynumspace\\
B_{\ell} &\triangleq& h_{\ell}(X^{\ellitoind(\ell,0)},Y_1^{\ellitoind(\ell,0)},Y_2^{\ellitoind(\ell,0)}).
\end{IEEEeqnarray}
Here,  $\ellitoind(\ell,i) = (\ell-1) m_n+i$ and
\begin{IEEEeqnarray}{rCl}
Z_{k,t} &\triangleq& \left\{\begin{array}{ll}
\imath_{P^{(\bar\vlflen)}_{B_{\ell(t)}},W_k}(X_{t}; Y_{k,t}) & \text{for } t \leq L_{\bar\vlflen}m_{\bar\vlflen}\\
  \imath_{P^*,W_k}(X_{t}; Y_{k,t}) & \text{for } t > L_{\bar\vlflen}m_{\bar\vlflen}
\end{array}\right.\\
\overline Z_{k,t} &\triangleq& \sum_{i=1}^t Z_{k,i}.
\end{IEEEeqnarray}
In view of Theorem~\ref{thm:simple_achiev_vlf}, we set for all integers $\bar \vlflen>0$ 
\begin{IEEEeqnarray}{rCl}
    \gamma_{\bar\vlflen} &\triangleq& C\bar  \vlflen - 2\bar \vlflen^{1/3}\log \bar\vlflen\label{eq:vlf_gammadef}\\
    q_{\bar\vlflen} &\triangleq& \frac{\bar \vlflen (\epsilon-\epsilon_{\bar\vlflen}^* - c_1/\bar\vlflen) -1}{\bar\vlflen - 1}\\
    M_{\bar\vlflen} &\triangleq& \lfloor \ee{\gamma_{\bar\vlflen} - \log \bar\vlflen} \rfloor.
\end{IEEEeqnarray}
Thus, we conclude that
\begin{IEEEeqnarray}{rCl}
  \IEEEeqnarraymulticol{3}{l}{q_{\bar\vlflen} + (1-q_{\bar\vlflen}) \Big((M_{\bar\vlflen} - 1)\ee{-\gamma_{\bar\vlflen}} +\epsilon^*(n_B,2^{L_{\bar\vlflen}})}\nonumber\\
  &&\quad\qquad{} + \max_k \pr{\tau_k = \tau_{\text{max}}}\Big)\nonumber\\
 \qquad &\leq& \frac{\bar\vlflen (\epsilon-\epsilon^*_{\bar\vlflen}- c_1/\bar\vlflen) -1}{\bar\vlflen - 1}\nonumber\\
 &&{} + \frac{\bar\vlflen (1-\epsilon + \epsilon^*_{\bar\vlflen}+ c_1/\bar\vlflen)}{\bar\vlflen - 1} \frac{1}{\bar\vlflen} + \epsilon_{\bar\vlflen}^*+ c_1/\bar\vlflen\IEEEeqnarraynumspace\\
  &=&\epsilon.
\end{IEEEeqnarray}
In Appendix~\ref{sec:Emaxtau_n_upper_bound}, we shall prove that there exists an integer  $\bar\vlflen_0\geq 0$ such that, for all $\bar\vlflen \geq \bar\vlflen_0$,
\begin{IEEEeqnarray}{rCl}
  \EBig{\max_k \tau_k} \leq \bar\vlflen.\label{eq:Emaxtau_n}
\end{IEEEeqnarray}
Here, $\tau_k$ is defined by \eqref{eq:tau_k_def_feed} with $\gamma$ replaced by $\gamma_{\bar \vlflen}$.
It then follows that
\begin{multline}
  (1-q_{\bar \vlflen})\Big(\EBig{\max_k \tau_k} + n_B\Big) \\\leq \underbrace{\frac{\bar\vlflen(1-\epsilon+\epsilon_{\bar\vlflen}^* + c_1/\bar\vlflen)}{\bar \vlflen - 1}(\bar \vlflen+ \kappa L_{\bar\vlflen})}_{\triangleq \vlflen_{\bar \vlflen}}.
\end{multline}
We conclude the proof by invoking Theorem~\ref{thm:simple_achiev_vlf} with $q = q_{\bar \vlflen}$, $\gamma = \gamma_{\bar\vlflen}$, and $M = M_{\bar \vlflen}$, which implies that there exists a sequence of $(\vlflen_{\bar\vlflen}, M_{\bar \vlflen},\epsilon)$-VLF codes for all $\bar \vlflen\geq \bar\vlflen_0$ satisfying
\begin{IEEEeqnarray}{rCl}
\log M_{\bar\vlflen} &\geq &C \bar \vlflen - 2\bar\vlflen^{1/3}\log \bar\vlflen - \log \bar\vlflen\\ &=& \frac{\vlflen_{\bar\vlflen} C}{1-\epsilon} + \mathcal{O}(\vlflen_{\bar\vlflen}^{1/3}\log \vlflen_{\bar\vlflen}).
\end{IEEEeqnarray}

Before proceeding to the proofs of \eqref{eq:asymp_vlf_tauk_vlflen} and \eqref{eq:Emaxtau_n}, we first introduce some notation.
It follows from Taylor's theorem and from Lemma~\ref{lem:diffIkzero} that there exist positive convergent sequences $\{\zeta_{\bar\vlflen}^{(1)}\}$ and $\{\zeta_{\bar\vlflen}^{(2)}\}$ such that
\begin{IEEEeqnarray}{rCl}
  \sum_k \eta_k I_k(P_b^{(\bar \vlflen)}) =  C - \bar\vlflen^{-2/3}\zeta^{(b)}_{\bar\vlflen}\label{eq:vlf_zeta_def}
\end{IEEEeqnarray}
for $b\in\{1,2\}$. Let also $\zeta^{(\text{max})}_{\bar\vlflen} \triangleq \max_{b\in \{0,1\}} \zeta^{(b)}_{\bar\vlflen}$ and let $\mu_{b,\bar\vlflen}$ for $b\in\{1,2\}$ be defined as
\begin{IEEEeqnarray}{rCl}
  \mu_{b,\bar\vlflen}\triangleq m_{\bar\vlflen}\left[I_1(P_b^{(\bar\vlflen)}) - I_2(P_b^{(\bar\vlflen)})\right] = \mathcal{O}(\bar\vlflen^{1/3}).
\end{IEEEeqnarray}
  We shall use the following decomposition for $t\geq \bar\vlflen_-$
\begin{IEEEeqnarray}{rCl}
  \overline Z_{k,t} = \overline Z_t  - \eta_{\bar k}(-1)^k \overline E_t.\label{eq:vlf_Z_decomp}
\end{IEEEeqnarray}
Here, 
\begin{IEEEeqnarray}{rCl}
Z_{t} &\triangleq& \sum_k\eta_k Z_{k,t} \\
\overline Z_{t} &\triangleq& \sum_{i=1}^t Z_{i} \\
  \overline E_t &\triangleq& \sum_{i=1}^t (Z_{1,i} - Z_{2,i}).
\end{IEEEeqnarray}
Some observations are in order. It follows from \eqref{eq:vlf_zeta_def} that $\{\overline Z_t-t C + \bar \vlflen^{-2/3} t \zeta_{\bar\vlflen}^{(\text{max})}\}$ is a sub-martingale and that $\{\overline Z_t-t C\}$ is a super-martingale with respect to the filtration $\{\sigma(\overline Z_t)\}_t$ (notice that $\sum_k \eta_k Z_{k,t} \leq C$ by concavity of mutual information and by the definition of $\{\eta_k\}$). Next, observe that $\overline E_t - \overline E_{\bar\vlflen_-}$ is a sum of i.i.d. RVs and that, by applying Lemma~\ref{lem:concentration_overall} as in the proof of Theorem~\ref{thm:asymp_no_variance_cond}, we obtain the following concentration inequality for $\overline E_{\bar\vlflen_-}$
\begin{IEEEeqnarray}{rCl}
    \prbig{|\overline E_{\bar\vlflen_-}| \geq \xi} 
    &\leq& 2\e{- \sqrt{\frac{2}{m_{\bar\vlflen}}} |\xi - \mu_{1,\bar\vlflen} + \mu_{2,\bar\vlflen}|_+ }.\label{eq:vlf_concentration}
\end{IEEEeqnarray}

\subsection{Proof of \eqref{eq:asymp_vlf_tauk_vlflen}}\label{sec:proof_of_tau_k_nplus}
As shown in the chain of inequalities \eqref{eq:prtauk_barnplus}--\eqref{eq:azuma_hoeffding3} in the top of the next page, \eqref{eq:asymp_vlf_tauk_vlflen} readily follows by an application of the Azuma-Hoeffding inequality (see Lemma~\ref{lem:azuma}).\begin{figure*}[!t]
\normalsize
\setcounter{MYtempeqncnt}{\value{equation}}
\setcounter{equation}{286}
\begin{IEEEeqnarray}{rCl}
\max_k \prBig{\tau_k = \bar\vlflen_+}
  &\leq& \max_k \prBig{\overline Z_{k,\bar\vlflen_+-1} < \gamma_{\bar\vlflen}}\label{eq:prtauk_barnplus}\\
&=&  \max_k \prBig{\min_k\Big\{\overline Z_{\bar\vlflen_-} -\eta_k (-1)^k \overline E_{\bar\vlflen_-} + (\overline Z_{k,\bar\vlflen_+-1} - \overline Z_{k,\bar\vlflen_-})\Big\} < \gamma_{\bar\vlflen}}\label{eq:thm7_decomp_used}\\
&\leq& \max_k \prbig{\overline Z_{\bar\vlflen_-} + \min_k\{\overline Z_{k,\bar\vlflen_+-1} - \overline Z_{k,\bar\vlflen_-}\} < \gamma_{\bar\vlflen}+\eta_{\text{max}} |\overline E_{\bar\vlflen_-}|}\label{eq:thm7_decomp_used2}\\
&\leq& \max_k \prbigg{\overline Z_{\bar\vlflen_-}  + \overline Z_{k,\bar\vlflen_+-1} - \overline Z_{k,\bar\vlflen_-} < \gamma_{\bar\vlflen} + \frac{1}{2}\bar\vlflen^{1/3}\log \bar\vlflen } + \mathcal{O}(1/\bar\vlflen)\label{eq:azuma_hoeffding1}\\   
  &\leq& 2 \ee{-\frac{\const \big|(\bar\vlflen_+-1)C-(\bar\vlflen_+-1)\bar\vlflen^{-2/3}\zeta_{\bar\vlflen}^{(\text{max})}-\gamma_{\bar\vlflen}-\frac{1}{2}\bar\vlflen^{1/3}\log \bar\vlflen\big|_+^2}{\bar\vlflen_+-1} }+\mathcal{O}(1/\bar\vlflen)\label{eq:azuma_hoeffding2}\IEEEeqnarraynumspace\\
        &=& \mathcal{O}(1/\bar\vlflen).\label{eq:azuma_hoeffding3}
\end{IEEEeqnarray}

\setcounter{equation}{\value{MYtempeqncnt}}
\hrulefill
\vspace*{4pt}
\end{figure*}\addtocounter{equation}{6} 
Here, \eqref{eq:thm7_decomp_used} follows from the decomposition \eqref{eq:vlf_Z_decomp}, \eqref{eq:thm7_decomp_used2} follows by setting $\eta_{\text{max}} = \max_k \eta_k$, \eqref{eq:azuma_hoeffding1} follows from \eqref{eq:vlf_concentration}, which implies that
\begin{IEEEeqnarray}{rCl}
  \pr{ \eta_{\text{max}} |E_{\bar\vlflen_-}| \geq \frac{1}{2} \bar\vlflen^{1/3} \log \bar\vlflen} \leq \mathcal{O}\farg{\frac{1}{\bar\vlflen}}.\label{eq:vlf_asymp_E_bound}
\end{IEEEeqnarray}
Finally, \eqref{eq:azuma_hoeffding2} follows from the Azuma-Hoeffding inequality (see Lemma~\ref{lem:azuma}) applied to the sub-martingale 
\begin{multline}
\Big\{\overline Z_{\min\{t,\bar\vlflen_-\}} + \indi{t>\bar\vlflen_-}(\overline Z_{k,t}-\overline{Z}_{k,\bar\vlflen_-}) \\-tC + t\bar \vlflen^{-2/3}\zeta^{(\text{max})}_{\bar\vlflen}\Big\}_{t=1}^{\bar\vlflen_+-1}.
\end{multline}

\subsection{Proof of \eqref{eq:Emaxtau_n}}\label{sec:Emaxtau_n_upper_bound}
The idea is to upper-bound $\E{\max_k \tau_k}$ via the following stopping time:
\begin{IEEEeqnarray}{rCl}
  \widetilde \tau \triangleq \inf\{t \geq \bar\vlflen_-: \overline Z_t \geq \widetilde\gamma_{\bar\vlflen}  \text{ or } t = \bar\vlflen_+\}\label{eq:tildetau}
\end{IEEEeqnarray}
where 
\begin{IEEEeqnarray}{rCl}
  \widetilde \gamma_{\bar\vlflen} &\triangleq& \gamma_{\bar\vlflen} + \bar\vlflen^{1/3}\log \bar\vlflen.\label{eq:vlf_tildegamma}
\end{IEEEeqnarray}
Intuitively, since $\widetilde \gamma_{\bar\vlflen} -\gamma_{\bar\vlflen} = \bar\vlflen^{1/3}\log \bar\vlflen$, the stopping time $\widetilde \tau$ is rarely smaller than $\max_k\tau_k$. In addition, the expectation of $\widetilde \tau$ can be upper-bounded by means of Doob's optional stopping theorem. This is simpler than upper-bounding the expectation of the maximum $\max_k \tau_k$ of the two stopping times. 
Formally, by using that $\max\tau_k\leq \vlflen_+$, we upper-bound $\E{\max_k\tau_k}$ in terms of $\E{\widetilde \tau}$ as follows:
\begin{IEEEeqnarray}{rCl}
  \EBig{\max_k\tau_k}\leq \E{\widetilde\tau} + \bar\vlflen_+ \prBig{\widetilde \tau < \max_k\tau_k }.   \label{eq:vlf_proof_Etau_second_term2}
\end{IEEEeqnarray}
In the remaining part of the proof, we demonstrate that there exist constants $c_2$ and $c_3$, which do not depend on $\bar\vlflen$, such that the following upper bounds hold for sufficiently large $\bar\vlflen$
\begin{IEEEeqnarray}{rCl}
  \E{\widetilde \tau} &\leq& \frac{\widetilde \gamma_{\bar \vlflen}}{C} + c_2\label{eq:vlf_asymp_ineq1}\\
  \bar\vlflen_+ \prBig{\widetilde \tau < \max_k\tau_k} &\leq& c_3.\label{eq:vlf_asymp_ineq2}
\end{IEEEeqnarray}
Consequently, we have that
\begin{IEEEeqnarray}{rCl}
  \EBig{\max_k \tau_k} \leq \frac{\widetilde \gamma_{\bar\vlflen}}{C} + c_2 + c_3 \leq \bar\vlflen
\end{IEEEeqnarray}
for all sufficiently large $\bar\vlflen$, as desired.

\paragraph*{Proof of \eqref{eq:vlf_asymp_ineq1}}
Applying Doob's optional stopping theorem\cite[Th.~10.10]{probability_with_martingales} to the sub-martingale $\big\{\overline Z_t - t C + \vlflen^{-2/3}t \zeta^{(\text{max})}_{\bar\vlflen} \big\}_t$ and the stopping time $\widetilde \tau$, we obtain
\begin{IEEEeqnarray}{rCl}
0 \leq   \E{\overline Z_{\widetilde \tau}}- \E{\widetilde \tau}\left(C- \bar\vlflen^{-2/3}\zeta^{(\text{max})}_\vlflen\right).
\end{IEEEeqnarray}
Hence, we conclude that there exists a constant $c_1$ such that, for all sufficiently large $\bar\vlflen$,
\begin{IEEEeqnarray}{rCl}
  \E{\widetilde \tau} \leq \frac{\E{\overline Z_{\widetilde \tau}}}{C}\left(1+  c_1  \bar\vlflen^{-2/3}\right).
\end{IEEEeqnarray}
In order to upper-bound $\E{\overline Z_{\widetilde \tau}}$, we shall use the decomposition
\begin{multline}
 \E{\overline Z_{\widetilde \tau}} \\=  \E{\overline Z_{\widetilde \tau}\indi{\overline Z_{\bar\vlflen_-} < \widetilde\gamma_{\bar\vlflen}}}+\E{\overline Z_{\bar\vlflen_-}\indi{\overline Z_{\bar\vlflen_-} \geq \widetilde\gamma_{\bar\vlflen}}}.\label{eq:decomposition_proof}
\end{multline}
This decomposition holds because \eqref{eq:tildetau} implies that $\widetilde \tau \in\{ \mathbb{\bar n}_-, \ldots, \mathbb{\bar n}_+\}$ and because $\overline Z_{\widetilde \tau} = \overline Z_{\mathbb{\bar n}_-}$ if and only if $\overline Z_{\bar\vlflen_-} \geq \widetilde\gamma_{\bar\vlflen}$.
The first term on the right-hand side of \eqref{eq:decomposition_proof} is trivially upper-bounded by $\widetilde \gamma_{\mathbb{\bar n}}$. The second term is upper-bounded by an application of the following lemma, which relates $\E{\overline Z_{\widetilde \tau}\indi{\overline Z_{\bar\vlflen_-} \geq \xi}}$ to the tail probability of $\overline Z_{\bar\vlflen_-}$. 
\begin{lemma}
  Suppose that $\pr{X\geq\xi} \leq \ee{-a |\xi-b|_+^2}$ for $a>0$, $b\in\mathbb{R}$, and $\xi \in\mathbb{R}$. Then, for all $c\geq b$, we have
  \begin{IEEEeqnarray}{rCl}
    \E{X \indi{X\geq c}} \leq (c  + \sqrt{\pi/a})  \ee{-a |c-b|_+^2}.
  \end{IEEEeqnarray}
  \label{lem:basic_lemma}
\end{lemma}
\begin{IEEEproof}
The result readily follows using integration and the upper bound $Q(x) \leq \e{-x^2/2}$ for all $x\geq 0$.
\end{IEEEproof}

In order to apply Lemma~\ref{lem:basic_lemma}, we need to upper-bound the tail probability $\pr{\overline Z_{\widetilde \tau} \geq \xi}$. To do so, we apply the Azuma-Hoeffding inequality (see Lemma~\ref{lem:azuma}) to the super-martingale $\{\overline Z_{t} - tC\}$ to obtain  for $\xi \geq \bar\vlflen_-C$
\begin{IEEEeqnarray}{rCl}
  \pr{\overline Z_{\bar\vlflen_-} \geq \xi} &\leq& \ee{-\frac{\const |\xi-\bar\vlflen_-C |_+^2}{ \bar\vlflen_-  }}.\label{eq:barZLm_geq_v}
\end{IEEEeqnarray}

Let $a_0$ be a constant not depending on $\vlflen$ such that $|Z_{k,t}| \leq a_0$ with probability one. It follows that there exists a positive constant $c_2$ such that
\begin{IEEEeqnarray}{rCl}
  \IEEEeqnarraymulticol{3}{l}{\E{\overline Z_{\widetilde \tau}}}\nonumber\\
   &=&  \E{\overline Z_{\widetilde \tau}\indi{\overline Z_{\bar\vlflen_-} < \widetilde\gamma_{\bar\vlflen}}}+\E{\overline Z_{\bar\vlflen_-}\indi{\overline Z_{\bar\vlflen_-} \geq \widetilde\gamma_{\bar\vlflen}}}\\
  &\leq& \widetilde\gamma_{\bar\vlflen} +a_0+ (\widetilde \gamma_{\bar\vlflen} + \const \sqrt{\bar\vlflen_-}) \ee{-\frac{\const|\widetilde \gamma_{\bar\vlflen} - \bar\vlflen_- C|_+^2 }{\bar\vlflen_-}}\label{eq:EZtau_upperbound1}\IEEEeqnarraynumspace\\
&\leq& \widetilde\gamma_{\bar\vlflen} + c_2\label{eq:EZtau_upperbound2}
\end{IEEEeqnarray}
for sufficiently large $\bar\vlflen$.
Here, \eqref{eq:EZtau_upperbound1} follows because $|Z_t|\leq a_0$ and from Lemma~\ref{lem:basic_lemma}, and \eqref{eq:EZtau_upperbound2} follows because  $\widetilde \gamma_{\bar\vlflen}-\bar\vlflen_- C = \mathcal{O}(\sqrt{\bar\vlflen}\log\bar\vlflen)$.

\paragraph*{Proof of \eqref{eq:vlf_asymp_ineq2}} 
We need to prove that $\prbig{\widetilde \tau < \max_k\tau_k } = \mathcal{O}(1/\vlflen)$ as $\vlflen \rightarrow \infty$. To establish this, observe that $\widetilde \tau < \max_k\tau_k$  implies $\min_k S_{k,\widetilde \tau}(X^{\widetilde \tau}, Y^{\widetilde \tau}_k; B^L) < \gamma_{\bar\vlflen}$ (see \eqref{eq:tau_k_def_feed}). Now, it follows from $\max_k \tau_k \leq \bar \vlflen_+$, from \eqref{eq:vlf_Skn}, and from  \eqref{eq:vlf_Z_decomp} that
\begin{IEEEeqnarray}{rCl}
  \IEEEeqnarraymulticol{3}{l}{\prBig{\widetilde \tau < \max_k\tau_k }}\nonumber\\
 \quad  &=& \prBig{\widetilde \tau < \max_k\tau_k, \widetilde \tau < \bar \vlflen_+} \\
   &\leq& \prBig{\min_k\Big\{ \overline Z_{\widetilde \tau}- \eta_{\bar k}(-1)^k \overline E_{\widetilde \tau} \Big\}< \gamma_{\bar\vlflen}, \widetilde \tau < \bar \vlflen_+} \label{eq:vlf_proof1}\IEEEeqnarraynumspace\\
  &\leq& \prBig{\min_k\Big\{ \widetilde\gamma_{\bar\vlflen}- \eta_{\bar k}(-1)^k \overline E_{\widetilde \tau} \Big\}< \gamma_{\bar\vlflen}} \label{eq:vlf_proof11}\\
   &\leq& \pr{  \eta_{\text{max}} |\overline E_{\widetilde \tau}| >\widetilde\gamma_{\bar\vlflen}- \gamma_{\bar\vlflen}} \label{eq:vlf_proof2}
\end{IEEEeqnarray}
where  \eqref{eq:vlf_proof11} follows from \eqref{eq:tildetau} because, in combination with $\widetilde\tau < \bar\vlflen_+$, the definition in \eqref{eq:tildetau} implies that $\overline Z_{\widetilde \tau} \geq \widetilde \gamma_{\bar\vlflen}$. We continue from \eqref{eq:vlf_proof2}  as follows
\begin{IEEEeqnarray}{rCl}
\IEEEeqnarraymulticol{3}{l}{\prBig{\widetilde \tau < \max_k\tau_k }}\nonumber\\
 &\leq& \pr{ \eta_{\text{max}} (|\overline E_{\bar\vlflen_-}|+|\overline E_{\widetilde \tau}- \overline E_{\bar\vlflen_-}|) >\bar\vlflen^{1/3}\log \bar\vlflen} \label{eq:vlf_proof3}\\
    &\leq& \pr{  \eta_{\text{max}} |\overline E_{\widetilde \tau}- \overline E_{\bar\vlflen_-}| >\frac{1}{2}\bar\vlflen^{1/3}\log \bar\vlflen} + \mathcal{O}(1/\bar\vlflen) \label{eq:vlf_proof4}\\
        &\leq& \pr{  \eta_{\text{max}} \max_{t\in\{\bar\vlflen_-,\ldots,\bar\vlflen_+\}} |\overline E_{t}- \overline E_{\bar\vlflen_-}| >\frac{1}{2}\bar\vlflen^{1/3}\log \bar\vlflen} \nonumber\\
        &&{}+ \mathcal{O}(1/\bar \vlflen) \label{eq:vlf_proof43}\\
  &\leq&  ( \bar\vlflen_+ - \bar\vlflen_-) \\
  &&{}\times \max_{t\in\{\bar\vlflen_-+ 1,\ldots,\bar\vlflen_+\}} \pr{\left|\frac{\overline E_{t}- \overline E_{\bar\vlflen_-}}{t-\bar\vlflen_-}\right| \geq \frac{\bar\vlflen^{1/3} \log \bar\vlflen}{2 \eta_{\text{max}} (t-\bar\vlflen_-)} }.\label{eq:vlf_union_bound}\IEEEeqnarraynumspace
\end{IEEEeqnarray}
Here,   \eqref{eq:vlf_proof3} follows by the triangle inequality and because $\widetilde\gamma_{\bar\vlflen} -\gamma_{\bar\vlflen} = \bar\vlflen^{1/3}\log \bar\vlflen$; \eqref{eq:vlf_proof4}  follows from \eqref{eq:vlf_asymp_E_bound}; finally, \eqref{eq:vlf_union_bound} follows from the union bound. We obtain the desired result  by applying Hoeffding inequality to $\overline E_t - \overline E_{\bar\vlflen_-}$ (indeed, $\overline E_t - \overline E_{\bar\vlflen_-}$ is a sum of bounded i.i.d. RVs with zero mean):
\begin{IEEEeqnarray}{rCl}
\IEEEeqnarraymulticol{3}{l}{\prBig{\widetilde \tau < \max_k\tau_k } } \nonumber\\
   \quad&\leq& 2( \bar\vlflen_+ - \bar\vlflen_-) \max_{t\in\{\bar\vlflen_-+ 1,\ldots,\bar\vlflen_+\}} \ee{- \frac{\const\bar\vlflen^{2/3}\log^2\bar\vlflen}{t-\bar\vlflen_-} }\label{eq:vlf_hoeffding}\IEEEeqnarraynumspace\\
  &\leq& \mathcal{O}(\sqrt{\bar\vlflen}\log \bar\vlflen) \ee{-\const  \bar\vlflen^{1/6}\log \bar\vlflen}\\
  &=& \mathcal{O}\farg{\frac{1}{\bar\vlflen}}.
\end{IEEEeqnarray}
We conclude that there exists a positive constant $c_3$ such that \eqref{eq:vlf_asymp_ineq2} holds.

\section{Proof of Lemma~\ref{lem:concentration_overall}}\label{app:concentration_proof}
For convenience, we set $\overline \mu\triangleq \mu_1-\mu_2$ and $\alpha \triangleq c\sqrt{\beta}$. We shall prove the desired result in \eqref{eq:concentration1} by mathematical induction. To do so, we start by using \eqref{eq:stab_lem_Xellb_ineq1}--\eqref{eq:stab_lem_Yell}  to prove the base case; namely that $\pr{Y_1 \geq v} \leq \e{-\alpha(v-\overline \mu)}$ and that $\pr{Y_1 \leq v} \leq \e{-\alpha(-v-\overline \mu)}$ for all $v\in\mathbb{R}$. Then, we prove the inductive step; namely that the statement $\pr{Y_{\ell-1}\geq v} \leq \e{-\alpha (v-\overline \mu)}$ implies $\pr{Y_{\ell}\geq v} \leq \e{-\alpha (v-\overline \mu)}$ for all integers $\ell\geq 2$ and $v \in\mathbb{R}$. The inductive step hinges on the assumptions in \eqref{eq:stab_lem_Xellb_ineq1}--\eqref{eq:concentration_cond}. Finally, \eqref{eq:concentration1} follows by applying an analogous argument to $\pr{Y_{\ell}\leq -v}$ and by mathematical induction. 

\paragraph*{Initial step} By using \eqref{eq:stab_lem_Xellb_ineq1}--\eqref{eq:stab_lem_Xellb_ineq2} and \eqref{eq:stab_lem_Yell}, we have
\begin{IEEEeqnarray}{rCl}
\pr{Y_1 \geq v} &=& \pr{X^{(2)}_1 \geq v}\\
&\leq& \e{- \beta |v-\mu_2|_+^2  }\\
&\leq& \e{- \alpha (v-\mu_2) + \frac{\alpha^2}{4\beta}  }\label{eq:last_ineq_concentration0}\\
&\leq& \e{- \alpha v}\label{eq:last_ineq_concentration}\\
&\leq& \e{- \alpha (v-\overline \mu)}.\label{eq:last_ineq_concentration2}
\end{IEEEeqnarray}
Here, \eqref{eq:last_ineq_concentration0} follows because $-\beta |x|_+^2+\alpha x - \alpha^2/(4\beta)  \leq 0$ for all $x\in\mathbb{R}$, \eqref{eq:last_ineq_concentration} follows because $\alpha^2/(4\beta) = c^2/4 \leq \sqrt{\pi}\e{c^2/4} =\frac{\alpha}{c}\sqrt{\pi/\beta}\e{c^2/4} \leq -\alpha\mu_2/c\leq -\alpha \mu_2$. Using the similar inequality $\alpha^2/(4\beta) \leq \alpha \mu_1$, we also have that
\begin{IEEEeqnarray}{rCl}
  \pr{Y_1 \leq v} &=& \pr{X^{(2)}_1 \leq v}\\
  &\leq& \e{-\beta |\mu_2 - v|_+^2}\label{eq:last_ineq_concentration_0}\\
  &\leq& \e{-\alpha(\mu_2 - v) + \frac{\alpha^2}{4\beta}}\\
  &\leq& \e{-\alpha(-v-\overline \mu)}\label{eq:base_step2}.
\end{IEEEeqnarray}

\paragraph*{Inductive step} We need to show that $\pr{Y_{\ell-1}\geq v} \leq \e{-\alpha (v-\overline \mu)}$ implies that $\pr{Y_{\ell}\geq v} \leq \e{-\alpha (v-\overline \mu)}$ for all integers $\ell\geq 2$. Suppose that 
\begin{IEEEeqnarray}{rCl}
\pr{Y_{\ell-1}\geq v} \leq \e{-\alpha (v-\overline \mu)}\label{eq:stab_proof_hyp}
\end{IEEEeqnarray}
 for some integer $\ell\geq 2$. Then, by relating $Y_{\ell-1}$ and $Y_{\ell}$ using \eqref{eq:stab_lem_Yell} and by applying \eqref{eq:stab_lem_Xellb_ineq1}--\eqref{eq:stab_lem_Xellb_ineq2}, we obtain
\begin{IEEEeqnarray}{rCl}
\IEEEeqnarraymulticol{3}{l}{\pr{Y_\ell\geq v}}\nonumber\\
&=& \EBig{\indi{Y_{\ell-1}\geq 0}\pr{X^{(2)}_\ell \geq v-Y_{\ell-1}|Y_{\ell-1}}\nonumber\\
&&{}\quad+\indi{Y_{\ell-1}< 0}\pr{X^{(1)}_\ell \geq v-Y_{\ell-1}|Y_{\ell-1}}}\\
&\leq& \EBig{\indi{Y_{\ell-1}\geq 0}\ee{- \beta \left|v-Y_{\ell-1}-\mu_2\right|_+^2}\nonumber\\
&&{}\quad+\indi{Y_{\ell-1}< 0}\ee{- \beta |v-Y_{\ell-1}-\mu_1|_+^2}}.\label{eq:stab_proof_Yell1}\IEEEeqnarraynumspace
\end{IEEEeqnarray}
To further upper-bound the right-hand side of \eqref{eq:stab_proof_Yell1}, we maximize over all distributions $Y_{\ell-1}$ satisfying \eqref{eq:stab_proof_hyp}:
\begin{IEEEeqnarray}{rCl}
\IEEEeqnarraymulticol{3}{l}{\pr{Y_\ell\geq v}}\nonumber\\
&\leq& \sup_{\substack{Z:\\ \pr{Z \geq v} \leq \ee{-\alpha (v- \overline \mu)}}}\nonumber\\
&& \EBig{\indi{Z\geq 0}\ee{-\beta |v-Z-\mu_2|_+^2} \nonumber\\
&&{}\qquad\qquad\qquad+\indi{Z< 0}\ee{- \beta\left|v-\mu_1\right|_+^2}}\IEEEeqnarraynumspace\label{eq:second_last_lem_eq}\\
&\leq& \sup_{\substack{Z:\\ \pr{Z \geq v} \leq \ee{-\alpha (v- \overline \mu)}}}\EBig{\ee{-\beta |v-Z-\mu_2|_+^2}} \label{eq:second_last_lem_eq2}\\
&=& \E{\ee{-\beta |v-\overline Z-\mu_2|_+^2}}.\label{eq:last_lem_eq}
\end{IEEEeqnarray}
Here, \eqref{eq:second_last_lem_eq2} follows since $\ee{-\beta|v-\mu_1|^2_+} \leq \ee{-\beta|v-z-\mu_2|_+^2}$ for all $z\geq \overline \mu$ and because $Z$ satisfying $\pr{Z\geq \overline \mu} = 1$ is a feasible solution to the maximization problem. Moreover, the RV $\overline Z$ has cumulative distribution function $F_{\overline Z}(z) = 1-\e{-\alpha|z-\overline \mu|_+}$ and \eqref{eq:last_lem_eq} thus follows because $\overline Z$ stochastically dominates all RVs $Z$ satisfying $\pr{Z\geq v} \leq \e{-\alpha (v- \overline \mu)}$ and because the function 
\begin{IEEEeqnarray}{rCl}
f(z) &\triangleq& \ee{-\beta|v-z-\mu_2|_+^2}
\end{IEEEeqnarray}
is monotonically nondecreasing in $z$, which imply that $\E{f(\overline Z)} \geq \E{f(Z)}$. 
  
To establish the inductive step, it remains to show that $\E{\ee{-\beta |v-\overline Z-\mu_2|_+^2}} \leq \e{-\alpha(v-\overline \mu)}$ for $v \in\mathbb{R}$. Observe that this inequality is trivially satisfied when $\overline \mu \geq v$ implying that we can focus only on the case $v \geq \overline \mu$.
Now, for $v\geq \overline \mu$, we have
\begin{IEEEeqnarray}{rCl}
 \IEEEeqnarraymulticol{3}{l}{ \E{\ee{- \beta |v-\overline Z-\mu_2|_+^2}} }\nonumber\\
 \qquad&\leq&  \alpha \int_{\overline \mu}^{v-\mu_2} \e{-\beta (v-\mu_2-z)^2-\alpha (z-\overline \mu)} \dd z \nonumber\\
 &&{} + \alpha \int_{v-\mu_2}^{\infty} \e{-\alpha (z-\overline \mu)} \dd z\label{eq:only_holds_for_vxy}\\
    &\leq&  \alpha \int_{-\infty}^{\infty} \e{-\left(\sqrt{\beta}(z -v+\mu_2)+ \frac{\alpha}{2\sqrt{\beta}} \right)^2 - \alpha(v-\mu_2-\overline \mu) + \frac{\alpha^2}{4\beta}} \dd z \nonumber\\
    &&{}+ \e{-\alpha(v-\mu_2-\overline \mu )}.\label{eq:algebraic_iden}
  \end{IEEEeqnarray}
  Here, \eqref{eq:only_holds_for_vxy} holds because $v-\mu_2  \geq \overline \mu $ (recall that $\mu_2 < 0$) and \eqref{eq:algebraic_iden} follows from the algebraic identity
\begin{multline}
\beta (v-\mu_2-z)^2 +\alpha (z-\overline \mu) \\
= \left(\sqrt{\beta}(z-v+\mu_2) + \frac{\alpha}{2\sqrt{\beta}} \right)^2 +\alpha(v-\mu_2-\overline \mu) - \frac{\alpha^2}{4\beta}.
\end{multline} 
Now, by algebraic manipulations, we obtain from \eqref{eq:algebraic_iden}
  \begin{IEEEeqnarray}{rCl}  
  \IEEEeqnarraymulticol{3}{l}{\E{\ee{- \beta |v-\overline Z-\mu_2|_+^2}}}\nonumber\\
      &\leq&  \alpha \e{\frac{\alpha^2}{4\beta}} \e{-\alpha(v-\mu_2-\overline \mu)} \int_{-\infty}^{\infty} \e{-\frac{1}{2} (\sqrt{2\beta}z)^2  } \dd z + \e{-\alpha(v-\mu_2-\overline \mu )}\IEEEeqnarraynumspace\\
 &=&  \left(1+\frac{\sqrt{\pi}\alpha \e{\frac{\alpha^2}{4\beta}} }{\sqrt{\beta }} \frac{1}{\sqrt{2\pi}} \int_{-\infty}^{\infty} \e{-\frac{1}{2} \xi^2  }  \dd \xi\right) \e{-\alpha(v-\mu_2-\overline \mu )}\\
&=&  \left(1+\frac{\sqrt{\pi}\alpha \e{\frac{\alpha^2}{4\beta}} }{\sqrt{\beta }}\right) \e{\alpha \mu_2}\e{-(v-\overline \mu )}\\
&\leq& \e{-\alpha\left(v-\overline \mu\right) }\label{eq:last_ineq2}
\end{IEEEeqnarray}
where \eqref{eq:last_ineq2} follows from the inequality $\log(1+z)\leq z$ and \eqref{eq:concentration_cond} because
 \begin{IEEEeqnarray}{rCl}
  1+\frac{\sqrt{\pi}\alpha \e{\frac{\alpha^2}{4\beta}} }{\sqrt{\beta }} 
  &\leq& \ee{\frac{\sqrt{\pi}\alpha \e{\frac{\alpha^2}{4\beta}} }{\sqrt{\beta }} }\\
  &=&\ee{\frac{\sqrt{\pi}\alpha\e{\frac{c^2}{4}} }{\sqrt{\beta }} }\\
  &\leq& \ee{ -\alpha \mu_2 }.
 \end{IEEEeqnarray}
Combining \eqref{eq:last_lem_eq} and \eqref{eq:last_ineq2}, we establish the inductive step, i.e., that $\pr{Y_{\ell-1}\geq v} \leq \e{-\alpha (v-\overline \mu)}$ implies $\pr{Y_{\ell}\geq v} \leq \e{-\alpha (v-\overline \mu)}$ for all integers $\ell\geq 2$ and $v \in\mathbb{R}$. An analogous argument shows that $\pr{Y_{\ell-1}\leq -v} \leq \e{-\alpha (v-\overline \mu)}$ implies that $\pr{Y_{\ell}\leq -v} \leq \e{-\alpha (v-\overline \mu)}$ for all integers $\ell\geq 2$ and $v \in\mathbb{R}$.

\bibliographystyle{IEEEtran}
\bibliography{cm_dmbc_feedback}

\begin{IEEEbiographynophoto}{Kasper Fløe Trillingsgaard}
 (S'12) received his B.Sc. degree in electrical engineering, his M.Sc. degree in wireless communications, and his Ph.D. degree in electrical engineering from Aalborg University, Denmark, in 2011, 2013, and 2017, respectively. He is currently a postdoctoral researcher at the same institution. He was a visiting student at New Jersey Institute of Technology, NJ, USA, in 2012 and at Chalmers University of Technology, Sweden, in 2014. His research interests are in the areas of information and communication theory. 
\end{IEEEbiographynophoto}

\begin{IEEEbiographynophoto}
{Wei Yang}(S'09--M'15) received the B.E. degree in communication engineering and M.E. degree in communication and information systems from the Beijing University of Posts and Telecommunications, Beijing, China, in 2008 and 2011, and the Ph.D. degree in Electrical Engineering from Chalmers University of Technology, Gothenburg, Sweden, in 2015. In the summers of 2012 and 2014, he was a visiting student at the Laboratory for Information and Decision Systems, Massachusetts Institute of Technology, Cambridge, MA. From 2015 to 2017, he was a postdoctoral research associate at Princeton University, Princeton, NJ. In Sep. 2017, he joined Qualcomm Research, San Diego, CA, where he is now a senior engineer.
\end{IEEEbiographynophoto}

\begin{IEEEbiographynophoto}{Giuseppe Durisi}
(S'02--M'06--SM'12) received the Laurea degree summa cum laude and the Doctor degree both from Politecnico di Torino, Italy, in 2001 and 2006, respectively.
From 2006 to 2010 he was a postdoctoral researcher at ETH Zurich, Zurich, Switzerland.
In 2010, he joined Chalmers University of Technology, Gothenburg, Sweden, where he is now professor and co-director of Chalmers information and communication technology Area of Advance.

Dr. Durisi is a senior member of the IEEE. He is the recipient of the 2013 IEEE ComSoc Best Young Researcher Award for the Europe, Middle East, and Africa Region, and is co-author of a paper that won a ``student paper award" at the 2012 International Symposium on Information Theory, and of a paper that won the 2013 IEEE Sweden VT-COM-IT joint chapter best student conference paper award.  In 2015, he joined the editorial board of the \textsc{IEEE Transactions on Communications} as associate editor.
From 2011 to 2014, he served as publications editor for the \textsc{IEEE Transactions on Information Theory}. His research interests are in the areas of communication theory, information theory, and machine learning.
\end{IEEEbiographynophoto}

\begin{IEEEbiographynophoto}{Petar Popovski}
 (S’97–A’98–M’04–SM’10–F’16) is a Professor of Wireless Communications with Aalborg University. He received the Dipl. Ing. degree in electrical engineering and the Magister Ing. degree in communication engineering from the "Sts. Cyril and Methodius" University, Skopje, Republic of Macedonia, in 1997 and 2000, respectively, and the Ph.D. degree from Aalborg University, Denmark, in 2004. He has over 300 publications in journals, conference proceedings, and edited books. He holds over 30 patents and patent applications. He received an ERC Consolidator Grant (2015), the Danish Elite Researcher award (2016), the IEEE Fred W. Ellersick prize (2016), and the IEEE Stephen O. Rice prize (2018). He is currently a Steering Committee Member of IEEE SmartGridComm and previously served as a Steering Committee Member of the \textsc{IEEE Internet of Things Journal}. He is also an Area Editor of the \textsc{IEEE Transactions on Wireless Communications}. His research interests are in the area of wireless communication and networking, and communication/information theory.
\end{IEEEbiographynophoto}

\end{document}